\documentclass[preprint,12pt]{elsarticle}

\usepackage{amssymb}
\usepackage{amsmath}

\usepackage{lineno}

\biboptions{compress}

\usepackage[table]{xcolor}
\usepackage{multirow}
\usepackage[scriptsize,tight,figbotcap]{subfigure}
\usepackage[cm]{fullpage}

\usepackage{bm}   
\usepackage{doi}
\hypersetup{
    colorlinks,
    linkcolor={red!50!black},
    citecolor={blue!50!black},
    urlcolor={blue!80!black}
}

\journal{Journal of Non-Newtonian Fluid Mechanics}

\begin{document}
\newcommand{\vf}[1]{\bm{\mathrm{#1}}}
\newcommand{\pd}[2]{\frac{\partial #1}{\partial #2}}

\begin{frontmatter}



\title{Thixotropic flow past a cylinder}


\author[dms,oc]{Alexandros Syrakos\corref{cor1}}
\ead{syrakos.alexandros@ucy.ac.cy}

\author[dms,oc]{Georgios C. Georgiou}
\ead{georgios@ucy.ac.cy}

\author[dmme]{Andreas N. Alexandrou}
\ead{andalexa@ucy.ac.cy}

\cortext[cor1]{Corresponding author}

\address[dms]{Department of Mathematics and Statistics, University of Cyprus, PO Box 20537, 1678 
Nicosia, Cyprus} \address[oc]{Oceanography Centre, University of Cyprus, PO Box 20537, 1678 Nicosia, 
Cyprus}
\address[dmme]{Department of Mechanical and Manufacturing Engineering, University of Cyprus, PO Box 
 20537, 1678 Nicosia, Cyprus}
 
\begin{abstract}
We study the flow of a thixotropic fluid around a cylinder. The rheology of the fluid is described 
by means of a structural viscoplastic model based on the Bingham constitutive equation, regularised 
using the Papanastasiou regularisation. The yield stress is assumed to vary linearly with the 
structural parameter, which varies from zero (completely broken structure) to one (fully developed 
skeleton structure), following a first-order rate equation which accounts for material structure 
break-down and build-up. The results were obtained numerically using the Finite Element Method. 
Simulations were performed for a moderate Reynolds number of 45, so that flow recirculation is 
observed behind the cylinder, but vortex shedding does not occur. The effects of the Bingham number 
and of the thixotropy parameters are studied. The results show that the viscous character of the 
flow can be controlled within certain limits through these parameters, despite the fact that the 
Reynolds number is fixed.
\end{abstract}


\end{frontmatter}

This is the accepted version of the article published in: Journal of Non-Newtonian Fluid Mechanics 
220 (2015) 44--56, \doi{10.1016/j.jnnfm.2014.08.008}

\textcopyright 2017. This manuscript version is made available under the CC-BY-NC-ND 4.0 license 
\url{http://creativecommons.org/licenses/by-nc-nd/4.0/}


\section{Introduction}
\label{sec: introduction}

\begin{figure}[tb]
\centering
 \noindent\makebox[\textwidth]{
  \subfigure {\includegraphics[scale=1.0]{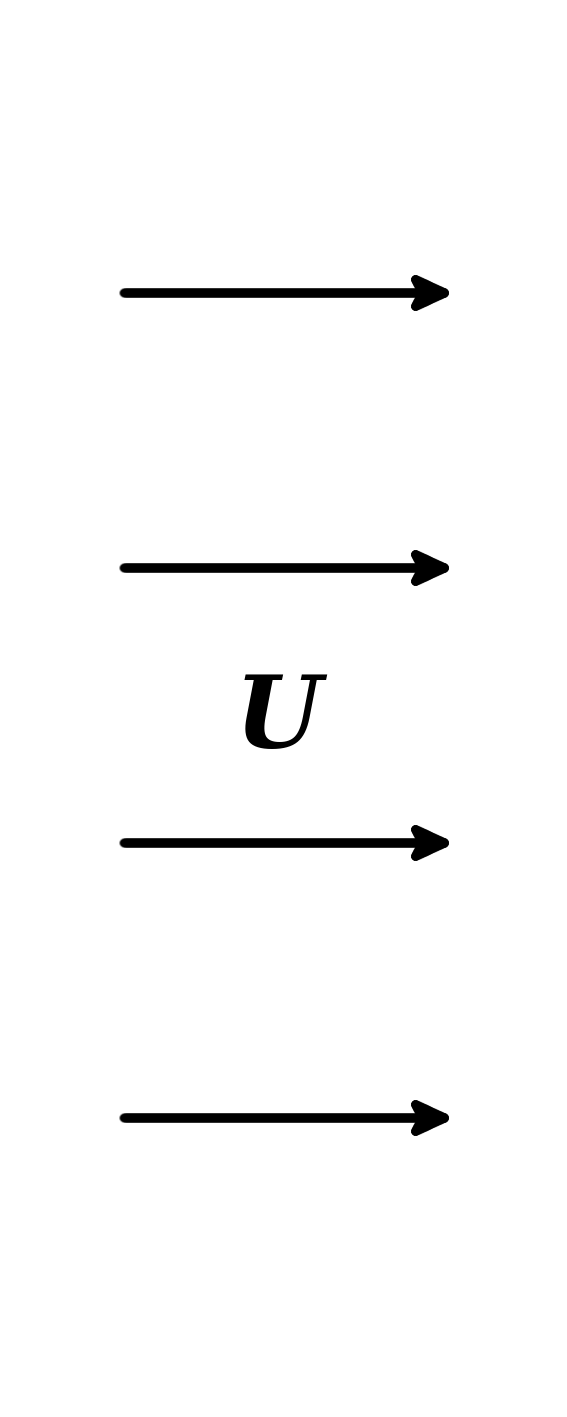}}
  \addtocounter{subfigure}{-1}
  \subfigure[{$Re = 1$}] {\label{sfig: Newtonian Re=1}
   \includegraphics[scale=1.0]{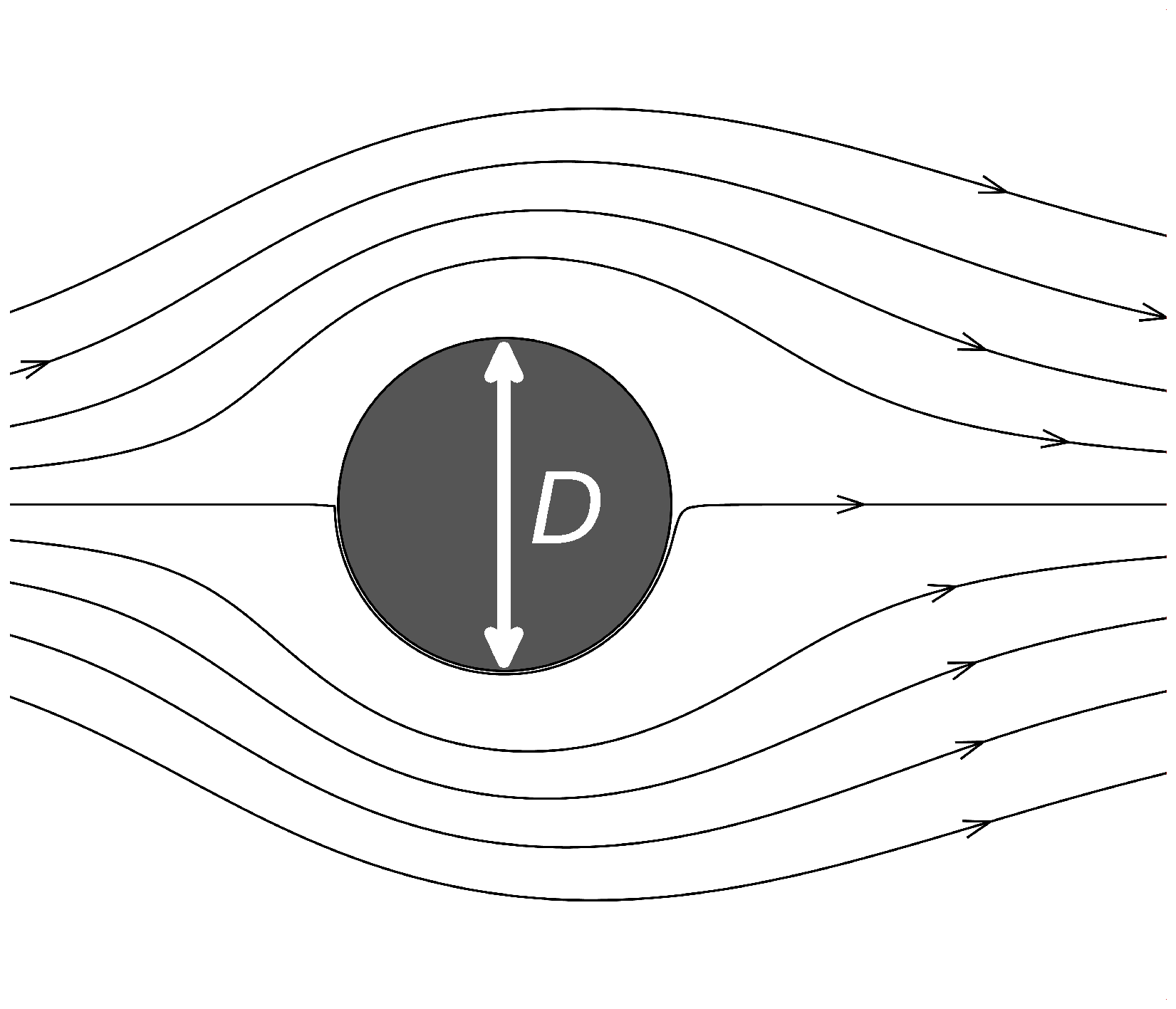}}
  \subfigure[{$Re = 20$}] {\label{sfig: Newtonian Re=20}
   \includegraphics[scale=1.0]{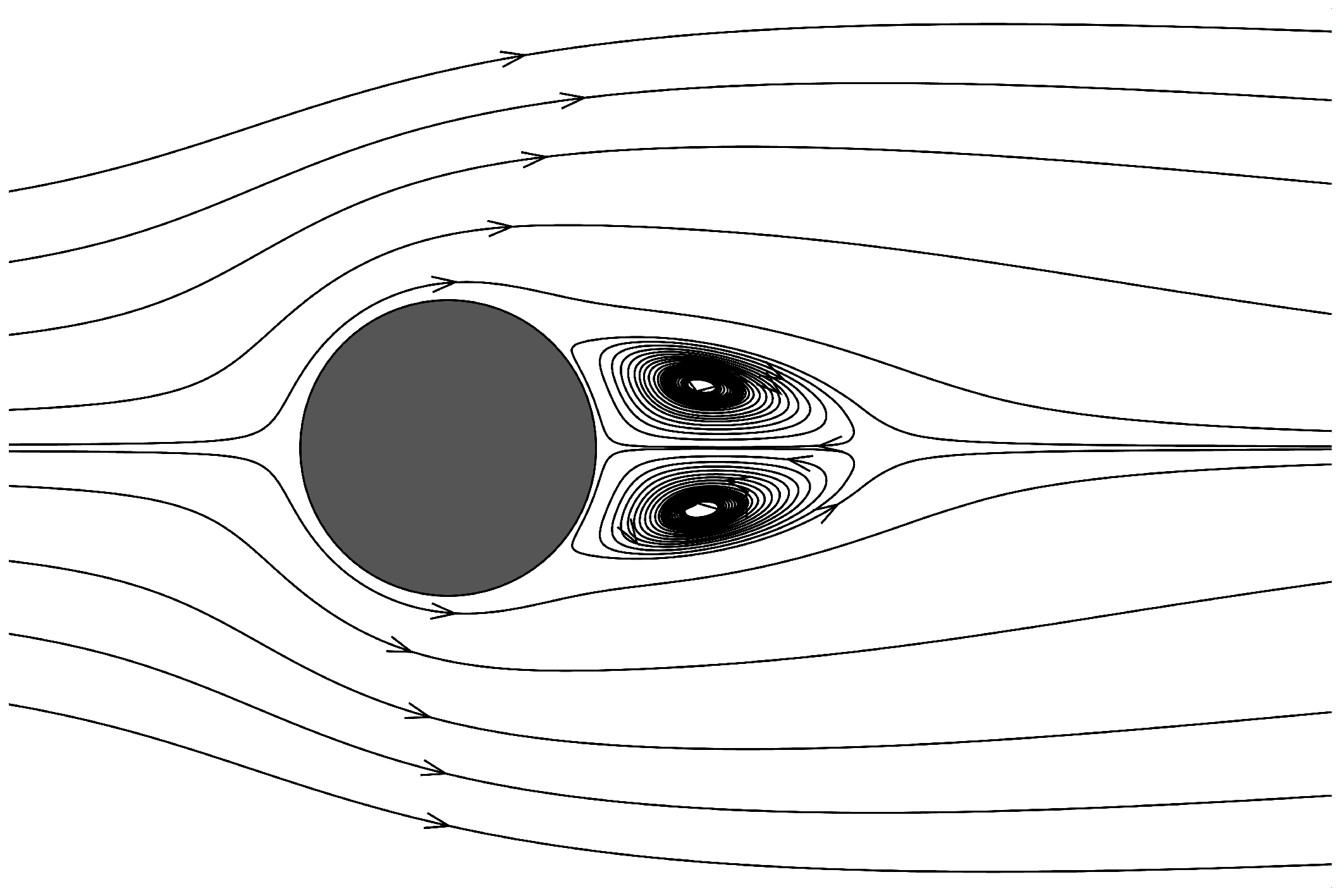}}
  \subfigure[{$Re = 100$}] {\label{sfig: Newtonian Re=100}
   \includegraphics[scale=1.0]{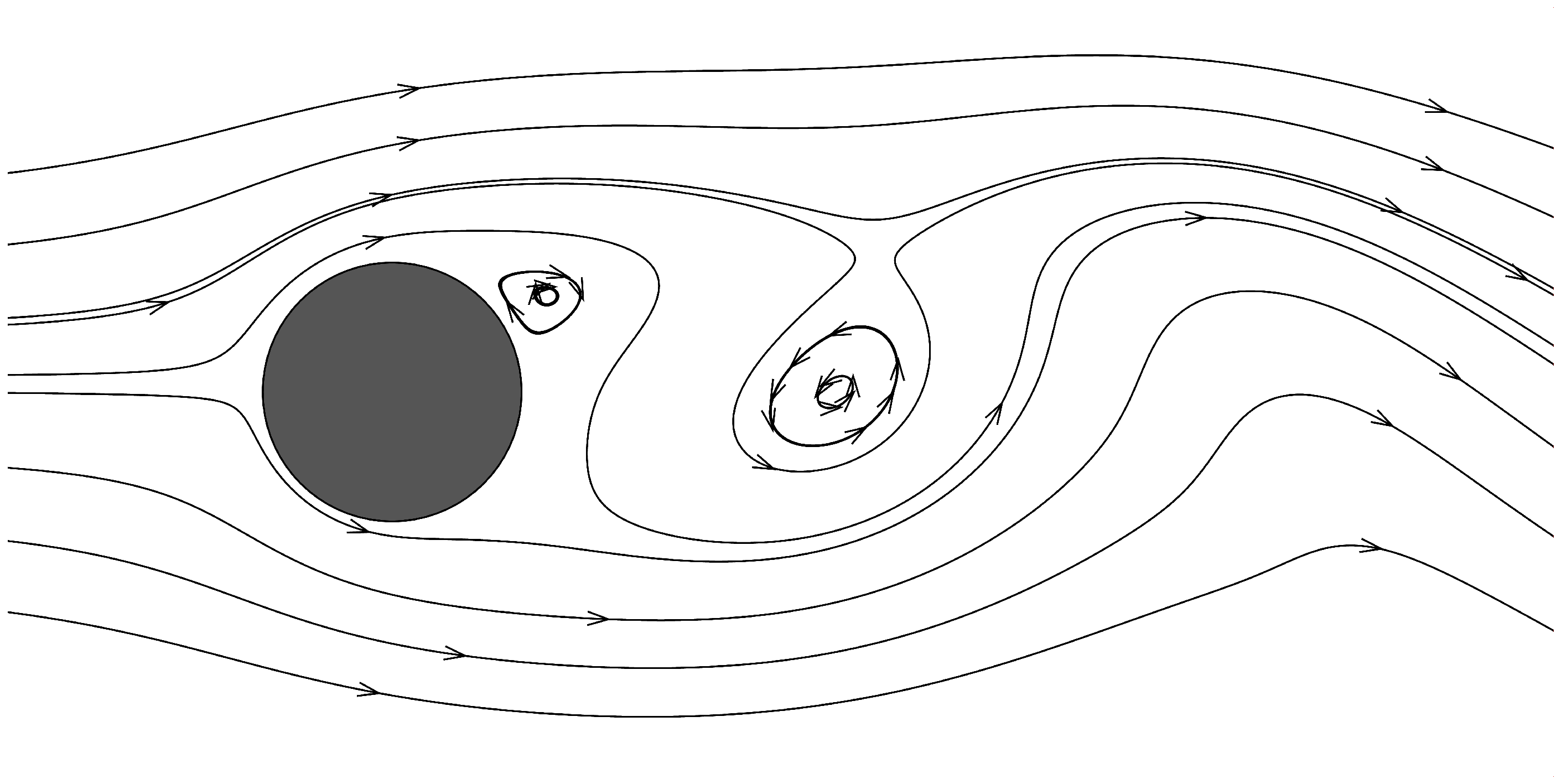}}
 }
 \caption{Individual streamlines (not equally spaced) plotted for three different flow fields 
corresponding to different flow regimes. From left to right: \subref{sfig: Newtonian Re=1} $Re = 1$, 
no recirculation, steady flow; \subref{sfig: Newtonian Re=20} $Re = 20$, flow separation and 
recirculation behind the cylinder, steady flow; and \subref{sfig: Newtonian Re=100} $Re = 100$, 
periodic flow, vortex shedding behind the cylinder (the streamlines correspond to a particular 
snapshot of the transient flow). The results were obtained with an in-house finite volume code, 
different from the one used in the rest of this paper, described in \cite{Syrakos_06b}.}
 \label{fig: newtonian regimes}
\end{figure}

The flow past a cylinder is of high theoretical and practical importance, and thus the Newtonian 
flow case has been studied extensively. The main results are included in a review paper by 
Williamson \cite{Williamson_96}. The characteristics of the flow field strongly depend on the 
Reynolds number, which is defined as $Re \equiv \rho U D / \mu$ where $\rho$ and $\mu$ are the 
fluid density and viscosity respectively, $U$ is the main stream velocity, and $D$ is the cylinder 
diameter (see Fig.\ \ref{fig: newtonian regimes}). Up to a Reynolds number of 5, the flow is 
characterised as creeping and no separation is exhibited (Fig.\ \ref{sfig: Newtonian Re=1}). 
However, in the range $5 \leq Re \leq 49$, flow separation is observed and a symmetric pair of 
recirculation bubbles appears behind the cylinder (Fig.\ \ref{sfig: Newtonian Re=20}), which 
increase in size with the Reynolds number. A further increase  in the Reynolds number ($49 \leq Re 
\leq 190$) causes the flow to become unsteady, with periodic vortex shedding behind the cylinder 
(Fig.\ \ref{sfig: Newtonian Re=100}). If the Reynolds number is further increased then transition 
to turbulence begins to take place. However, in the present study we limit ourselves to the laminar 
flow regime.

Viscoplastic flow past a cylinder is less studied, but there do exist a number of published results. 
Most of these results concern creeping flow, and are relatively recent, with the exception of the 
early theoretical work of Adachi and Yoshioka \cite{Adachi_73}. About a decade ago computational 
studies began to appear on creeping viscoplastic flow around a cylinder. There are two basic 
configurations: flow in an infinite medium \cite{Deglo_03, Tokpavi_08} and flow between two parallel 
plates \cite{Zisis_02, Roquet_03, Mitsoulis_04}. The two configurations are equivalent when the 
yielded zone which surrounds the cylinder does not extend up to the plates. There exist also studies 
of creeping viscoplastic flow around cylinders of non-circular cross section \cite{Putz_10, 
Nirmalkar_12}, but this is beyond the scope of the present work. Flow at non-zero Reynolds numbers 
was studied recently by Mossaz et al.\ \cite{Mossaz_10, Mossaz_12_com}, and to the best of our 
knowledge these are the only published results. 

Usually, viscoplasticity is accompanied by other rheological phenomena such as thixotropy and/or 
viscoelasticity. Such behaviour was exhibited by the materials used in the few available 
experimental studies of viscoplastic flow around a cylinder \cite{Tokpavi_09, Mossaz_12_exp}. A 
recent computational study which includes the effects of viscoplasticity, thixotropy and 
viscoelasticity was conducted by Fonseca et al.\ \cite{Fonseca_13}. That study is limited to 
creeping flow. The aim of the present work is to isolate the effects of thixotropy and yield stress 
in the case of non-zero Reynolds number flows which may exhibit recirculation bubbles behind the 
cylinder.

The rest of the paper is organised as follows: In Section \ref{sec: equations} the equations which 
govern the flow are presented. In Section \ref{sec: method} the domain, boundary conditions and 
finite element solution method are described. In Section \ref{sec: results} the results are 
presented, and the paper ends with some conclusions in Section \ref{sec:conclusions}.

\section{Governing equations}
\label{sec: equations}

The flow is governed by the continuity and momentum equations for incompressible, constant-density  
flow:
\begin{equation} \label{eq: continuity}
 \nabla \cdot \vf{u} \;=\; 0
\end{equation}
\begin{equation} \label{eq: momentum}
 \rho \left( \pd{\vf{u}}{t} \;+\; \vf{u} \cdot \nabla \vf{u} \right) \;=\; -\nabla p \;+\; \nabla 
 \cdot \vf{\tau}
\end{equation}
where $\vf{u} = (u,v)$ is the velocity vector, $p$ is the pressure, $\rho$ is the density (a 
constant) and $\vf{\tau}$ is the deviatoric stress tensor.

In the present work, the model material under study is assumed to have an internal structure which 
governs the behaviour of the material and can change over time. The components of the material form 
a structure, which is capable of withstanding loads up to a limit without permitting flow. Flow 
occurs when the magnitude of the deviatoric stress tensor $\tau \equiv (\frac{1}{2}\vf{\tau} \!:\! 
\vf{\tau})^{1/2}$ exceeds a threshold value, the yield stress $\tau_0$. The yield stress is assumed 
to depend on the current state of the structure of the material. It is also assumed that, when the 
material flows, i.e. the yield stress is exceeded, then the structure starts to break down. This is 
a time-dependent process, with the rate of breakdown being proportional to the rate of shear, and 
will lead to partial or complete breakdown of the structure. Partial breakdown of the structure 
reduces the ability of the material to sustain loads, i.e. it reduces the yield stress. The 
structure also has a tendency to recover with time, thus increasing the yield stress. This tendency 
is independent of the shear rate. These two mechanisms compete with each other.

It is customary to associate the state of the structure with a dimensionless variable, denoted 
$\lambda$ here, which equals 1 when the material is completely structured and 0 when the structure 
is completely broken. All intermediate states correspond to $0 < \lambda < 1$. Our model material is 
assumed to follow a simple relationship between the structure parameter $\lambda$ and the yield 
stress:
\begin{equation} \label{eq: tau(lambda)}
 \tau_0(x,y,t) \;=\; \lambda(x,y,t) \, \tau_y
\end{equation}
where $\tau_0(x,y,t)$ is the current yield stress of the material at point $(x,y)$ at time $t$, 
$\lambda(x,y,t)$ is the instantaneous value of the structure parameter at that point, and $\tau_y$ 
is the maximum yield stress, when the material is fully structured ($\lambda = 1$). According to the 
above equation, the thixotropic material is no longer viscoplastic when the structure is completely 
broken ($\lambda = 0 \Rightarrow \tau_0 = 0$). More complicated constitutive relations can be 
constructed which assume that the rest of the rheological parameters are also functions of 
$\lambda$; indeed several such relations have been suggested in the literature, and the review paper 
by Mewis and Wagner \cite{Mewis_2009} contains a list with many of them. By using the simple Eq.\ 
(\ref{eq: tau(lambda)}) we isolate the effects of thixotropy on the most important parameter, the 
yield stress.

At stresses higher than the yield stress, when the material flows, it is assumed that a linear 
relationship is exhibited between the stress and the rate of strain, i.e.\ the material is assumed 
to be of the Bingham type:
\begin{equation} \label{eq: constitutive Bingham}
 \vf{\tau} \;=\; \left( \frac{\tau_0}{\dot{\gamma}} \;+\; \mu \right) \vf{\dot{\gamma}} \;=\; 
                 \left( \frac{\lambda\,\tau_y}{\dot{\gamma}} \;+\; \mu \right) \vf{\dot{\gamma}}
\end{equation}
where $\mu$ is the plastic viscosity, $\vf{\dot{\gamma}} \equiv \nabla \vf{u} + (\nabla \vf{u})^T$ 
is the rate-of-strain tensor, and $\dot{\gamma} \equiv (\frac{1}{2}\vf{\dot{\gamma}} \!:\! 
\vf{\dot{\gamma}})^{1/2}$ is its magnitude. This constitutive equation applies only where the 
material flows, whereas in unyielded regions ($\tau \leq \tau_0$) there is no deformation. This is a 
simplified version of the models of Tiu and Boger \cite{Tiu_1974} and Houska \cite{Houska_1980, 
Sestak_1987}, who assumed a fluid of the Herschel-Bulkley type and that both the yield stress and 
the plastic viscosity are linear or affine functions of $\lambda$.

As already mentioned, the evolution of the structure is due to two mechanisms: a shear-driven 
breakdown mechanism, and a recovery mechanism. In the present work, this evolution is described by 
the following simple equation, which was originally proposed by Moore \cite{Moore_1959} and is a 
simplified version of the model used in \cite{Alexandrou_09}:
\begin{equation} \label{eq: lambda evolution lagrangian}
 \frac{D\lambda}{Dt} \;=\; \underbrace{\alpha\,(1-\lambda)}_{\text{rate of recovery}} \;-\; 
                           \underbrace{\beta\,\lambda\,\dot{\gamma}}_{\text{rate of breakdown}}
\end{equation}
where $D\lambda/Dt$ is the rate of change of $\lambda$ within a particle of the material which moves 
with the flow. The recovery term is proportional to $(1-\lambda)$, so that recovery ceases when the 
structure has fully recovered ($\lambda=1)$, and the constant of proportionality is the recovery 
parameter $\alpha$, with units of [time]$^{-1}$. The breakdown term is proportional to $\lambda$, so 
that at complete breakdown ($\lambda=0$) there is no more breakdown, and also to the magnitude 
$\dot{\gamma}$ of the rate of strain tensor, since breakdown is assumed to be a shear-driven 
mechanism. The constant of proportionality is the breakdown parameter $\beta$, which is 
dimensionless. In an Eulerian frame of reference, Eq.\ (\ref{eq: lambda evolution lagrangian}) is 
written as
\begin{equation} \label{eq: lambda evolution eulerian}
 \pd{\lambda}{t} \;+\; \vf{u} \cdot \nabla\lambda \;=\; \alpha\,(1-\lambda) \;-\; \beta\,\lambda\,\dot{\gamma}
\end{equation}

\begin{figure}[tb]
 \centering
 \includegraphics[scale=1.0]{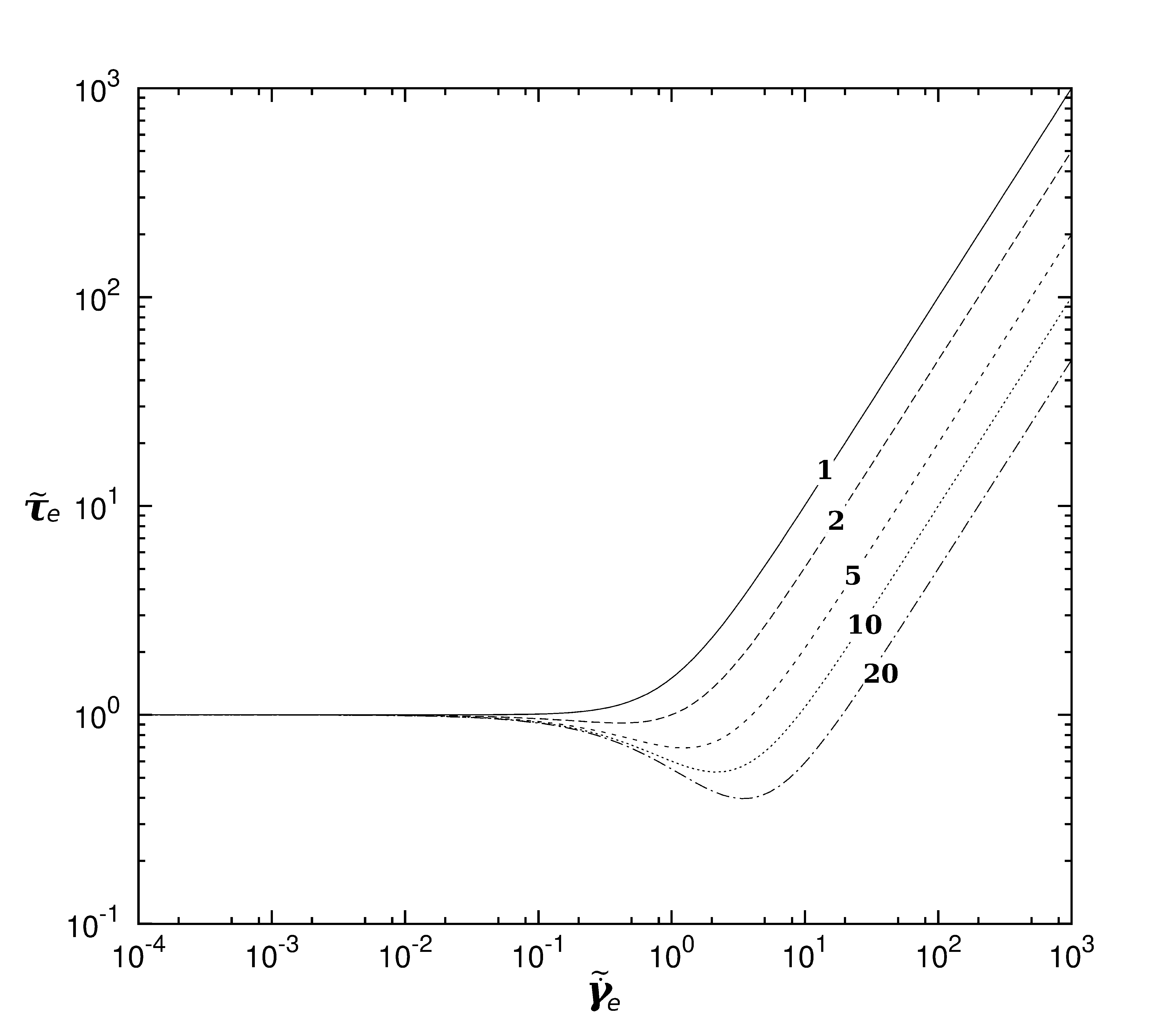}
 \caption{Non-dimensional shear stress versus non-dimensional strain rate at equilibrium, at 
various Bingham numbers, which are indicated on each curve, for $\tilde{\alpha} = \beta$.}
 \label{fig: tau_e model}
\end{figure}

The structure of a particle which experiences a constant strain rate $\dot{\gamma}_e$ will 
eventually reach an equilibrium where $D\lambda/Dt=0$ and the rate of breakdown is matched by the 
rate of build-up. The equilibrium value of $\lambda$ can be found from Eq.\ (\ref{eq: lambda 
evolution lagrangian}):
\begin{equation} \label{eq: lambda_e}
 \lambda_e \;=\; \frac{1}{1+\frac{\beta}{\alpha}\dot{\gamma}_e}
\end{equation}
By substituting Eq.\ (\ref{eq: lambda_e}) into Eq.\ (\ref{eq: constitutive Bingham}) we get the 
equilibrium value of the magnitude of the stress as a function of the strain rate:
\begin{equation} \label{eq: tau_e}
 \tau_e \;=\; \frac{\tau_y}{1+\frac{\beta}{\alpha}\dot{\gamma}_e} \;+\; \mu\,\dot{\gamma}_e
\end{equation}
If we dedimensionalise the stresses by $\tau_y$, and the strain rate and $\alpha$ by $U/D$ then the 
above equation can be written in non-dimensional form as
\begin{equation} \label{eq: tau_e nd}
 \tilde{\tau}_e \;=\; \frac{1}{1+\frac{\beta}{\tilde{\alpha}}\tilde{\dot{\gamma}}_e} \;+\;
                      \frac{1}{Bn}\,\tilde{\dot{\gamma}}_e
\end{equation}
where the tilde (\textasciitilde) denotes dedimensionalised quantities, and $Bn$ is the Bingham  
number, defined by
\begin{equation} \label{eq: Bn}
 Bn \;\equiv\; \frac{\tau_y D}{\mu U}
\end{equation}
The function $\tilde{\tau}_e$ (Eq.\ (\ref{eq: tau_e nd})) is plotted against 
$\tilde{\dot{\gamma}}_e$ in Figure \ref{fig: tau_e model}, for the case $\tilde{\alpha} = \beta$. 
Since, according to Eq. (\ref{eq: constitutive Bingham}), the magnitude of the total stress is due 
to two components (the yield stress $\tau_{0e}$ and the viscous part $\mu \dot{\gamma}_e$), the 
shapes of the curves shown in Fig.\ \ref{fig: tau_e model} reflect the relative balance between 
these two components at each value of the shear rate $\dot{\gamma}_e$. In particular, at low shear 
rates the viscous component $\mu \dot{\gamma}_e$ is small and the total stress $\tau_e$ is 
approximately equal to the yield stress $\tau_{0e}$, which in turn is nearly equal to $\tau_y$ since 
at low shear rates the breakdown is nearly zero and the material is nearly fully structured 
($\lambda_e \approx 1$, Eq.\ (\ref{eq: lambda_e})). Thus at low rates of strain all curves in Fig.\ 
\ref{fig: tau_e model} converge to $\tilde{\tau}_e = 1$. Then, if $\dot{\gamma}_e$ is progressively 
increased, the structure breakdown rate also increases and thus $\lambda_e$ is reduced below 1 and 
the yield stress $\tau_{0e}$ falls, according to Eq.\ (\ref{eq: tau(lambda)}). As long as 
$\tau_{0e}$ dominates the total stress $\tau_e$, this causes also the total stress to fall, which 
can be seen in Fig.\ \ref{fig: tau_e model} where $\tilde{\tau}_e$ falls below 1, especially at 
higher Bingham numbers where the yield stress is more dominant. However, as $\dot{\gamma}_e$ is 
increased further the viscous part $\mu\dot{\gamma}_e$ of the stress starts to become significant, 
and quickly becomes the dominant stress component: $\tau_e \approx \mu \dot{\gamma}_e$. Thus at 
large values of $\dot{\gamma}_e$, $\tilde{\tau}_e$ increases again, eventually with a slope of 
$1/Bn$, as can be seen from Eq.\ (\ref{eq: tau_e nd}). At some point in between the stress reaches 
its minimum value. Therefore each of the curves of Figure \ref{fig: tau_e model} exhibits a minimum 
at some value of $\tilde{\dot{\gamma}}_e$. Experimental studies \cite{Beris_08} show that real 
thixotropic materials may exhibit this sort of behaviour.

It may be noticed that within an unyielded region $\dot{\gamma}$ = 0 and thus there is only recovery 
(no structure breakdown), and Eq.\ (\ref{eq: lambda evolution lagrangian}) becomes:
\begin{equation} \label{eq: lambda evolution unyielded}
 \frac{D\lambda}{Dt} \;=\; \alpha\,(1-\lambda)
\end{equation}
Assuming that the structure parameter has a value $\lambda_0$ at time $t_0=0$, this ordinary 
differential equation can be solved to calculate the value of $\lambda$ as a function of time:
\begin{equation} \label{eq: lambda recovery}
 \frac{1-\lambda}{1-\lambda_0} \;=\; \frac{1}{e^{\alpha t}} \;=\; \frac{1}{e^{\tilde{\alpha} \tilde{t}}}
\end{equation}
where $\tilde{\alpha} = \alpha / (U/D)$ and $\tilde{t} = t / (D/U)$ are the the nondimensional 
recovery parameter and time, respectively. Therefore, every $t=1/\alpha$ time units (or 
$\tilde{t}=1/\tilde{\alpha}$ nondimensional time units), the structure recovers by a factor of $e 
\approx 2.72$ (the degree of structure breakdown $1-\lambda$ becomes $e$ times smaller). Thus 
$1/\alpha$ can be regarded as a characteristic time scale for structure recovery.

\begin{figure}[p]
 \centering
  \subfigure[{$Re = 40; Bn = 0$}] {\label{sfig: Re=40 Bn=0}
   \includegraphics[scale=1.0]{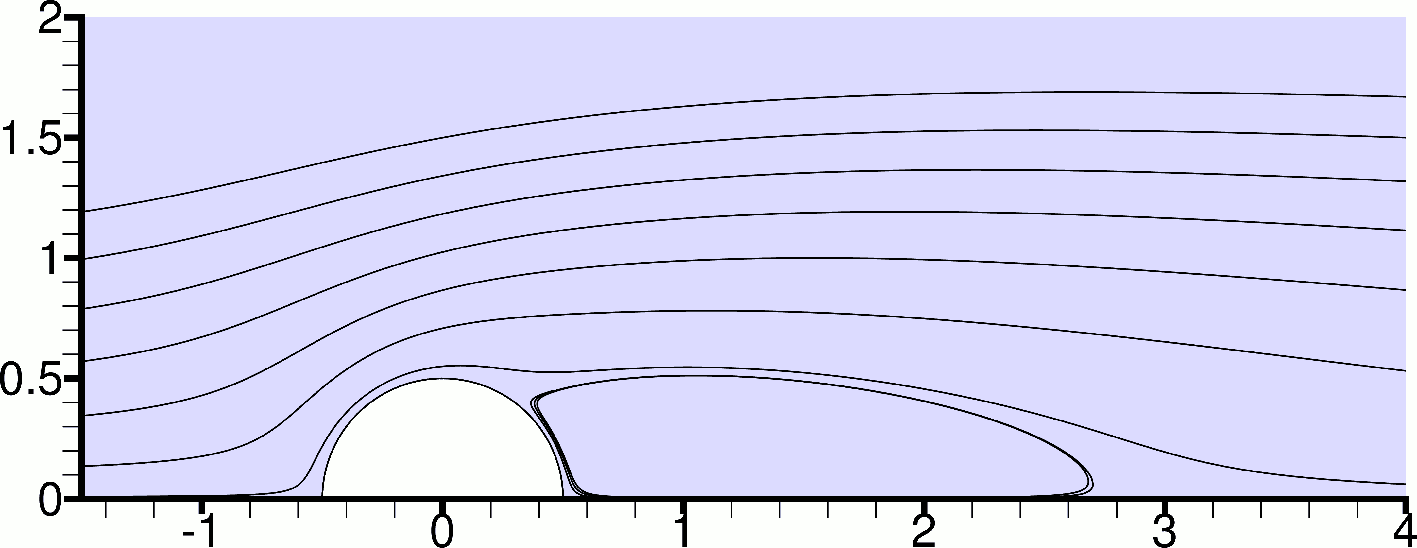}} \\
  \subfigure[{$Re = 40; Bn = 0.1$}] {\label{sfig: Re=40 Bn=0.1}
   \includegraphics[scale=1.0]{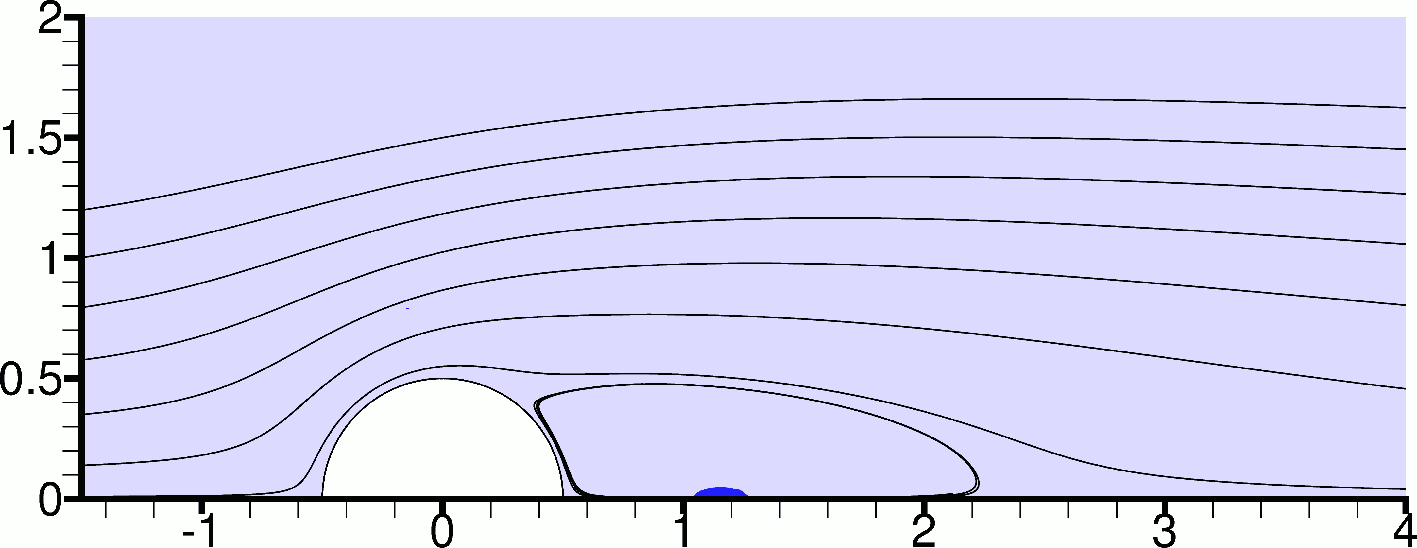}} \\
  \subfigure[{$Re = 40; Bn = 0.4$}] {\label{sfig: Re=40 Bn=0.4}
   \includegraphics[scale=1.0]{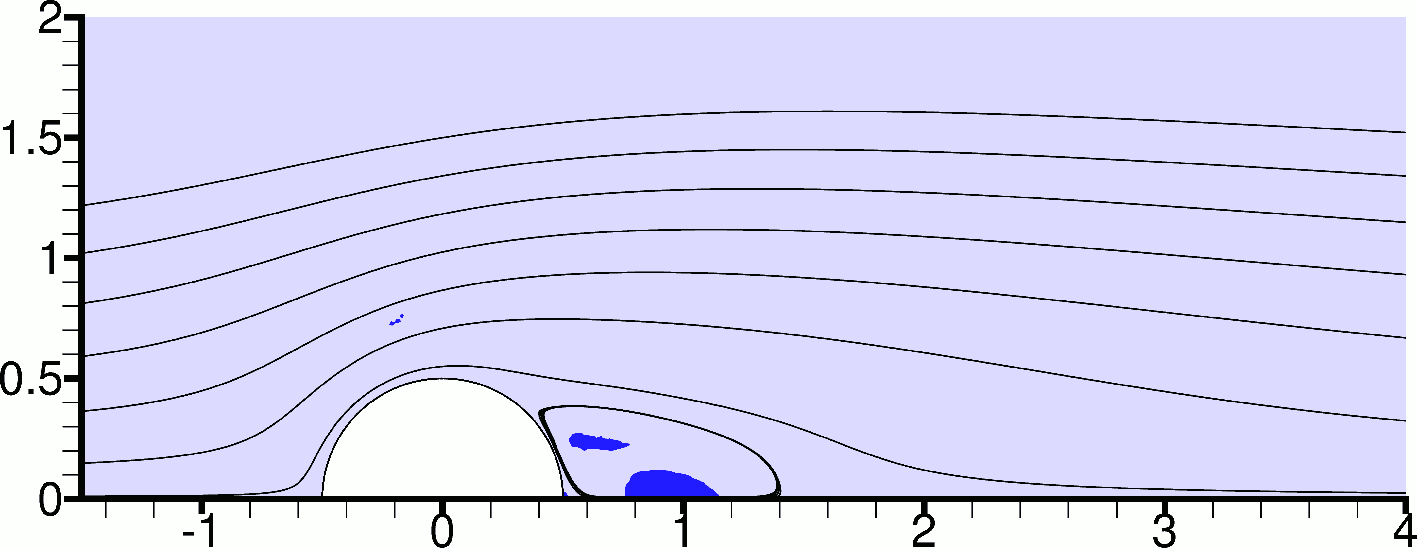}} \\
  \subfigure[{$Re = 40; Bn = 0.5$}] {\label{sfig: Re=40 Bn=0.5}
   \includegraphics[scale=1.0]{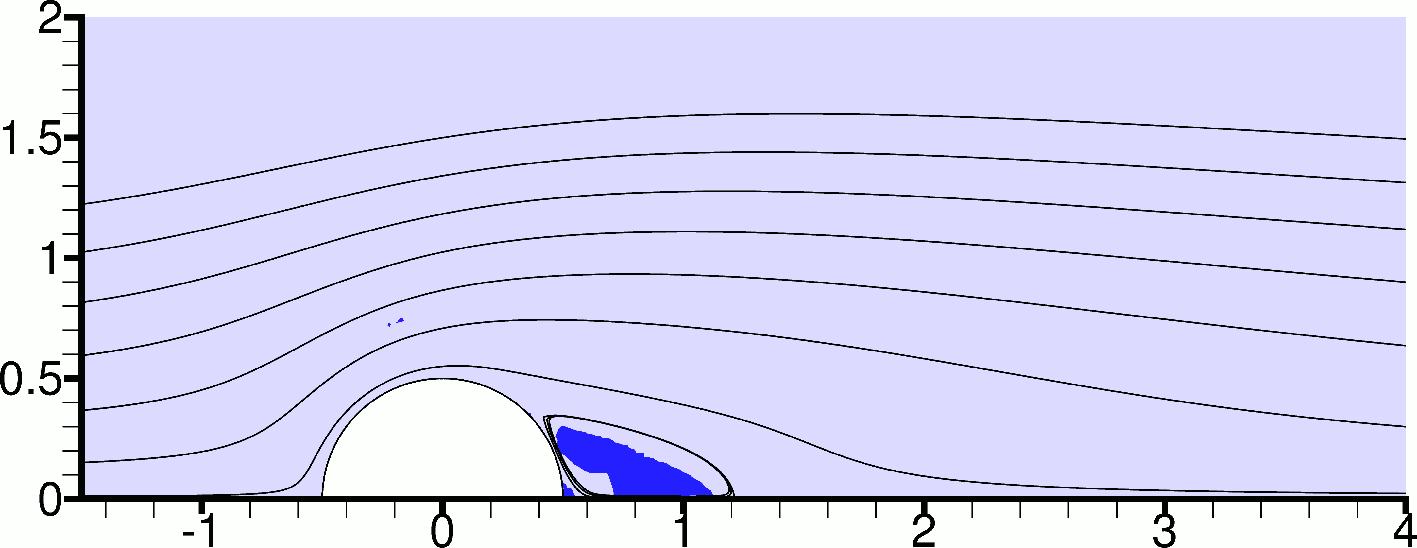}} \\
  \subfigure[{$Re = 40; Bn = 1.0$}] {\label{sfig: Re=40 Bn=1}
   \includegraphics[scale=1.0]{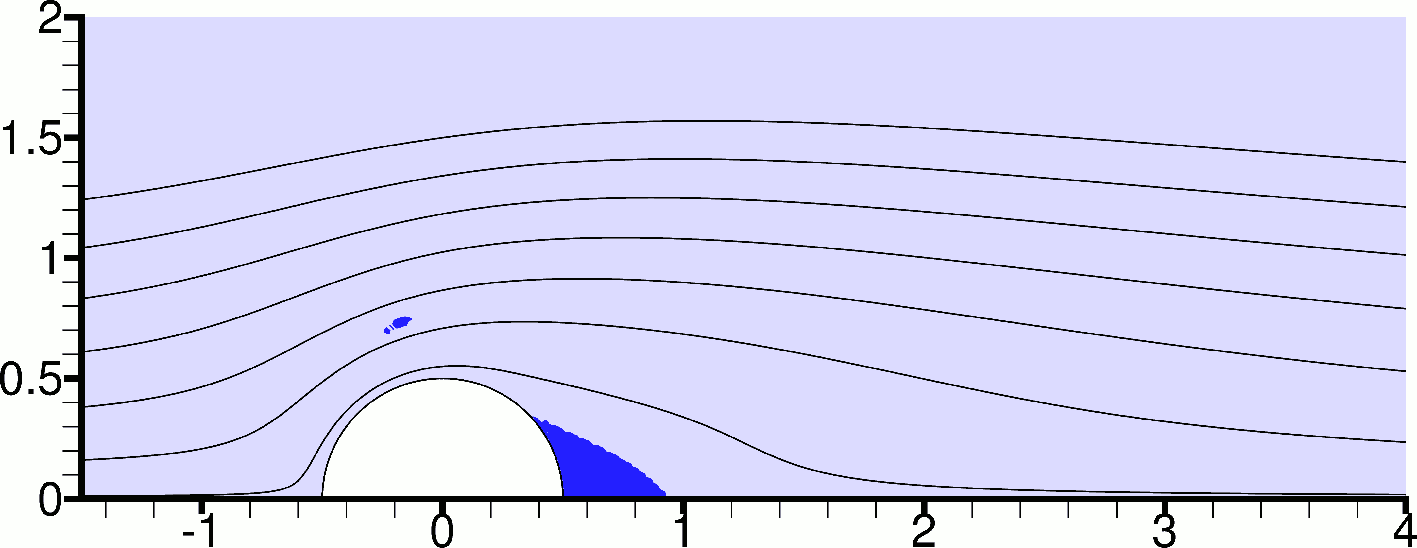}} \\
  \subfigure[{$Re = 69; Bn = 1.0$}] {\label{sfig: Re=69 Bn=1}
   \includegraphics[scale=1.0]{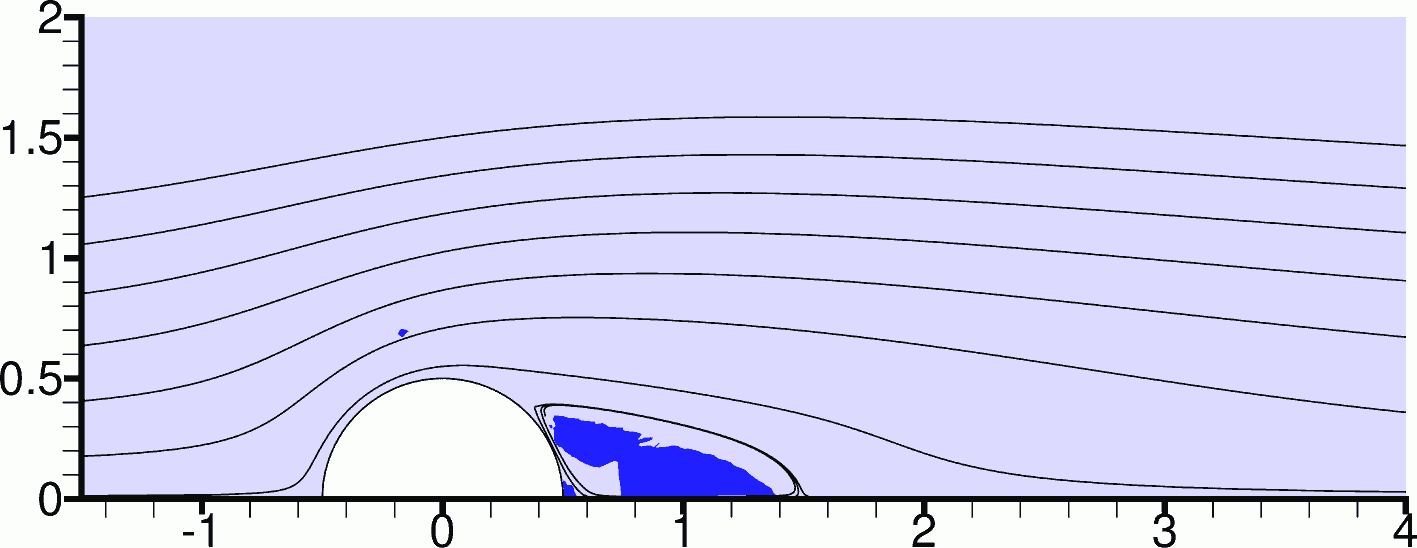}} \\
  \subfigure[{$Re = 72; Bn = 1.0$}] {\label{sfig: Re=72 Bn=1}
   \includegraphics[scale=1.0]{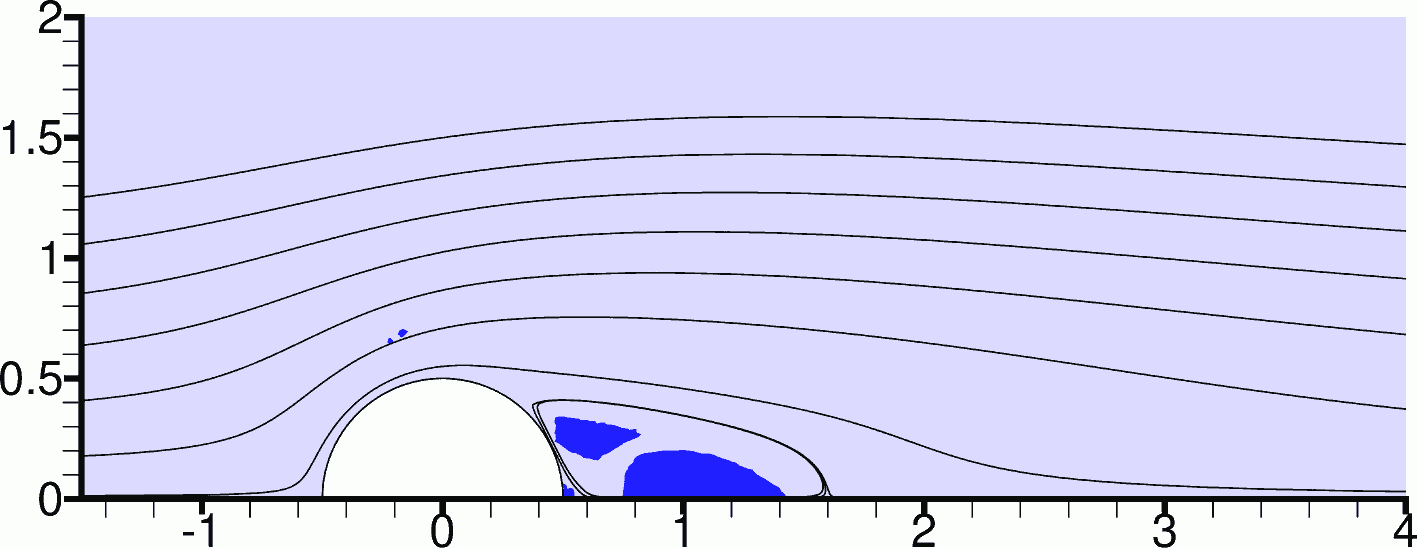}}
 \caption{Selected streamlines and unyielded areas ($\tau \leq \tau_0$, dark blue regions) shown 
for various combinations of Bingham and Reynolds numbers, for Bingham flow (no thixotropy, 
steady-state results).}
 \label{fig: Bn vs Re}
\end{figure}

Equations (\ref{eq: continuity}) -- (\ref{eq: momentum}) can be written in nondimensional form, by 
scaling lengths by $D$, time by $D/U$, velocities by $U$, and stress and pressure by $\tau_y$. We 
thus obtain the following nondimensional forms:
\begin{equation} \label{eq: continuity nd}
 \tilde{\nabla} \cdot \tilde{\vf{u}} \;=\; 0
\end{equation}
\begin{equation} \label{eq: momentum nd}
 Re \left( \pd{\tilde{\vf{u}}}{\tilde{t}} \;+\; \tilde{\vf{u}} \cdot \tilde{\nabla} \tilde{\vf{u}} 
 \right) \;=\; Bn \left( -\tilde{\nabla} \tilde{p} \;+\; \tilde{\nabla} \cdot \tilde{\vf{\tau}} 
 \right)
\end{equation}
where
\begin{equation} \label{eq: constitutive nd}
 \tilde{\vf{\tau}} \;=\;
 \left( \frac{\lambda}{\tilde{\dot{\gamma}}} \;+\, \frac{1}{Bn} \right) \tilde{\dot{\vf{\gamma}}}
\end{equation}
It may be seen that if the $Bn$ and $Re$ numbers are increased together so that the ratio $Re / Bn$ 
is kept constant then the term $1/Bn$ in Eq.\ (\ref{eq: constitutive nd}) becomes less and less 
important. Eventually it can be seen in Eq.\ (\ref{eq: momentum nd}) that the flow is influenced 
only by the $Bn$ number outside the parentheses on the right hand side, and not by the Bingham 
number which is implicit in the definition of the stress from equation (\ref{eq: constitutive nd}). 
In this case the flow field is actually governed by the ratio $Re / Bn$, as can be seen by dividing 
Eq.\ (\ref{eq: momentum nd}) by $Bn$. Therefore, the Reynolds and Bingham numbers have opposite 
effects on the flow, so that increasing one of them has a similar effect as decreasing the other. 
To illustrate this, we jump ahead for the moment and without yet discussing the numerical method we 
present in Figure \ref{fig: Bn vs Re} results for Bingham flow without thixotropy, chosen for 
validation purposes because they match corresponding results in Ref.\ \cite{Mossaz_10}. Moving from 
top to bottom, in Figures \ref{sfig: Re=40 Bn=0} -- \ref{sfig: Re=40 Bn=1} the Bingham number is 
increased while the Reynolds number is held fixed at $Re = 40$. In the Newtonian case (Figure 
\ref{sfig: Re=40 Bn=0}) the recirculation bubbles are fairly large, but as the Bingham number is 
increased they reduce in size and at $Bn = 1$ (Figure \ref{sfig: Re=40 Bn=1}) the flow has become 
creeping, without separation. Increasing the Bingham number also causes the formation of unyielded 
zones (shown shaded) behind the cylinder, which are initially detached from the cylinder and are 
thus moving with the flow (rigid body motion) but beyond a certain Bingham number they merge and get 
attached to the cylinder surface, becoming motionless (Figure \ref{sfig: Re=40 Bn=1}). One can also 
notice very small unyielded regions above the cylinder (and below, in the symmetric part of the 
images which is not shown) which grow as $Bn$ is increased. The reader who is familiar with the 
creeping ($Re = 0$) viscoplastic flow around a cylinder will notice that increasing the Bingham 
number makes the flow field more similar to the creeping flow case - see the relevant references in 
Section \ref{sec: introduction}.

In the rest of the figures, \ref{sfig: Re=40 Bn=1} -- \ref{sfig: Re=72 Bn=1}, the Bingham number is 
held fixed while the Reynolds number is increased. As a result, the flow phenomena are reversed: The 
unyielded regions detach from the cylinder and brake apart, while the recirculation bubbles reappear 
and grow.

\section{Numerical method}
\label{sec: method}

The Bingham constitutive equation (\ref{eq: constitutive Bingham}) is discontinuous, separating the 
material into yielded / unyielded zones whose boundaries are not known a priori. This complicates 
the numerical solution of the flow. To avoid numerical difficulties we used the customary 
Papanastasiou regularisation scheme \cite{Papanastasiou_87}. In fact this approach has been adopted 
by all the previous studies on viscoplastic flow past a cylinder mentioned in Section \ref{sec: 
introduction} except for that of Roquet and Saramito \cite{Roquet_03}, who used an augmented 
Lagrangian approach, thus directly solving the original Bingham constitutive equation. According to 
Tokpavi et al. \cite{Tokpavi_08} the difference between the results produced by the two methods is 
rather small. Thus the constitutive equation used here, applicable throughout the material, is the 
following:
\begin{equation} \label{eq: constitutive Papanastasiou}
 \vf{\tau} \;=\; \left[ \frac{\tau_0 (1-e^{-m\dot{\gamma}})}{\dot{\gamma}} \;+\; \mu \right] 
 \vf{\dot{\gamma}}
\end{equation}
where the constant $m$ is the regularisation parameter, which should be large enough so that the 
original Bingham equation is approximated adequately. We note that in the present study we follow 
the usual practice of identifying the unyielded regions as those regions where $\tau < \tau_0 = 
\lambda \, \tau_y$ or equivalently $\tilde{\tau} < \lambda$; see Burgos et al.\ \cite{Burgos_99} 
for more information. We note that regularisation implies that unyielded regions are approximated 
by small, but non-zero rate of strain. The maximum rate of strain in an unyielded region can be 
calculated by writing Eq.\ (\ref{eq: constitutive Papanastasiou}) in terms of tensor magnitudes and 
substituting $\tau = \tau_0 = \lambda \tau_y$. Then, after some manipulation and 
dedimensionalisation the following equation is obtained:
\begin{equation} \label{eq: g0}
 \tilde{\dot{\gamma}}_0 \;-\; \lambda \: Bn \: e^{-M \tilde{\dot{\gamma}}_0} \;=\; 0
\end{equation}
where $\tilde{\dot{\gamma}}_0$ is the non-dimensional rate of strain at the surfaces where $\tau = 
\tau_0$, assumed to be the yield surfaces; the rate of strain is smaller than that in the interior 
of the unyielded regions. $M$ is the non-dimensional regularisation parameter $M = mU/D$. Eq.\ 
(\ref{eq: g0}) has an analytic solution:
\begin{equation}
 \tilde{\dot{\gamma}}_0 \;=\; \frac{1}{M}W(\lambda \!\cdot\! Bn \!\cdot\! M)
\end{equation}
where $W$ is the Lambert W function \cite{Corless_1996, You_2008}. Just to give an idea of the order 
of $\tilde{\dot{\gamma}}_0$ for the range of parameters used in the present study, we give some 
examples for $\lambda=1$: For \{$Bn$=0.5, $M$=1000\}, $\tilde{\dot{\gamma}}_0 \approx 4.7\cdot 
10^{-3}$; for \{$Bn$=0.5, $M$=10000\}, $\tilde{\dot{\gamma}}_0 \approx 6.6\cdot 10^{-4}$; for 
\{$Bn$=5, $M$=1000\}, $\tilde{\dot{\gamma}}_0 \approx 6.6\cdot 10^{-3}$; and for \{$Bn$=5, 
$M$=10000\}, $\tilde{\dot{\gamma}}_0 \approx 8.7\cdot 10^{-4}$. The value $\dot{\gamma}_0$ is about 
the smallest rate of strain that the regularised model can predict with relative accuracy - for 
smaller values the error increases significantly.

\begin{figure}[tb]
 \centering
  \subfigure[] {\label{sfig: mesh far}
   \includegraphics[scale=0.90]{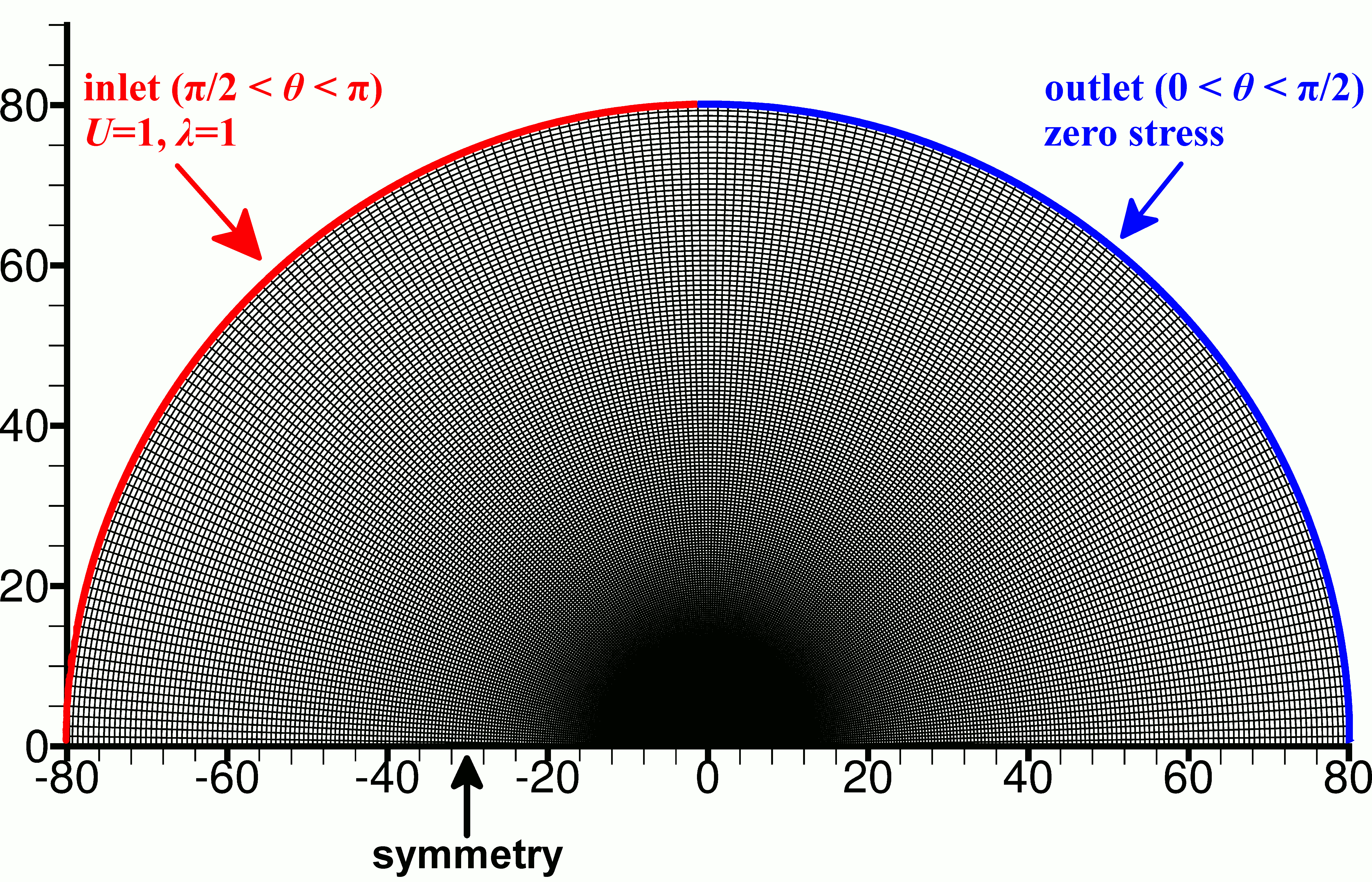}}
  \subfigure[] {\label{sfig: mesh close}
   \includegraphics[scale=0.90]{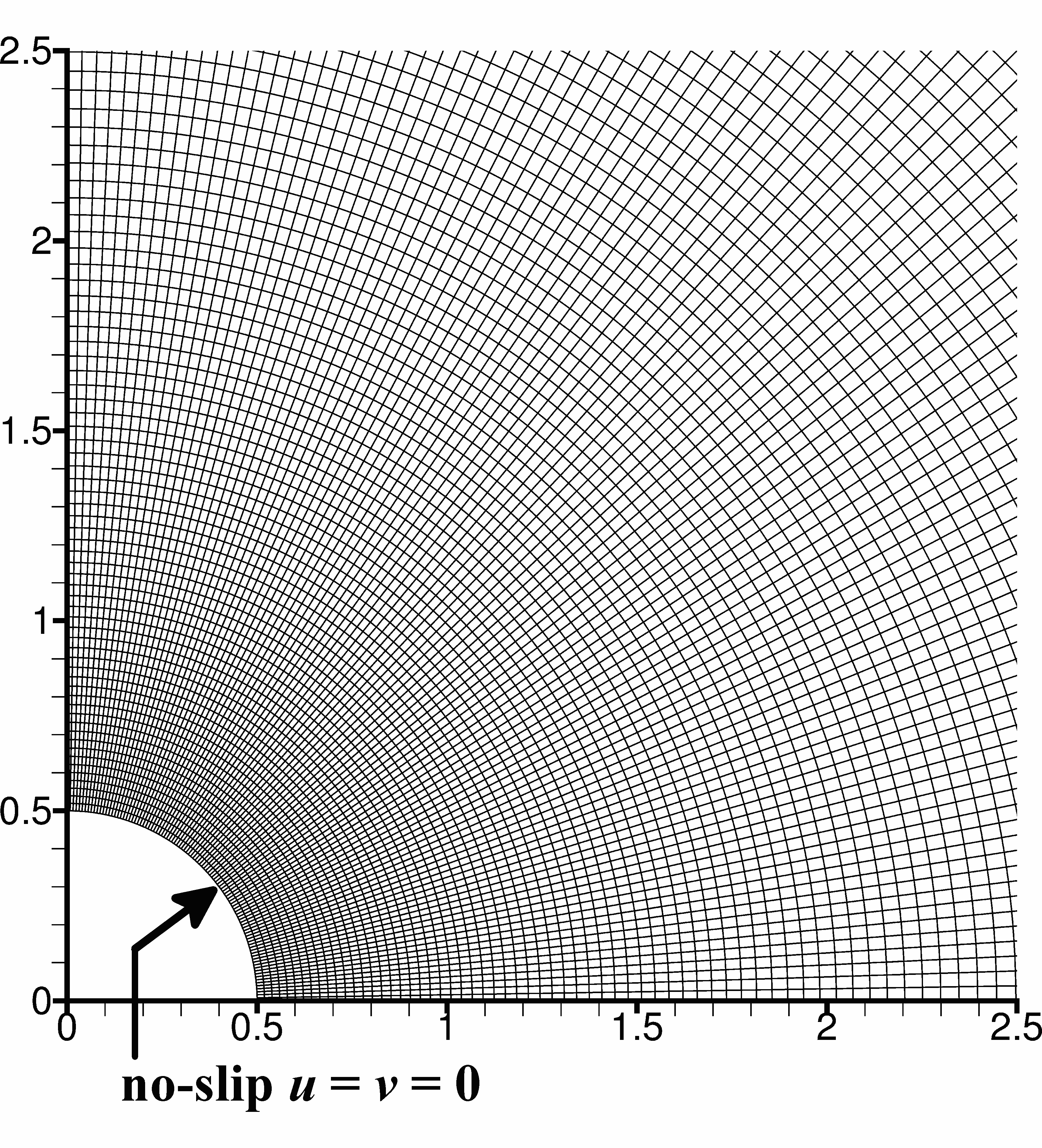}}
 \caption{Finite element mesh used, and boundary conditions. On the right, a detail of the mesh 
near the cylinder is shown.}
 \label{fig: mesh}
\end{figure}

The governing equations were solved numerically using a mixed Galerkin finite element method. In the 
present work the focus is on subcritical flow where vortex shedding is not present. Therefore, the 
flow field is symmetric, with the plane of symmetry being parallel to the main flow, and only half 
of the domain needs to be modelled. The computational domain is shown in Figure \ref{fig: mesh}. The 
cylinder, of radius $R_c = 0.5$, is centred at point (0,0). Modelling the flow domain up to an outer 
radius of $R_o = 80$ was considered sufficient for the simulation of unconfined flow, based on the 
observations of Mossaz et al.\ \cite{Mossaz_10}. The boundary conditions are also illustrated in 
Figure \ref{fig: mesh}. A no-slip condition (zero velocity) is applied to the cylinder surface. The 
left half of the outer circumference of the computational domain is an inlet, where fluid flows 
horizontally into the domain with a velocity $U$ = 1. The incoming fluid is fully structured 
($\lambda_i = 1$). The right half of the outer circumference of the domain is an outlet, where a 
zero-stress condition is applied. Finally, the bottom of the domain is a symmetry plane and 
therefore $v = 0$ and $\tau_{xy} = 0$.

Figure \ref{fig: mesh} also shows the computational mesh used. The mesh consists of 100 $\times$ 150 
mixed finite elements, along the circumferential and radial directions, respectively. Therefore each 
element spans four grid cells, ordered in a $2\times 2$ fashion. This mesh is more refined than that 
used by Mossaz et al.\ \cite{Mossaz_10}, who obtained reliable results that we used for validation 
of our own results. Each problem is solved as a time-dependent problem until a time of $t = 120$, 
which was observed to be sufficient to obtain the steady-state solution. The initial condition is 
that $u$ = $v$ = 0 and $\lambda$ = 1 at $t$ = 0. The temporal discretisation scheme employed is the 
implicit Euler scheme. The Newton-Raphson procedure is used to solve the resulting non-linear 
algebraic system within each time step, and in each Newton-Raphson step the linear system which 
arises is solved using the freely available sparse frontal solver MUMPS \cite{MUMPS:1, MUMPS:2}. 
Eq.\ (\ref{eq: lambda evolution eulerian}) is solved individually within the Newton-Raphson 
iterative scheme.

A complexity arises from the fact that the transport equation for $\lambda$, Eq.\ (\ref{eq: lambda 
evolution eulerian}), does not contain any diffusion terms. Therefore, without any special 
treatment, it is anticipated that its numerical solution will produce a $\lambda$ field which 
contains spurious oscillations, and in fact the reader can indeed, in some cases, observe such 
oscillations in the Figures presented in the next Section. These non-physical oscillations spoil the 
aesthetics of the solution, but the mean field is unaffected. Although there do exist more elaborate 
discretisation schemes which prevent the appearance of such oscillations (e.g.\ \cite{Brooks_82}), 
in the present work for simplicity we adopted a more ``crude'' approach of allowing the 
oscillations, and just setting the value of $\lambda$ equal to 0 or 1 whenever a negative value or a 
value greater than 1 is produced, respectively.

\section{Numerical results}
\label{sec: results}

In the present Section, the results will be presented in terms of the non-dimensional form of the 
variables, but, for simplicity, tildes will be dropped from the corresponding variable names.

The equations that govern the flow involve many parameters: the Reynolds number $Re$, the Bingham 
number $Bn$, and the thixotropy parameters $\alpha$ and $\beta$. Hence, in order to get the complete 
picture, one has to assign different values to each of these parameters and repeat the simulations. 
This produces a prohibitively large number of numerical experiments that have to be performed. For 
practical reasons therefore we confined our study to the regime where the flow is symmetric, which 
requires that the Reynolds number is small enough. Numerous studies agree that, for Newtonian flow, 
the onset of periodic vortex shedding occurs at a critical Reynolds number of $Re \approx 47$ (see 
\cite{Williamson_96, Kumar_06, Sivakumar_06} and references therein). It was therefore decided to 
fix the Reynolds number at a value of $Re = 45$ which is close to, but smaller than, this critical 
Reynolds number in order to have a relatively high Reynolds number but at the same time excluding 
the possibility of vortex shedding irrespective of the values of the other parameters ($Bn, \alpha, 
\beta$). This is because the effect of the Bingham number is to increase the viscous character of 
the flow, thus increasing the critical Reynolds number beyond the value of 47; thixotropy on the 
other hand recovers some of the inertial character of the flow, but it cannot make the flow more 
inertial than in the Newtonian case, that is it cannot reduce the critical Reynolds number below 47.

So, in the following the Reynolds number is fixed at $Re = 45$, while two values are assigned to the 
Bingham number: a low value of $Bn = 0.5$, and a medium value of $Bn = 5$. As will be shown later in 
this Section, with these two choices of Bingham number the results span both steady-state regimes: 
Recirculating flow and creeping flow, respectively. Then the effect of the thixotropy parameter 
$\alpha$ is investigated by holding $\beta$ fixed at $\beta = 0.05$ and varying $\alpha$ in the set 
$\alpha \in \{0.01, 0.05, 0.10\}$. Also, the effect of $\beta$ is investigated by holding $\alpha$ 
fixed at $\alpha = 0.05$ and varying $\beta$ in the set $\beta \in \{0.01, 0.05, 0.10\}$. From these 
results, one can also obtain a feel about the effect of $Re$ on the flow, because, as discussed in 
Section \ref{sec: equations}, the flow characteristics are approximately governed by the ratio $Re / 
Bn$. Increasing $Bn$, increasing $\alpha$, or reducing $\beta$ would make the flow more ``viscous'' 
- that is, it would increase the magnitude of the viscous stresses, thus increasing their 
significance relative to momentum (inertial) fluxes - causing an effect similar to decreasing the 
Reynolds number; and the opposite action on these parameters would make the flow more ``inertial'', 
causing an effect similar to increasing the Reynolds number. A difference is that, whereas the 
effect of $Re$ and $Bn$ on the viscous character of the flow is independent of time, the effect of 
$\alpha$ and $\beta$ is not, because it takes time for structural changes to occur and viscous 
stresses to increase or decrease due to thixotropy.

\begin{figure}[tb]
 \centering
  \subfigure[Newtonian] {\label{sfig: Re=45 newtonian}
   \includegraphics[scale=1.00]{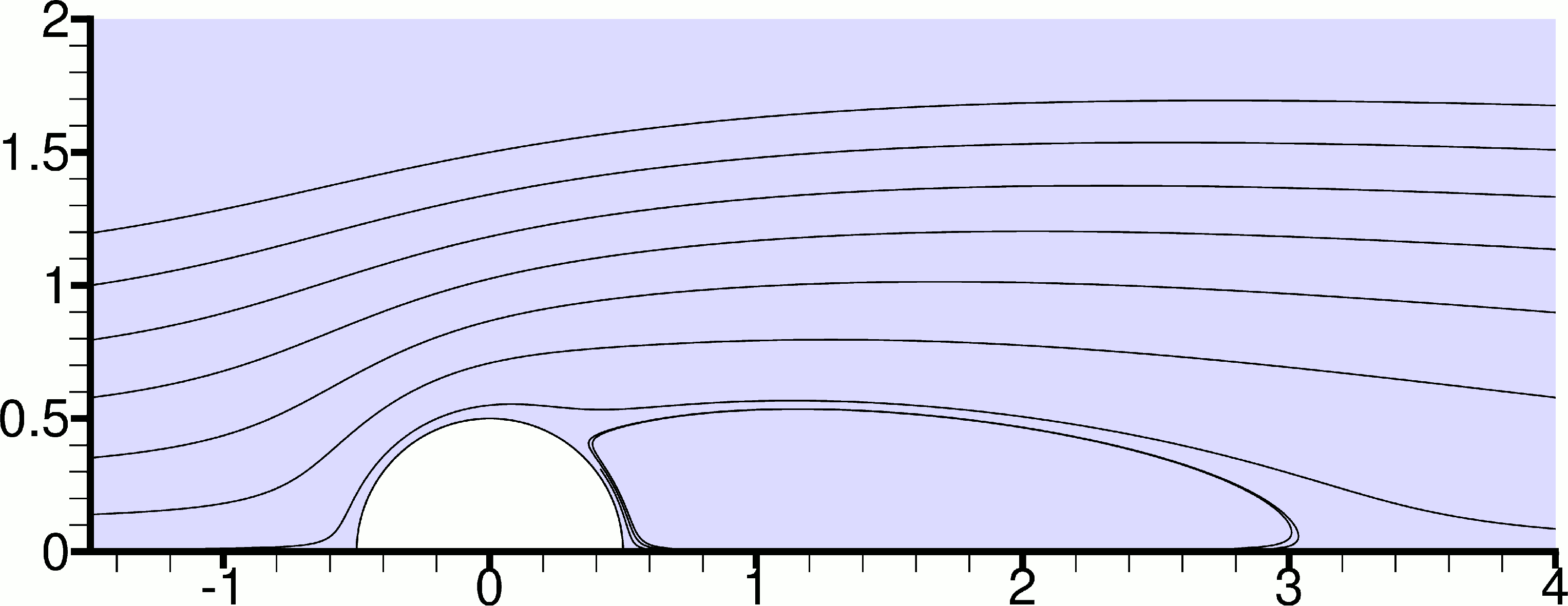}} \\
  \subfigure[$Bn = 0.5$] {\label{sfig: Re=45 Bn=0.5}
   \includegraphics[scale=1.00]{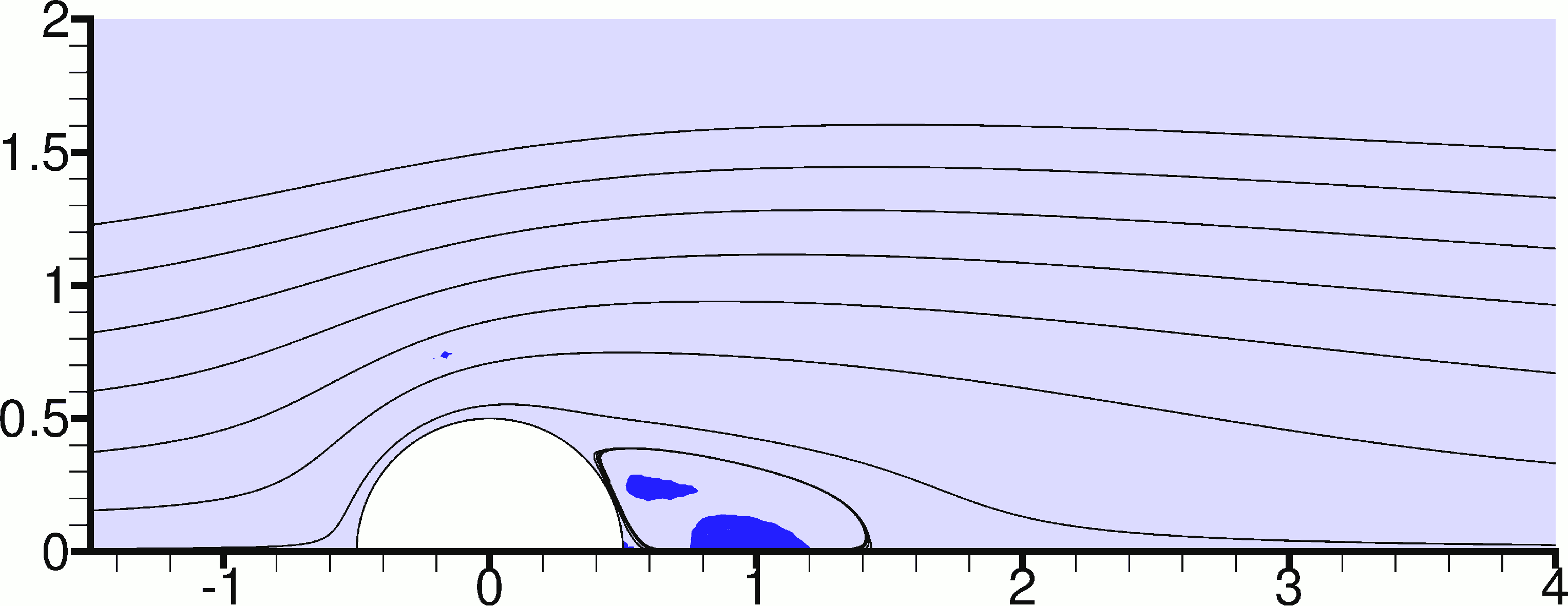}} \\
  \subfigure[$Bn = 5$] {\label{sfig: Re=45 Bn=5}
   \includegraphics[scale=1.00]{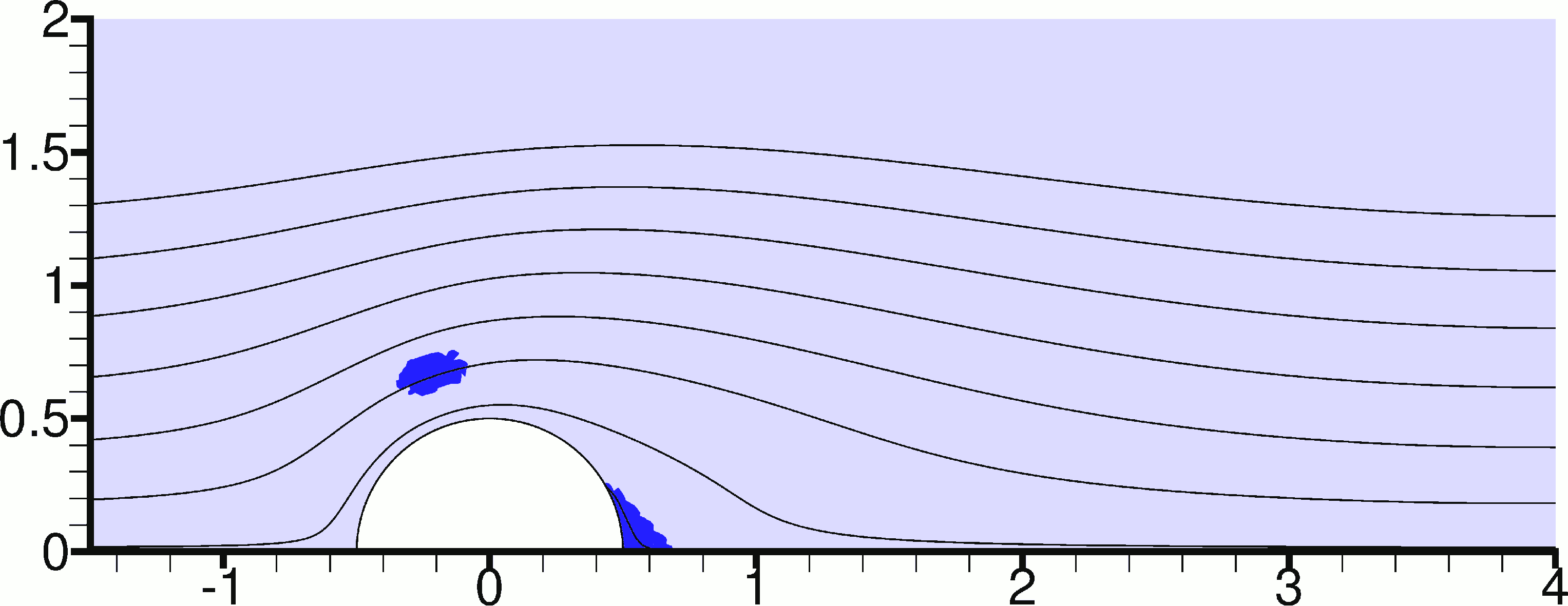}}
 \caption{Individual streamlines and unyielded zones (shaded) for flow at $Re = 45$ and selected 
$Bn$ numbers, without thixotropy. The steady state is shown.}
 \label{fig: Re=45 no thixotropy}
\end{figure}

The effect of thixotropy will be studied in comparison with the base cases of $Re = 45$ and $Bn \in 
\{0.5,5\}$ without thixotropy, which are plotted in Figure \ref{fig: Re=45 no thixotropy}. In this 
and subsequent figures, only half of the physical domain is shown, and we will follow the convention 
of referring only to the flow features that are visible in the figures. Thus, for example, we will 
refer to one recirculation bubble, although there is another identical bubble in the symmetric part 
of the physical domain. With this convention in mind, we make the following observations from Figure 
\ref{fig: Re=45 no thixotropy}: Newtonian flow exhibits a very large recirculation bubble, which is 
anticipated since the flow is close to the onset of vortex shedding. At a Bingham number of 0.5 the 
recirculation bubble has decreased in size, and a pair of unyielded zones appears behind the 
cylinder, detached from its surface, and moving at small but non-zero velocities with the flow. 
Actually, there are two more unyielded zones, which are very small and are difficult to see: one 
touching the back of the cylinder at the symmetry plane, and one above the cylinder and slightly 
upstream of it. At a Bingham number of 5, the recirculation bubble has disappeared, and the 
unyielded zones behind the cylinder have merged into a single zone which is in contact with the 
cylinder and is therefore motionless, and whose size is rather small. The unyielded zone above the 
cylinder can be seen to have grown considerably in size. In general, the flow field shown in Figure 
\ref{sfig: Re=45 Bn=5} looks a lot more like the flow field of creeping viscoplastic flow, which is 
described in references such as \cite{Deglo_03, Tokpavi_08, Zisis_02, Roquet_03, Mitsoulis_04}. Thus 
although the Reynolds number is fixed at $Re = 45$, the choice of two Bingham numbers $Bn = 0.5$ and 
$Bn = 5$ allows the investigation of two different flow regimes.

\begin{figure}[tb]
 \centering
  \subfigure[] {\label{sfig: Bn=0.5 yield lines close}
   \includegraphics[scale=1.08]{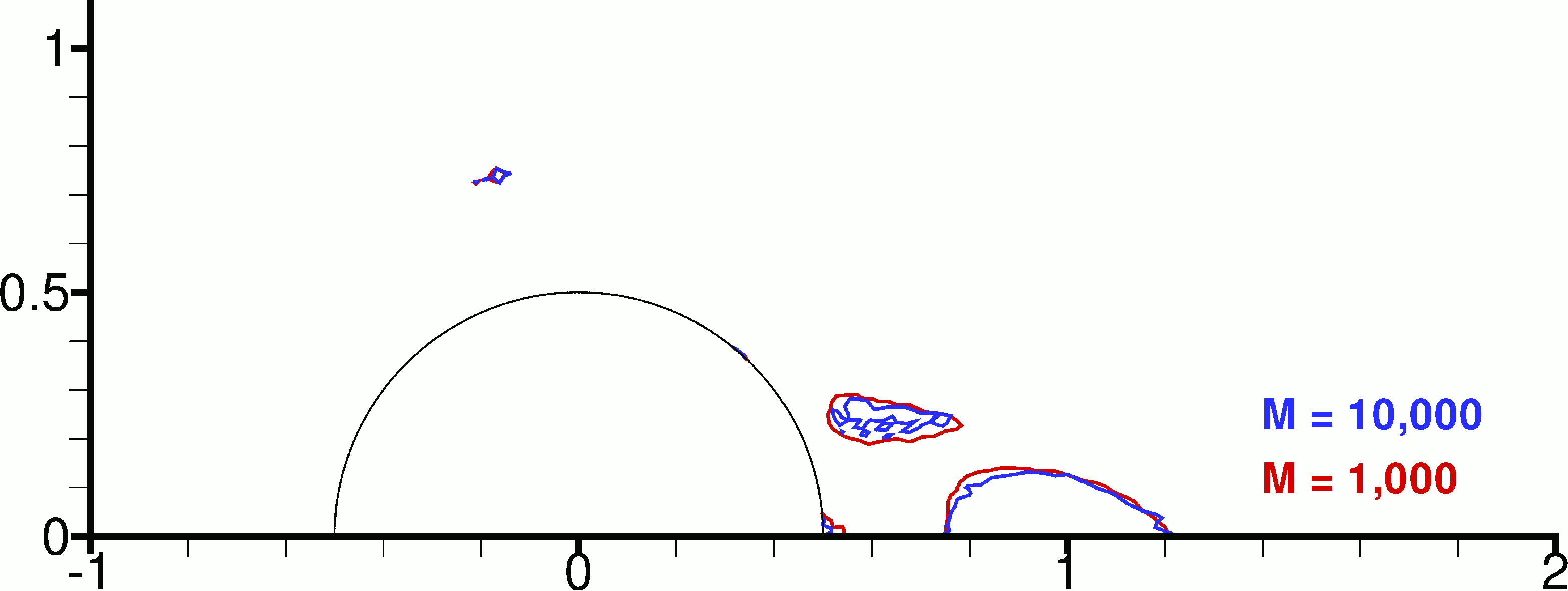}}
  \subfigure[] {\label{sfig: Bn=0.5 yield lines far}
   \includegraphics[scale=1.08]{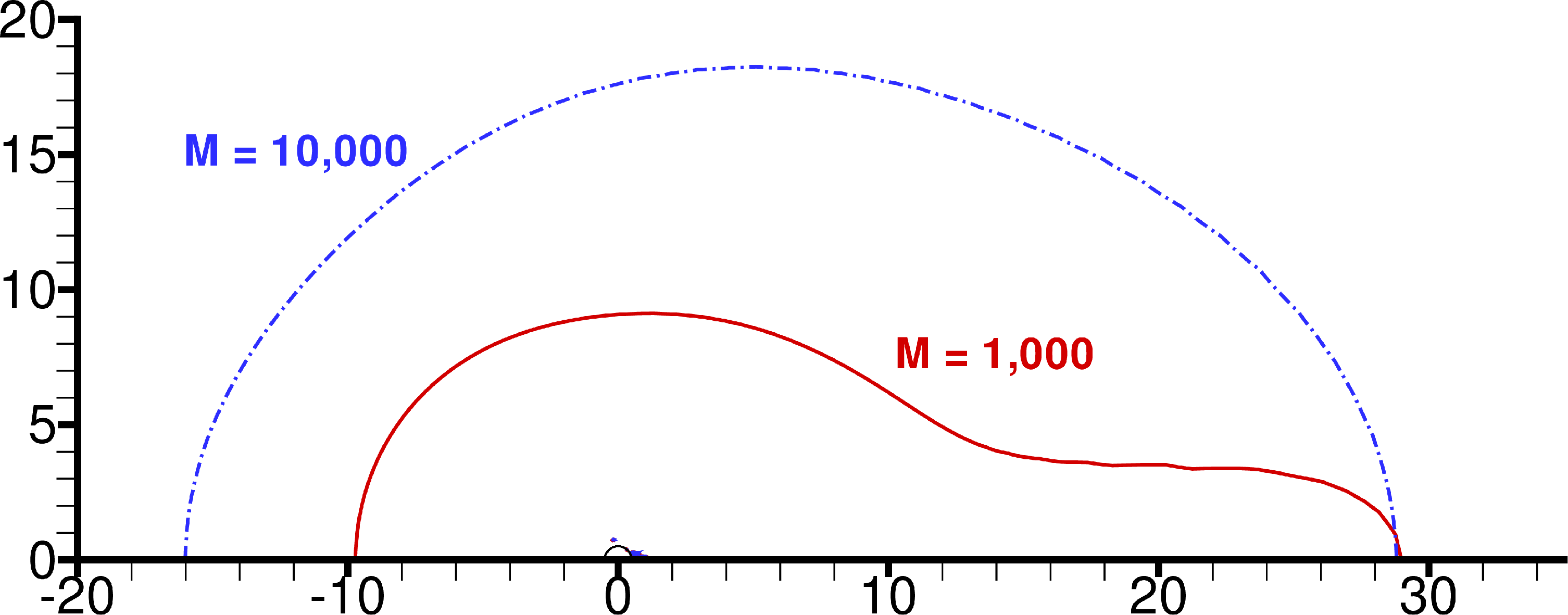}}
  \subfigure[] {\label{sfig: Bn=0.5 stress contours close}
   \includegraphics[scale=1.08]{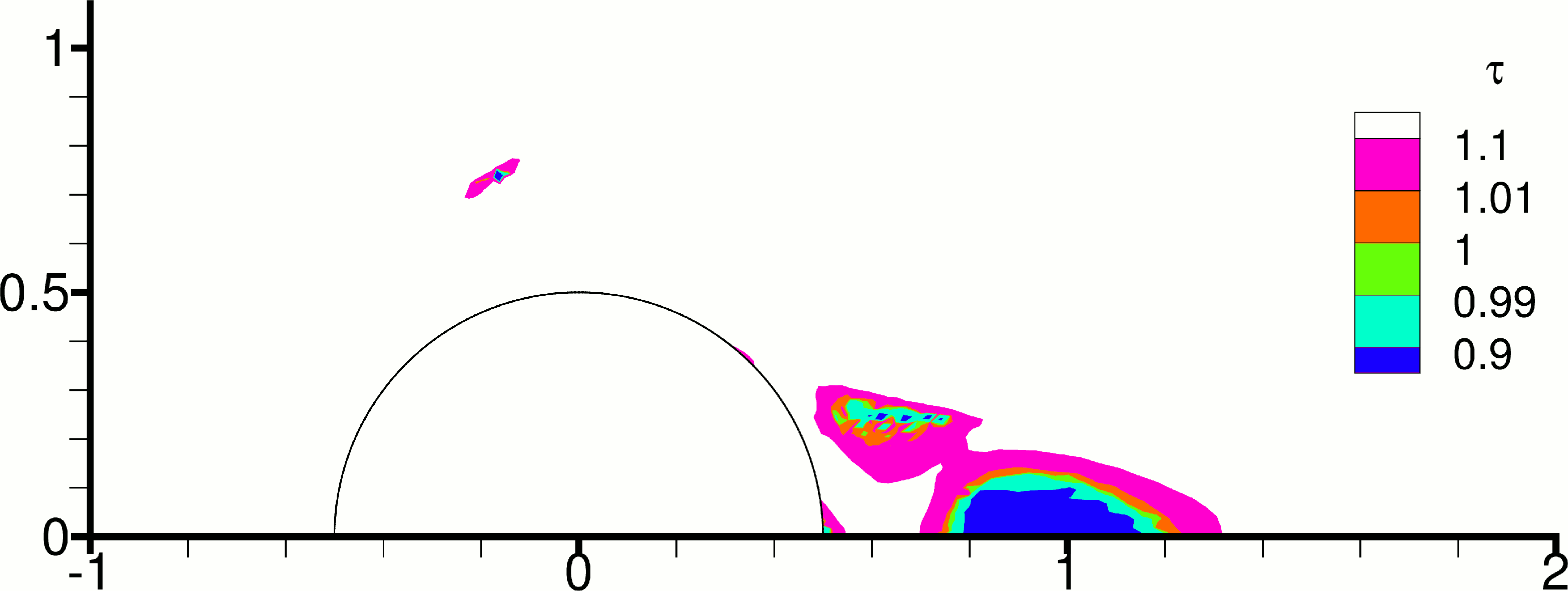}}
  \subfigure[] {\label{sfig: Bn=0.5 stress contours far}
   \includegraphics[scale=1.08]{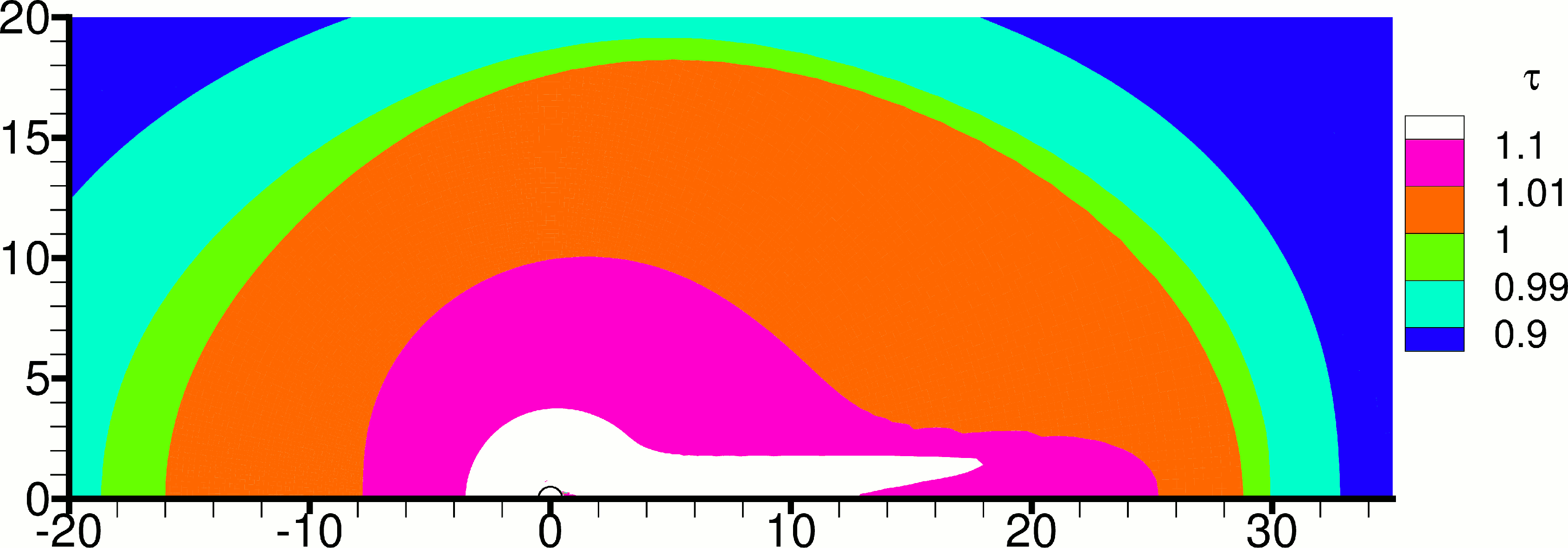}}
 \caption{For the $Re=45$, $Bn=0.5$ case without thixotropy, the top row shows yield lines 
(approximated by $\tau = 1$) computed with different regularisation parameters, $m$ = 1000 and $m$ 
= 10000, while the bottom row shows contours of nondimensional stress $\tau$ computed with $m$ = 
10000.}
 \label{fig: evaluation of M Bn=0.5}
\end{figure}

\begin{figure}[tb]
 \centering
  \subfigure[] {\label{sfig: Bn=5 yield lines close}
   \includegraphics[scale=1.08]{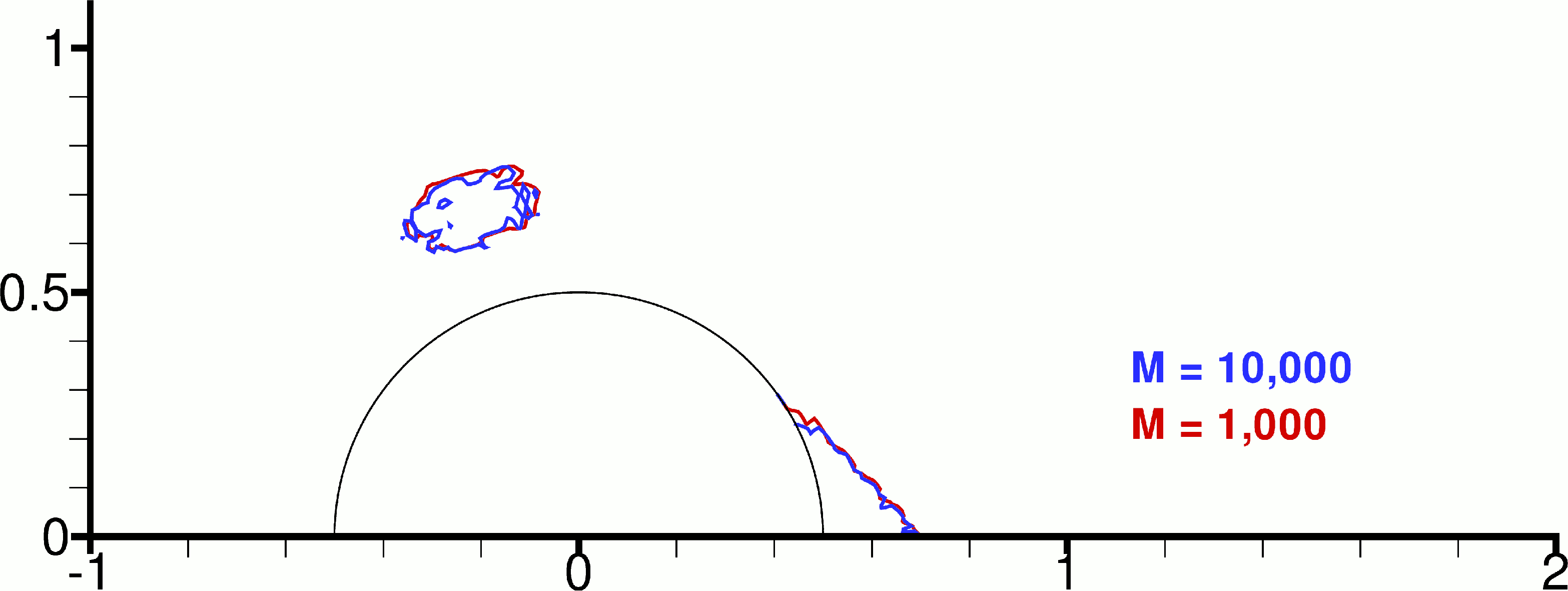}}
  \subfigure[] {\label{sfig: Bn=5 yield lines middle}
   \includegraphics[scale=1.08]{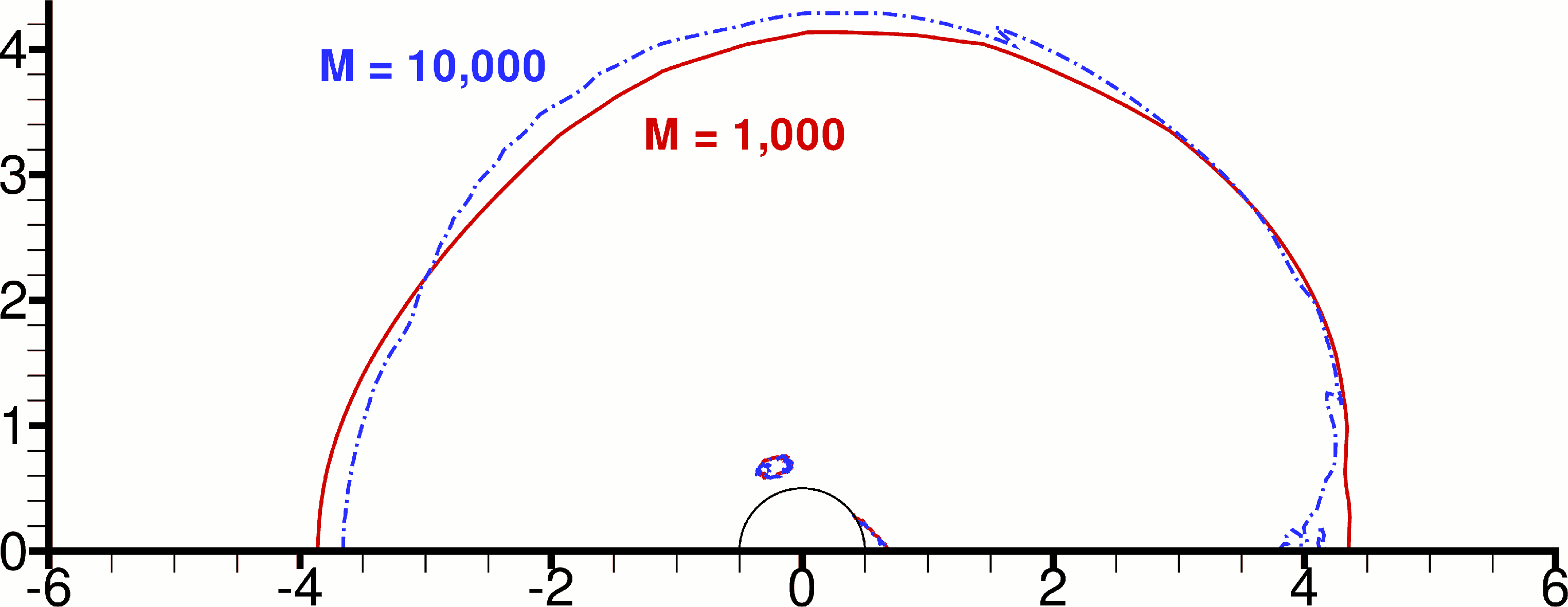}}
  \subfigure[] {\label{sfig: Bn=5 stress contours close}
   \includegraphics[scale=1.08]{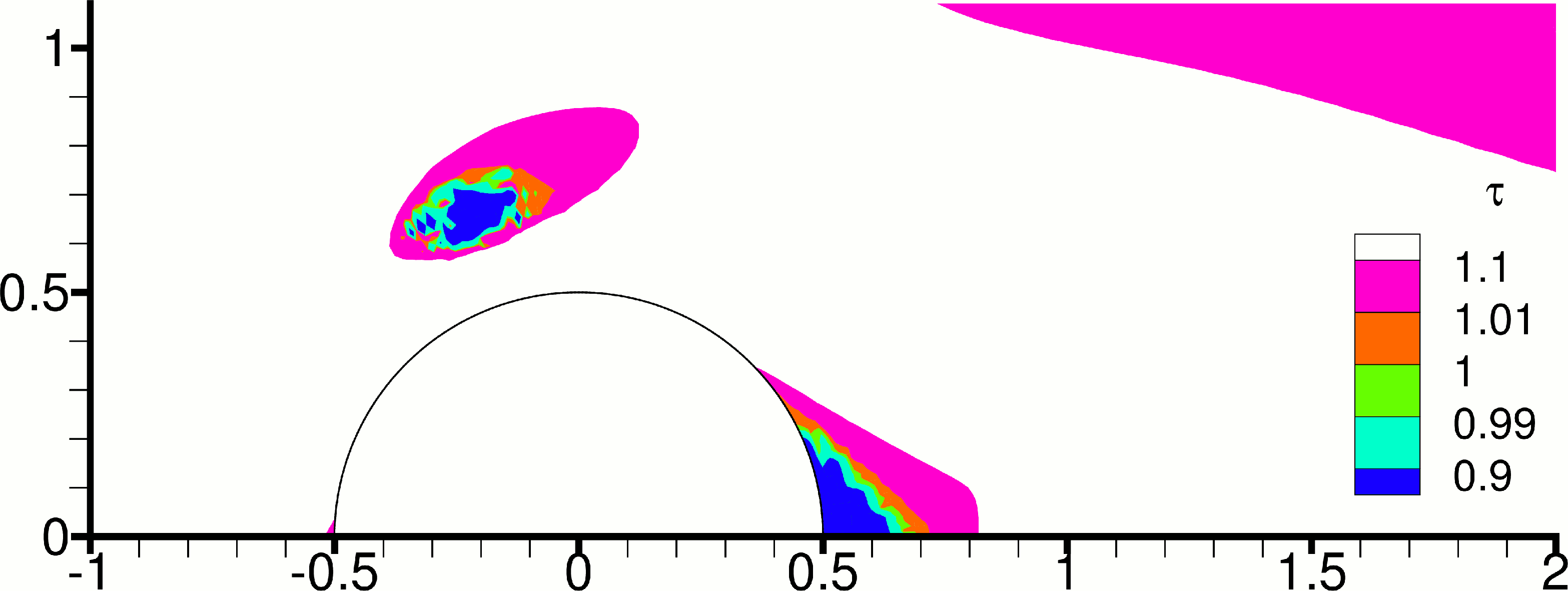}}
  \subfigure[] {\label{sfig: Bn=5 stress contours middle}
   \includegraphics[scale=1.08]{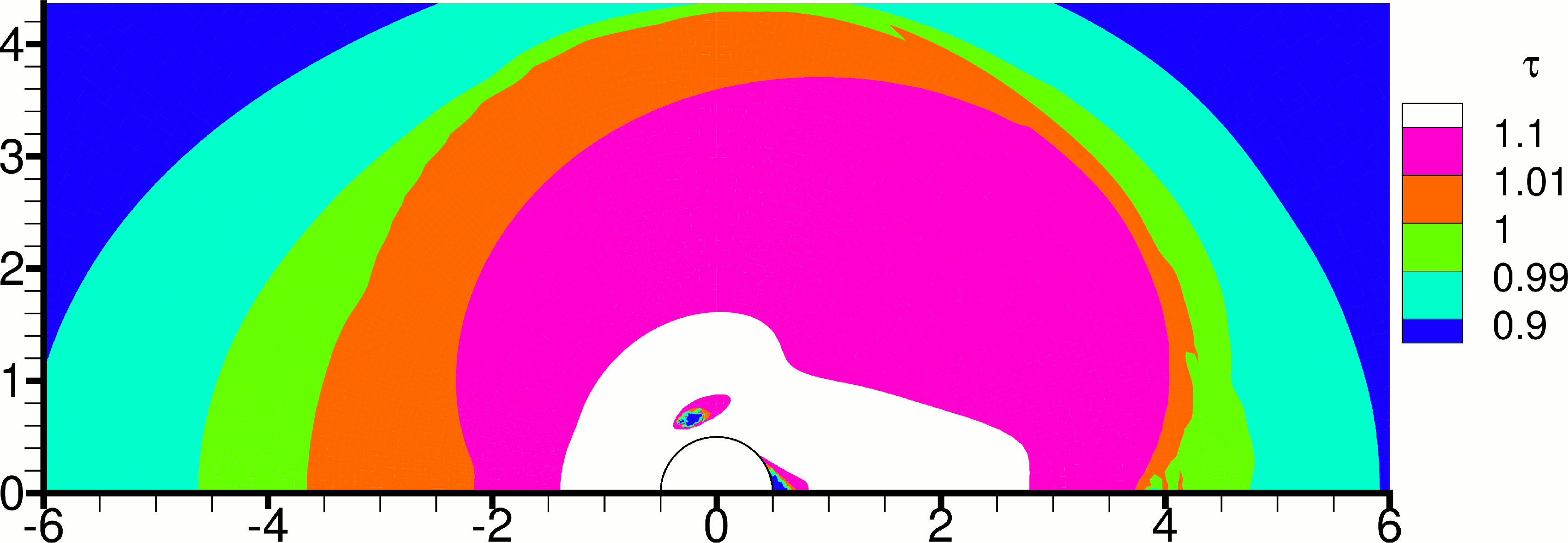}}
 \caption{For the $Re=45$, $Bn=5$ case without thixotropy, the top row shows yield lines 
approximated by $\tau = 1$) computed with different regularisation parameters, $m$ = 1000 and $m$ = 
10000, while the bottom row shows contours of nondimensional stress $\tau$ computed with $m$ = 
10000.}
 \label{fig: evaluation of M Bn=5}
\end{figure}

\begin{figure}[tb]
 \centering
  \subfigure[$Bn=0.5$] {\label{sfig: Bn=0.5 stress over cylinder}
   \includegraphics[scale=1.00]{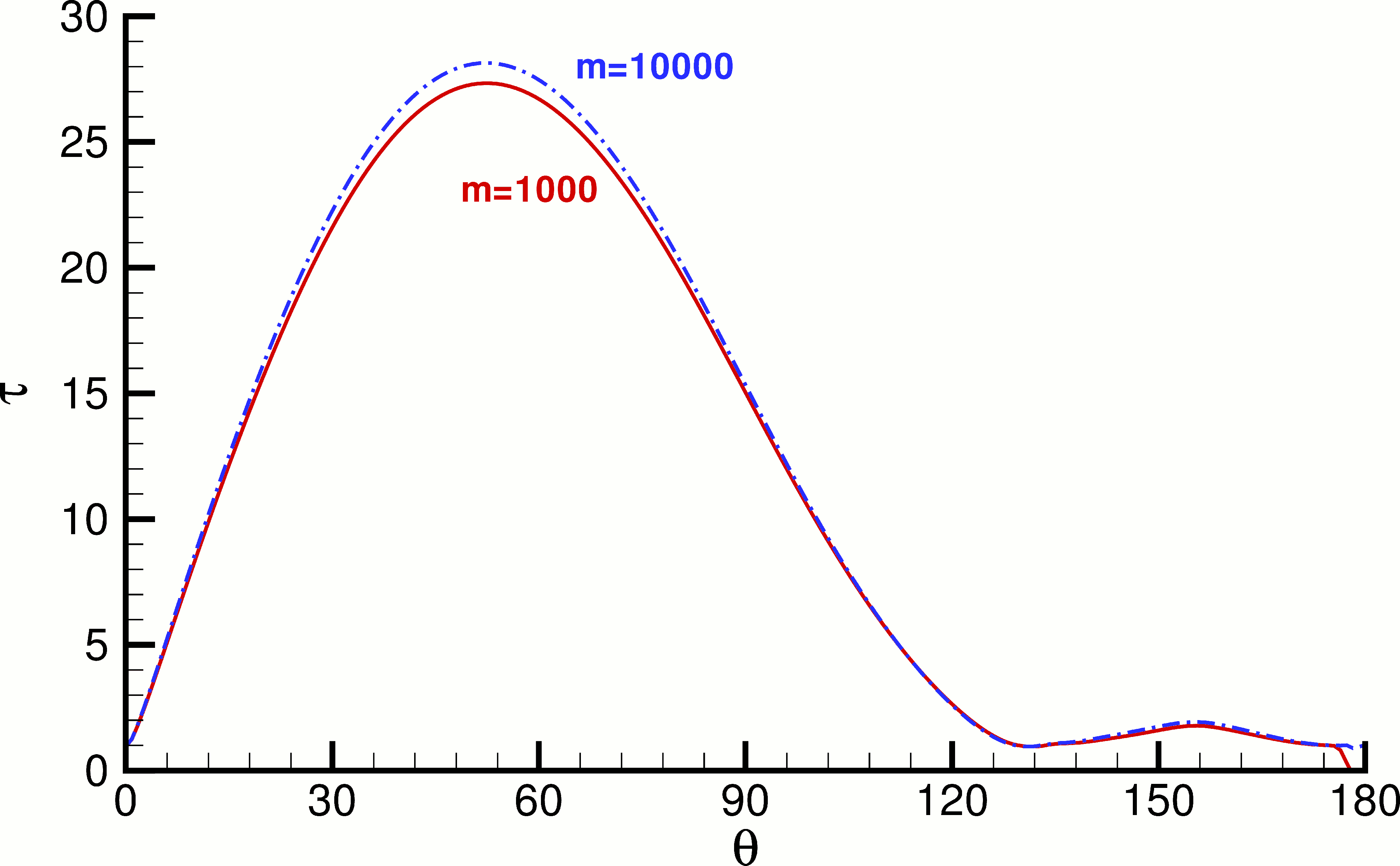}}
  \subfigure[$Bn=5$] {\label{sfig: Bn=5 stress over cylinder}
   \includegraphics[scale=1.00]{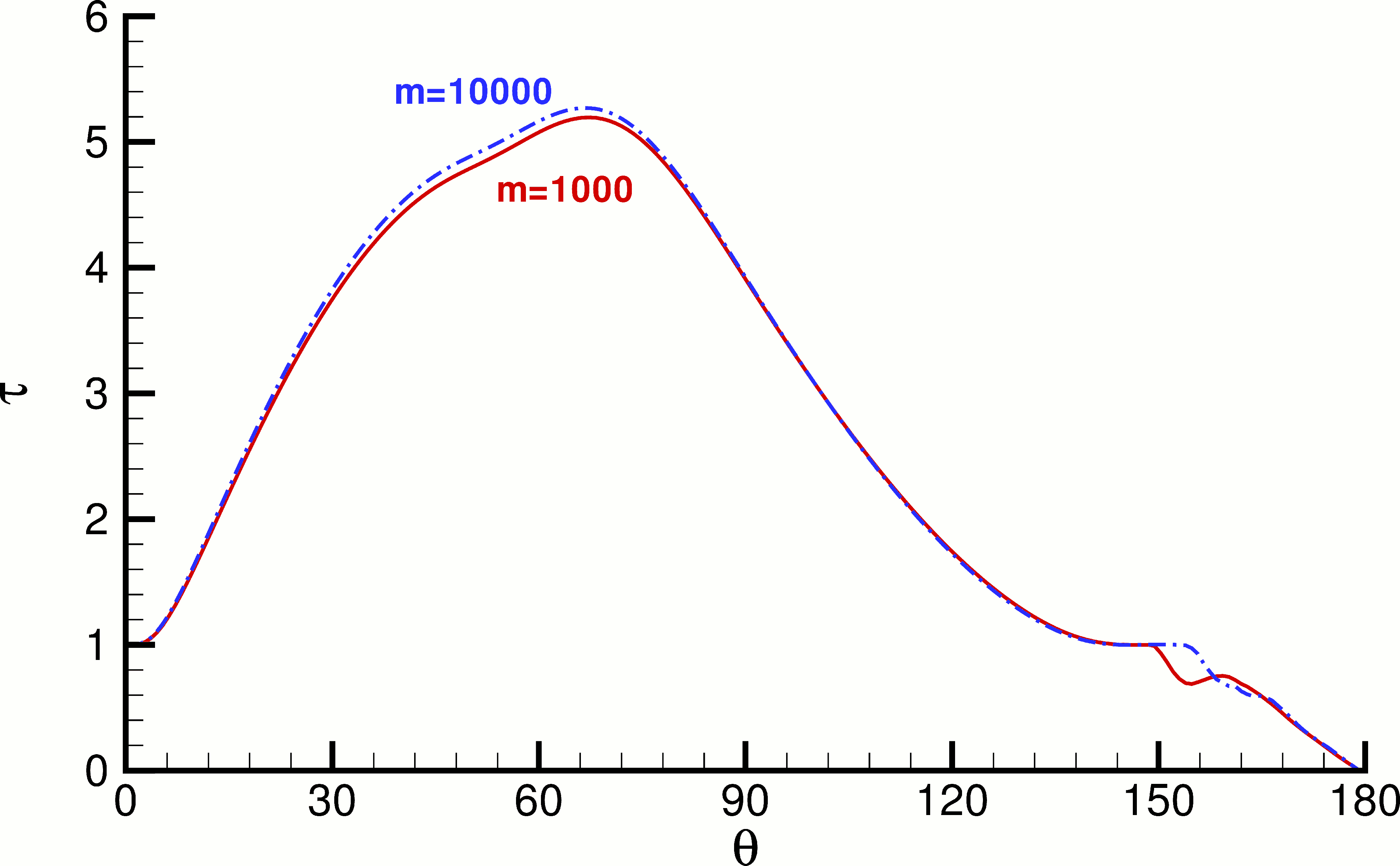}}
 \caption{Non-dimensional stress distributions over the cylinder surface for $Bn=0.5$ (left) and 
$Bn=5$ (right), without thixotropy. $\theta$ is the angle measured clockwise in degrees, starting 
from the upstream edge of the cylinder ($\theta = 0^{\circ}$).}
 \label{fig: stress over cylinder}
\end{figure}

Before proceeding to the presentation of the thixotropic results, it is useful to discuss the choice 
of the regularisation parameter $m$. Figures \ref{fig: evaluation of M Bn=0.5} and \ref{fig: 
evaluation of M Bn=5} provide a comparison of the results with two different values of this 
parameter, $m = 1000$ and $m = 10000$, both close to, and far from, the cylinder. Figures \ref{sfig: 
Bn=0.5 yield lines close} and \ref{sfig: Bn=5 yield lines close} show that, concerning the yield 
lines, estimated using $\tau = 1$, both values of $m$ produce essentially the same results near the 
cylinder, the main difference being that the results with $m = 10000$ are not smooth. This 
observation, that the predicted yield lines lose their smoothness when $m$ is large, has been noted 
also in other studies (e.g. \cite{Burgos_99b, Mitsoulis_01}). On the other hand, Figure \ref{sfig: 
Bn=0.5 yield lines far} shows that, for $Bn = 0.5$, the different $m$ parameters produce very 
different yield surfaces far from the cylinder: the yield surface produced with $m = 10000$ is 
located much farther away from the cylinder than that produced with $m = 1000$. Figure \ref{sfig: 
Bn=0.5 stress contours far} sheds some light into this: it can be seen that between the two yield 
surfaces (that computed with $m = 1000$ and that computed with $m = 10000$) the stress field changes 
by less than one percent. Therefore, if changing $m = 1000$ to $m = 10000$ causes a one percent 
change in the stress field, then this change in the location of the yield line is expected. The 
problem lies in the very gradual variation of the stress field near the yield stress, which is a 
problematic situation for predicting the yield surfaces with regularisation methods, as pointed out 
by Frigaard and Nouar \cite{Frigaard_05}. Close to the cylinder (Fig.\ \ref{sfig: Bn=0.5 stress 
contours close}) the variation of stress is much more rapid and the problem is not manifested. In 
the $Bn = 5$ case the stress varies relatively rapidly both close to the cylinder (Fig. \ref{sfig: 
Bn=5 stress contours close}) and far from it (Fig.\ \ref{sfig: Bn=5 stress contours middle}), so 
the results with $m = 1000$ and $m = 10000$ are similar even far from the cylinder (Fig.\ 
\ref{sfig: Bn=5 yield lines middle}). Another test is performed by comparing the stress magnitude 
distributions along the cylinder surface in Fig.\ \ref{fig: stress over cylinder}, and also the 
drag coefficients obtained with the different values of $m$ in Table \ref{table: Cd(m)}. The drag 
coefficient is defined as
\begin{equation} \label{eq: drag coefficient}
 C_D \;\equiv\; \frac{F_D}{\frac{1}{2}\rho U^2 D}
\end{equation}
where $F_D$ is the total horizontal force on the cylinder (due to both pressure and shear stress). 
The force $F_D$ equals twice the value computed for half the cylinder in our half-domain. Table 
\ref{table: Cd(m)} shows that, for $Bn = 0.5$, using $m = 10000$ instead of $m = 1000$ results in a 
relative difference of 2.77 \% for the drag coefficient, which is small but not negligible. For $Bn 
= 5$ the relative difference is much smaller, 0.45 \%. These observations agree with those of 
previous studies such as \cite{Tsamopoulos_96, Burgos_99}, where it was suggested that lower values 
of $m$ can be used with higher values of $Bn$. See however \cite{Syrakos_14} for a counterexample. 
In the present study it was decided to use a value of $m$ = 1000, because the focus is close to the 
cylinder, and $m$ = 1000 produces smoother results.

\begin{table}[tb]
\caption{Values of drag coefficient computed with different values of regularisation parameter $m$, 
for Bingham flow without thixotropy.}
\label{table: Cd(m)}
\begin{center}
\begin{tabular}{l|cc} \hline
                  &   $Bn$ = 0.5 &    $Bn$ = 5 \\ \hline
$m$ = 1000        &       1.7789 &      4.7947 \\
$m$ = 10000       &       1.8296 &      4.8162 \\
\text{difference} &      2.77 \% &     0.45 \% \\ \hline
\end{tabular}
\end{center}
\end{table}

\begin{figure}[!b]
 \centering
  \subfigure[$t=1$] {\label{sfig: Bn=0.5 t=1}
   \includegraphics[scale=1.00]{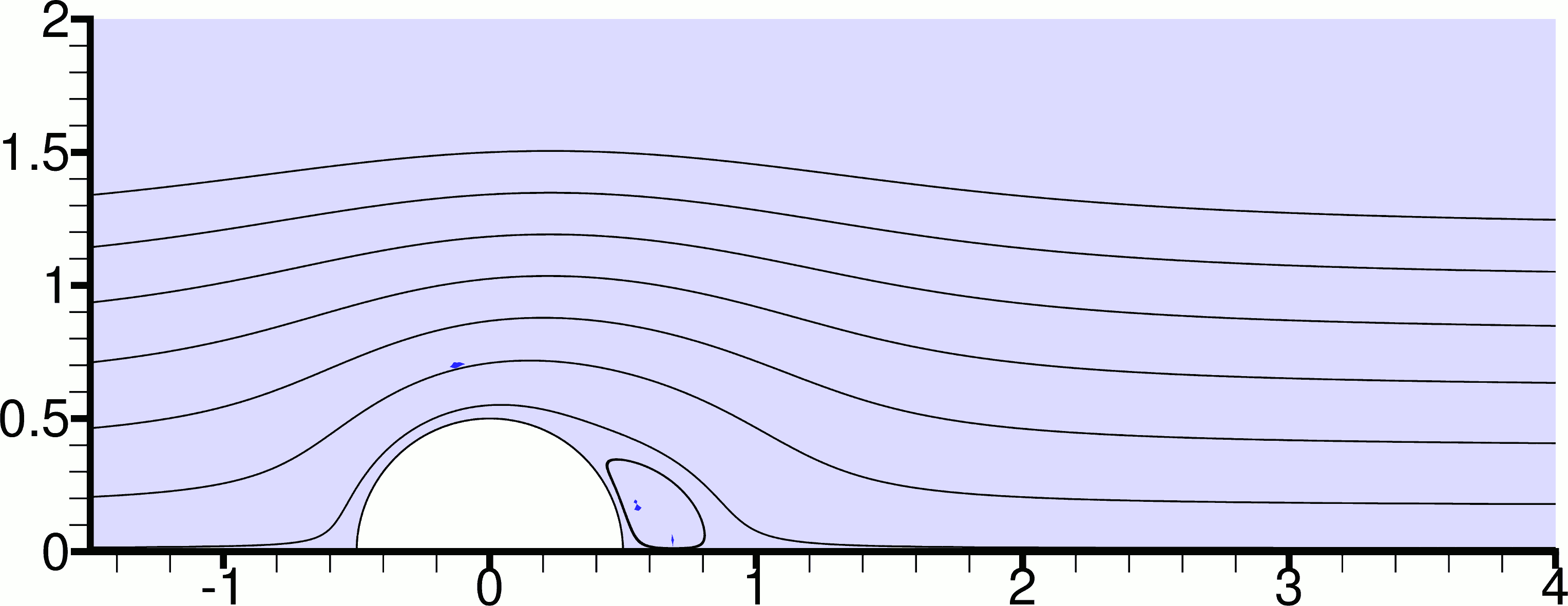}}
  \subfigure[$t=1$] {\label{sfig: Bn=0.5 t=1 l}
   \includegraphics[scale=1.00]{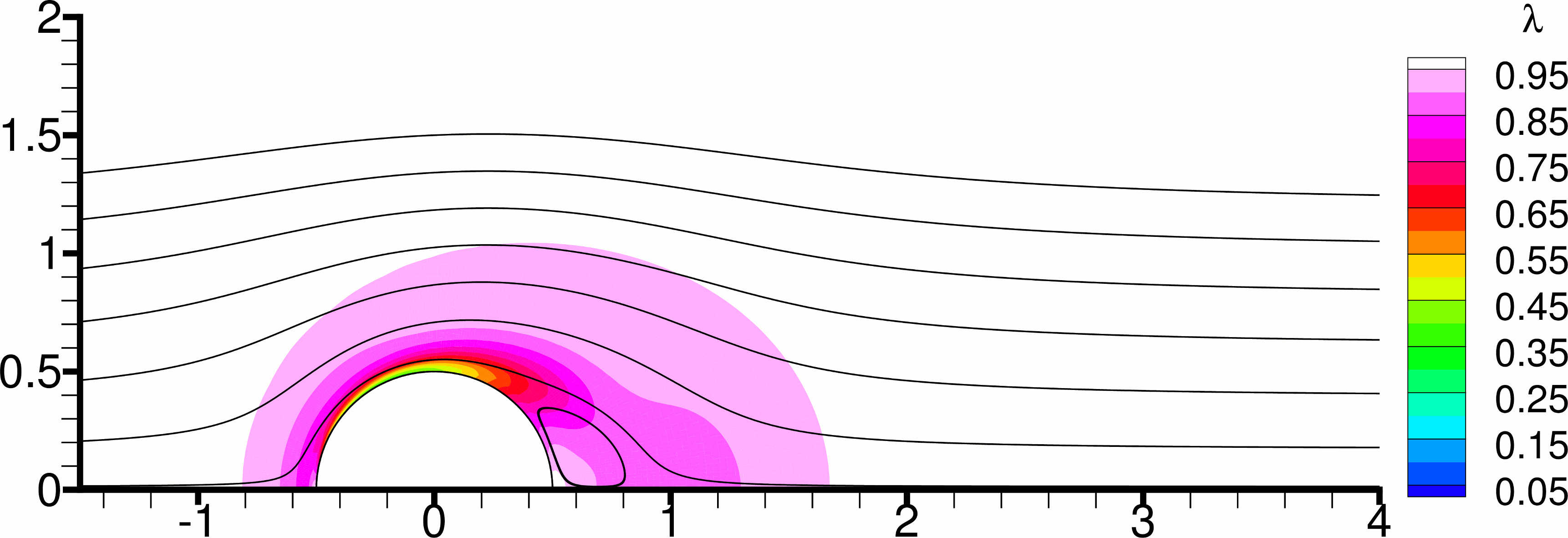}}
  \subfigure[$t=5$] {\label{sfig: Bn=0.5 t=5}
   \includegraphics[scale=1.00]{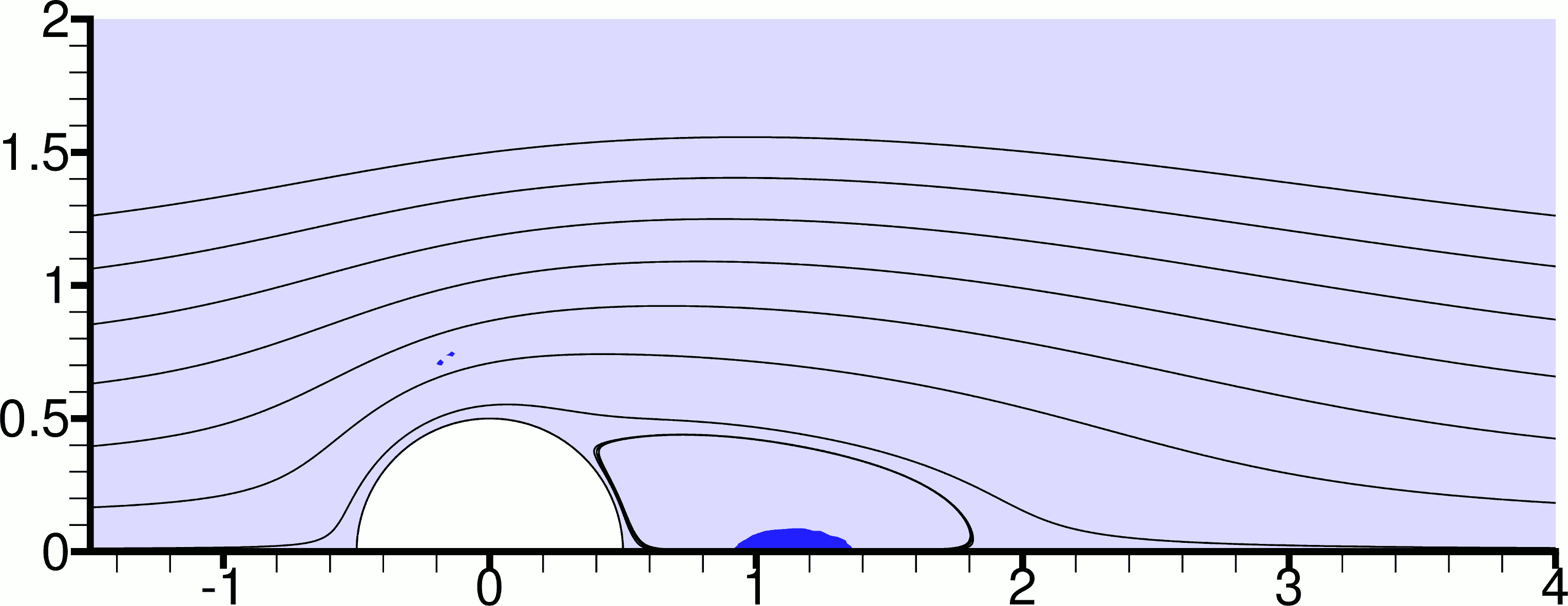}}
  \subfigure[$t=5$] {\label{sfig: Bn=0.5 t=5 l}
   \includegraphics[scale=1.00]{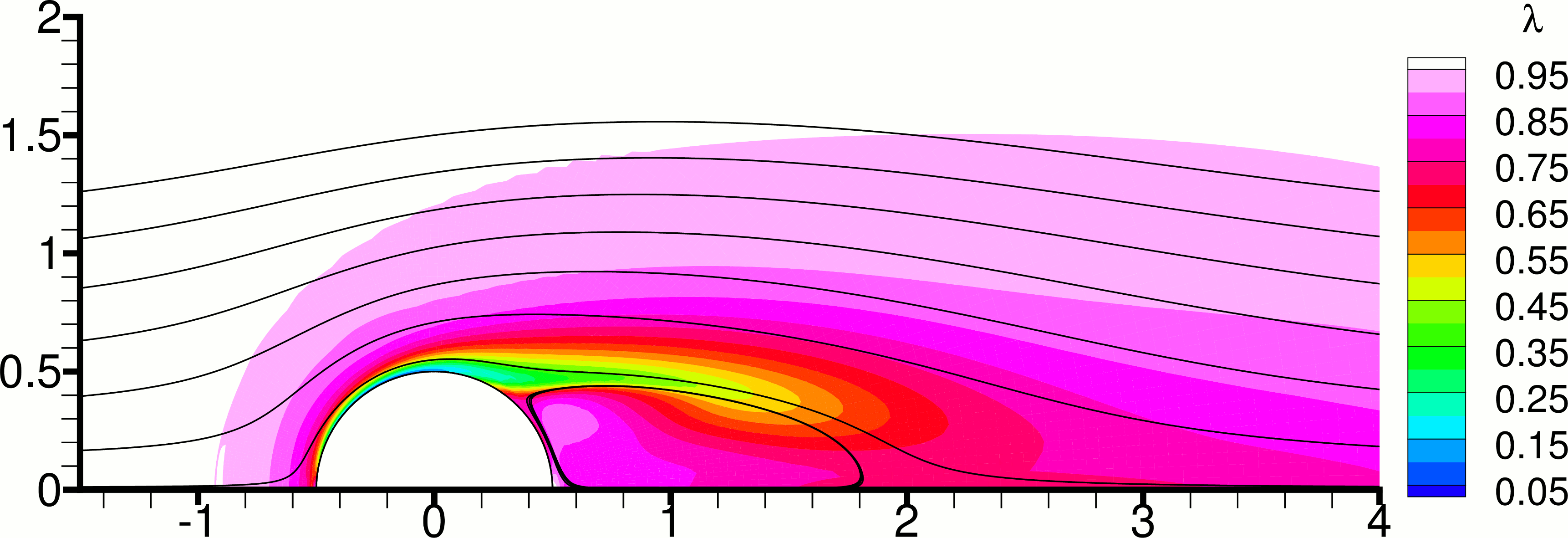}}
  \subfigure[$t=25$] {\label{sfig: Bn=0.5 t=25}
   \includegraphics[scale=1.00]{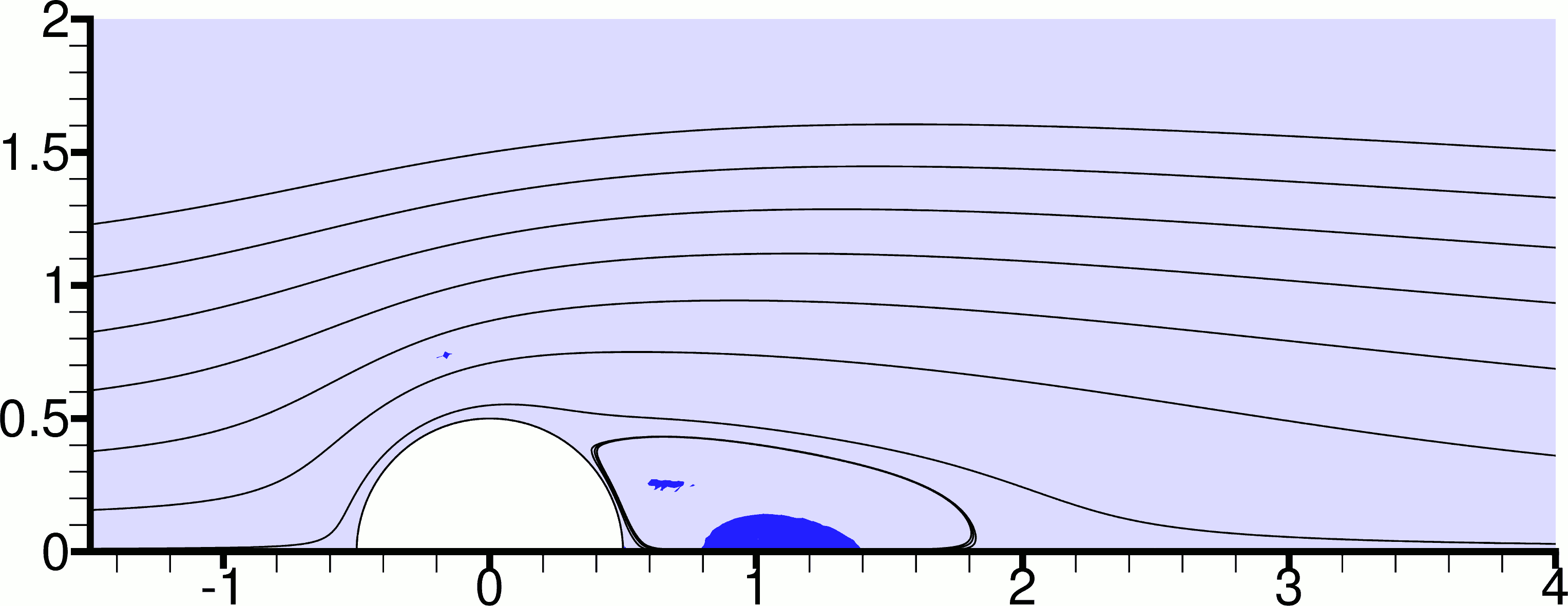}}
  \subfigure[$t=25$] {\label{sfig: Bn=0.5 t=25 l}
   \includegraphics[scale=1.00]{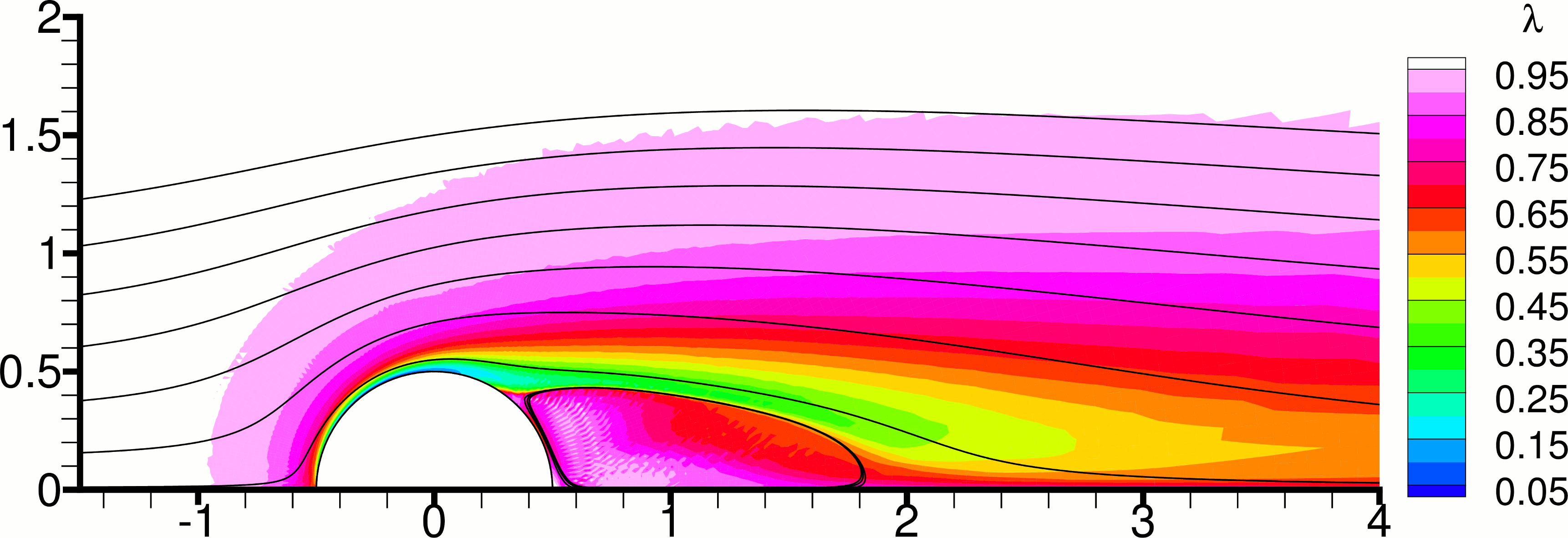}}
 \caption{Snapshots of the temporal evolution of the flow field at times $t$ = 1, 5 and 25 for $Re$ 
= 45, $Bn$ = 0.5, $\alpha = \beta$ = 0.05. The left figures show selected streamlines and unyielded 
zones (shaded), and the right figures show contours of $\lambda$.}
 \label{fig: transient Bn=0.5}
\end{figure}

\begin{figure}[tb]
 \centering
  \subfigure[$t=1$] {\label{sfig: Bn=5 t=1}
   \includegraphics[scale=1.00]{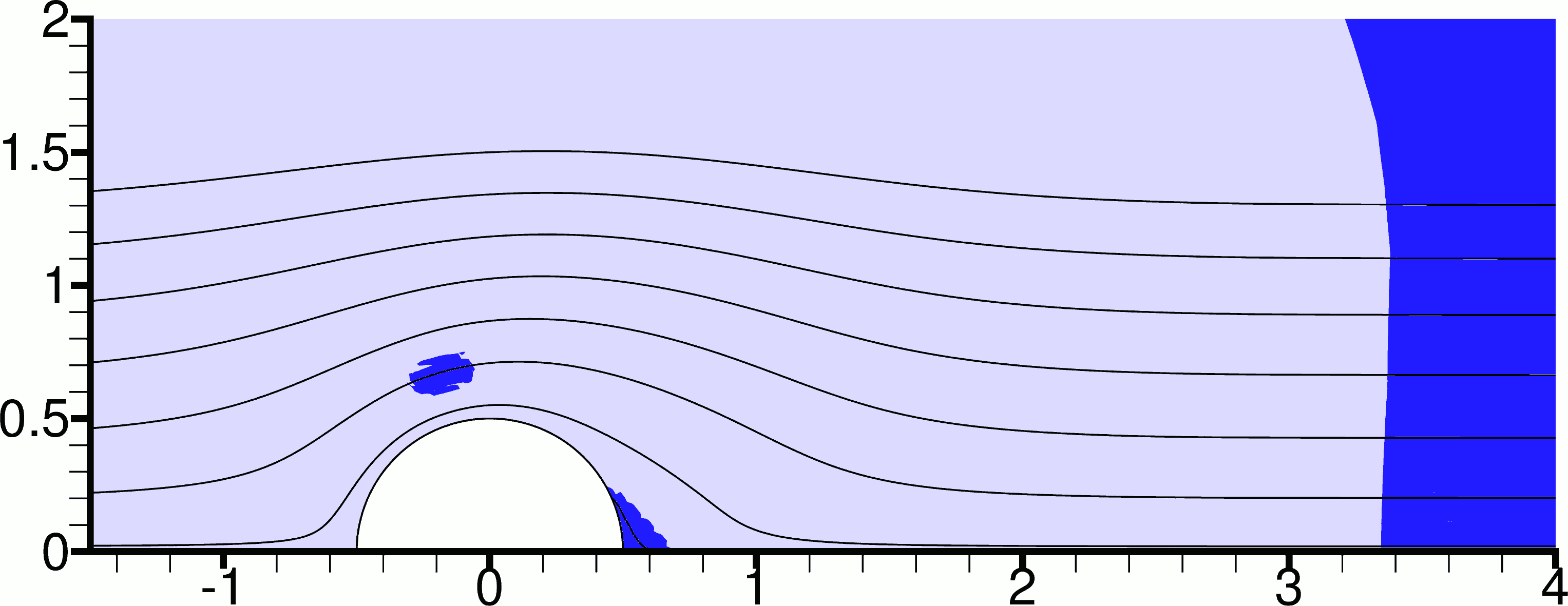}}
  \subfigure[$t=1$] {\label{sfig: Bn=5 t=1 l}
   \includegraphics[scale=1.00]{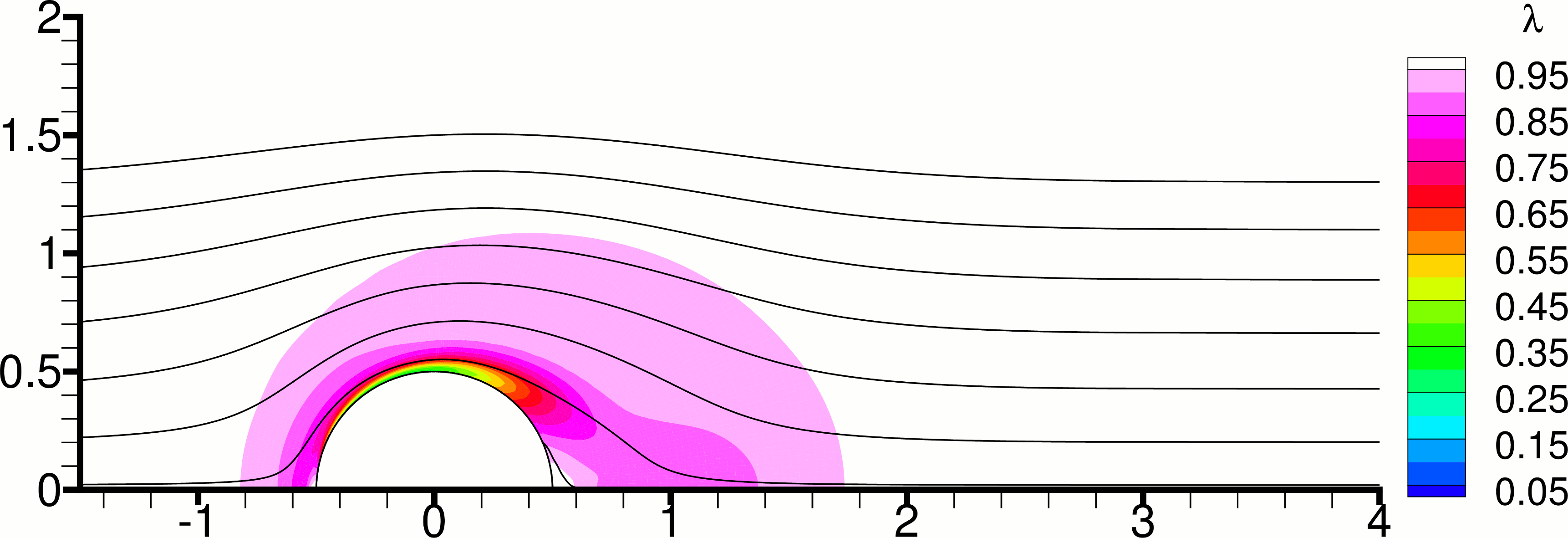}}
  \subfigure[$t=5$] {\label{sfig: Bn=5 t=5}
   \includegraphics[scale=1.00]{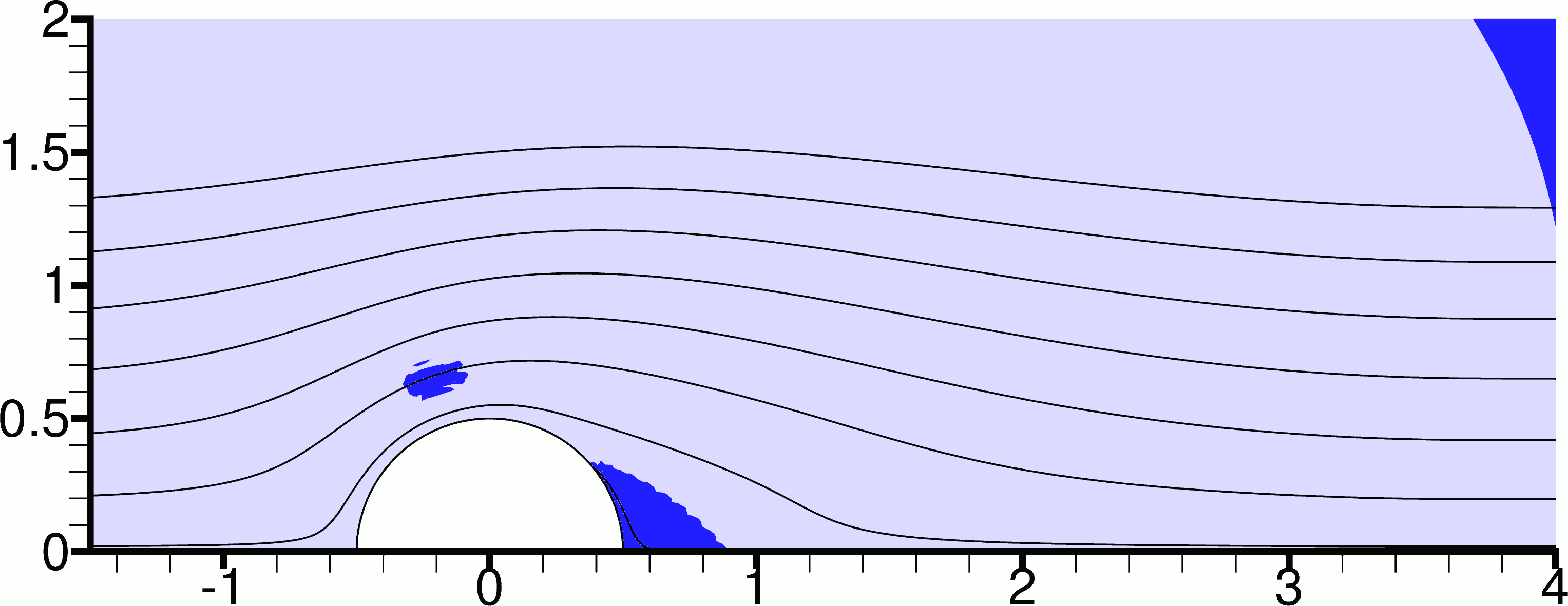}}
  \subfigure[$t=5$] {\label{sfig: Bn=5 t=5 l}
   \includegraphics[scale=1.00]{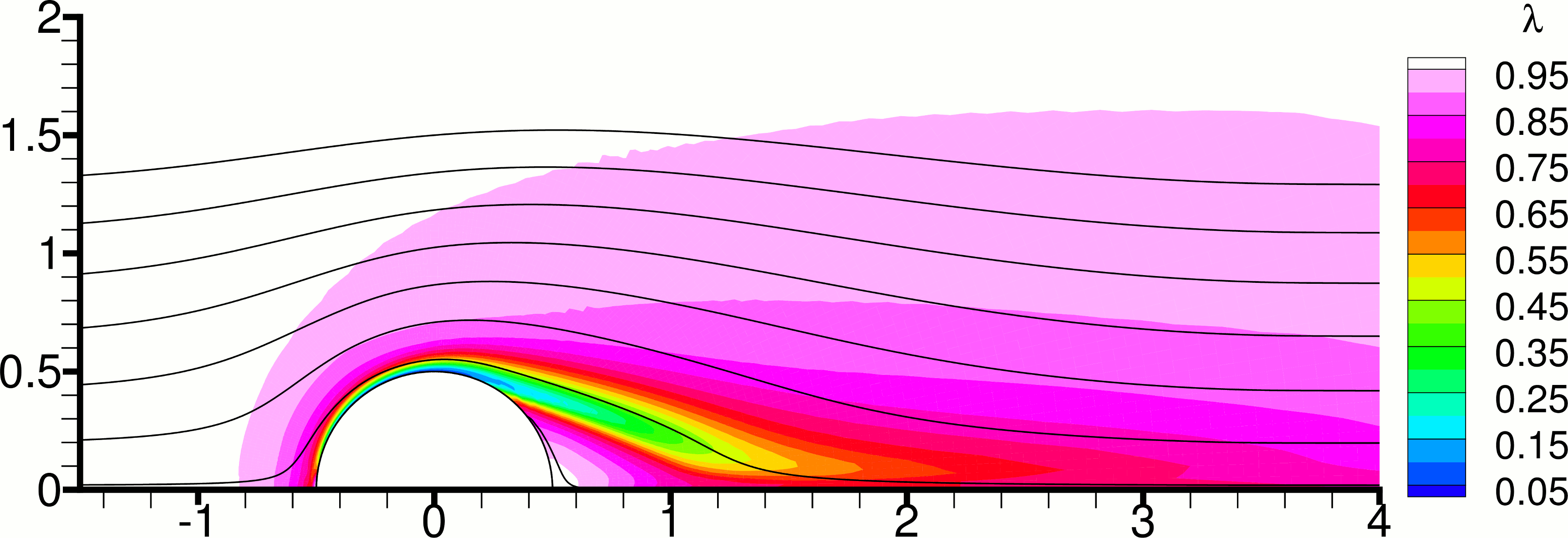}}
  \subfigure[$t=25$] {\label{sfig: Bn=5 t=25}
   \includegraphics[scale=1.00]{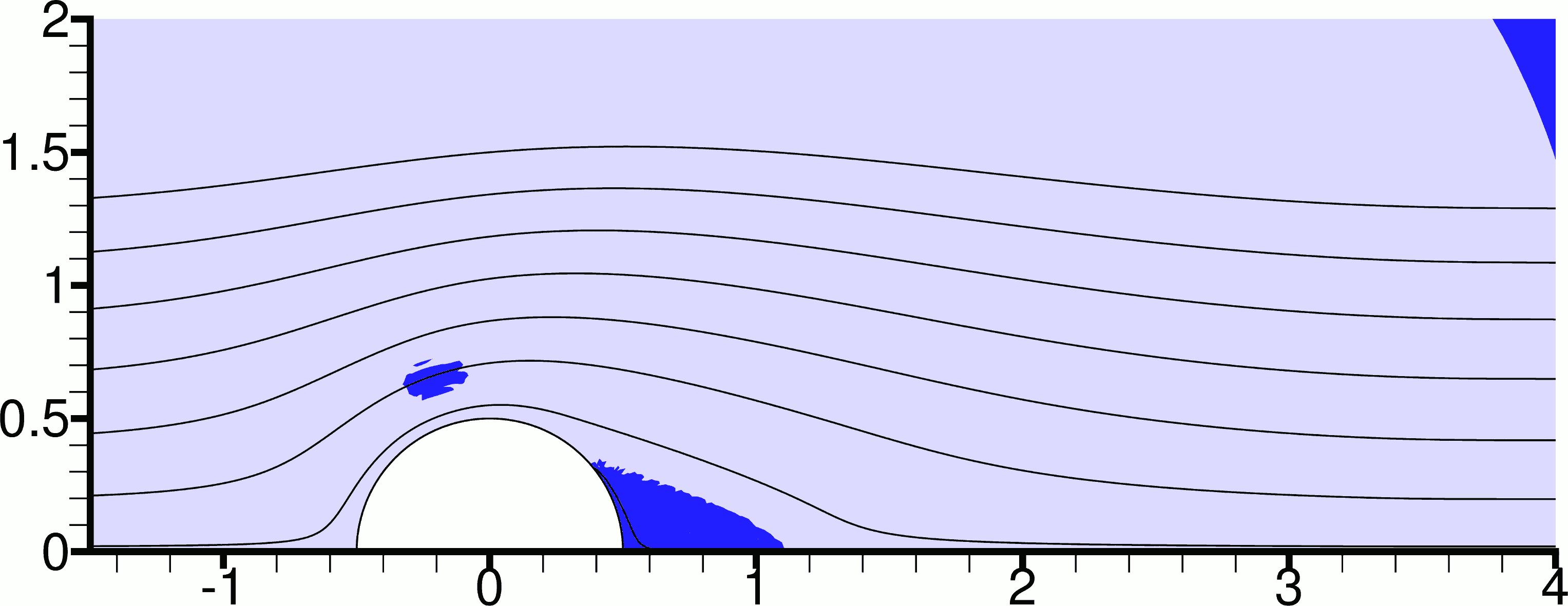}}
  \subfigure[$t=25$] {\label{sfig: Bn=5 t=25 l}
   \includegraphics[scale=1.00]{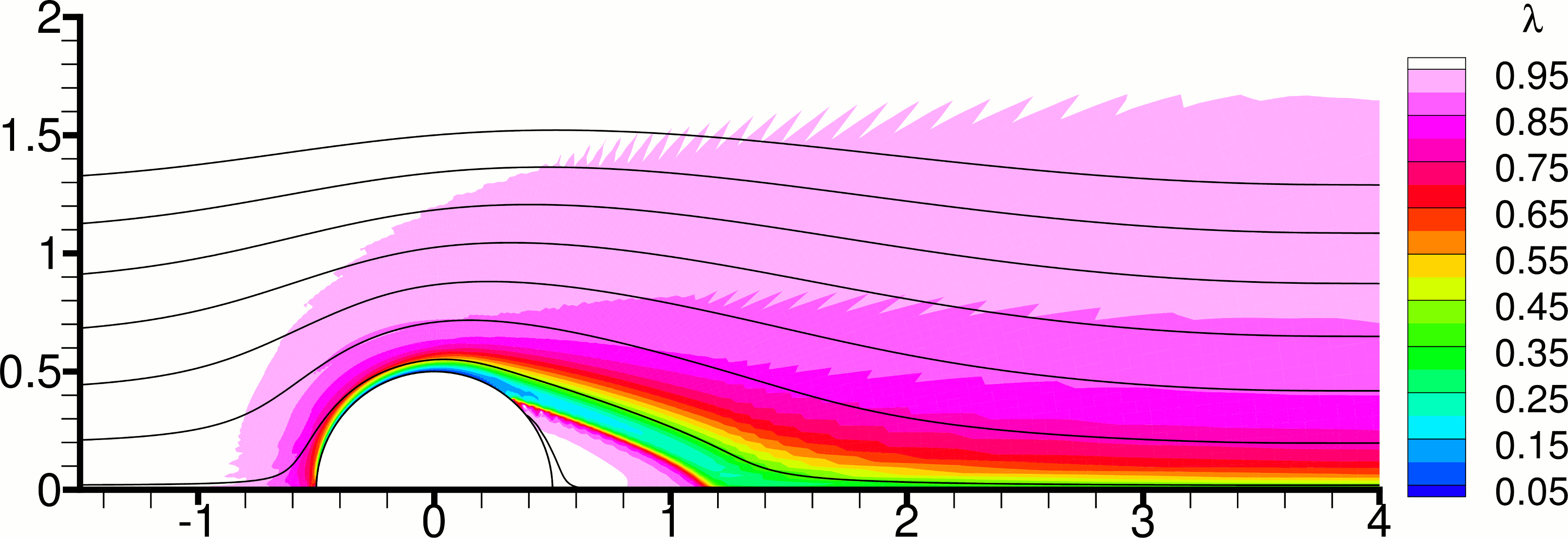}}
 \caption{Snapshots of the temporal evolution of the flow field at times $t$ = 1, 5 and 25 for $Re$ 
= 45, $Bn$ = 5, $\alpha = \beta$ = 0.05. The left figures show selected streamlines and unyielded 
zones (shaded), and the right figures show contours of $\lambda$.}
 \label{fig: transient Bn=5}
\end{figure}

\begin{figure}[tb]
 \centering
  \subfigure[$\alpha=0.01$] {\label{sfig: Bn=0.5 a=0.01}
   \includegraphics[scale=1.00]{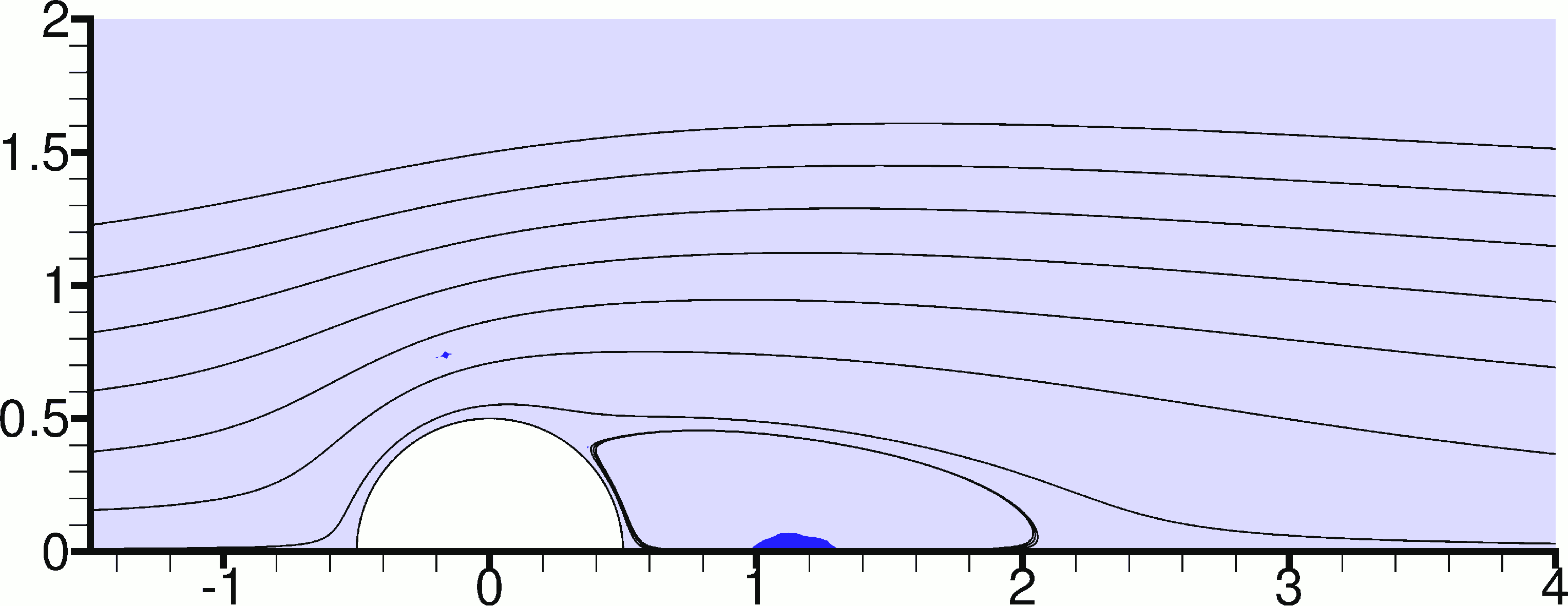}}
  \subfigure[$\alpha=0.01$] {\label{sfig: Bn=0.5 a=0.01 l}
   \includegraphics[scale=1.00]{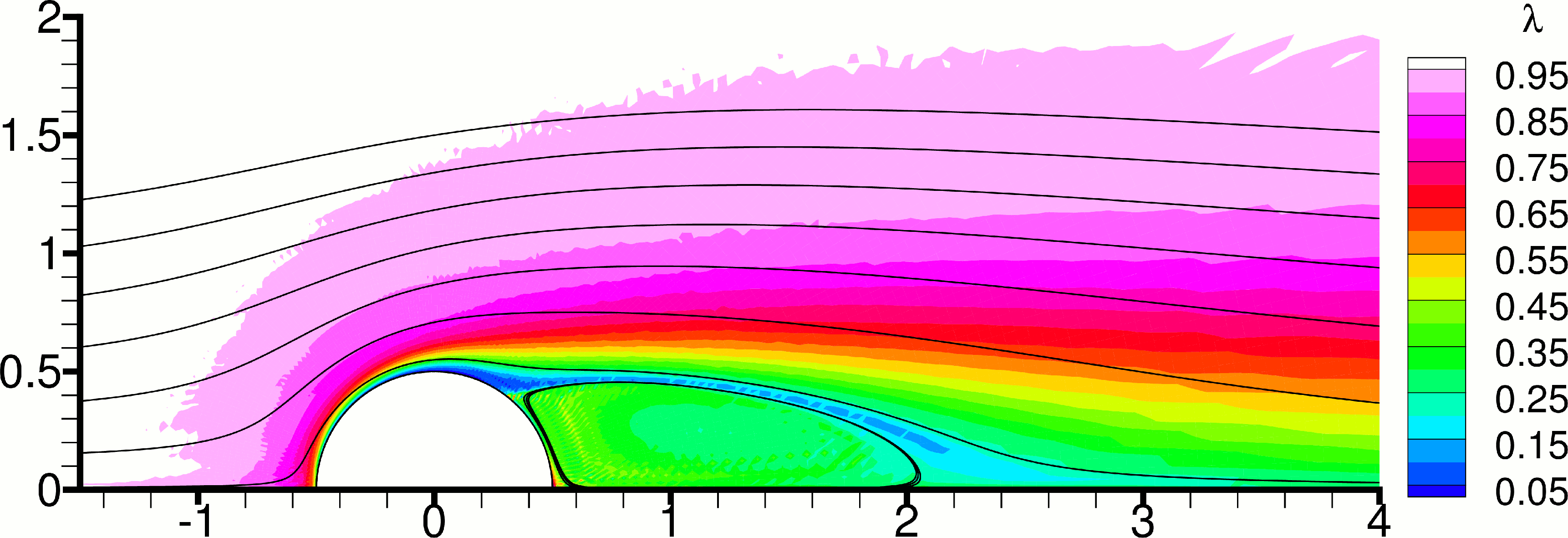}}
  \subfigure[$\alpha=0.05$] {\label{sfig: Bn=0.5 a=0.05}
   \includegraphics[scale=1.00]{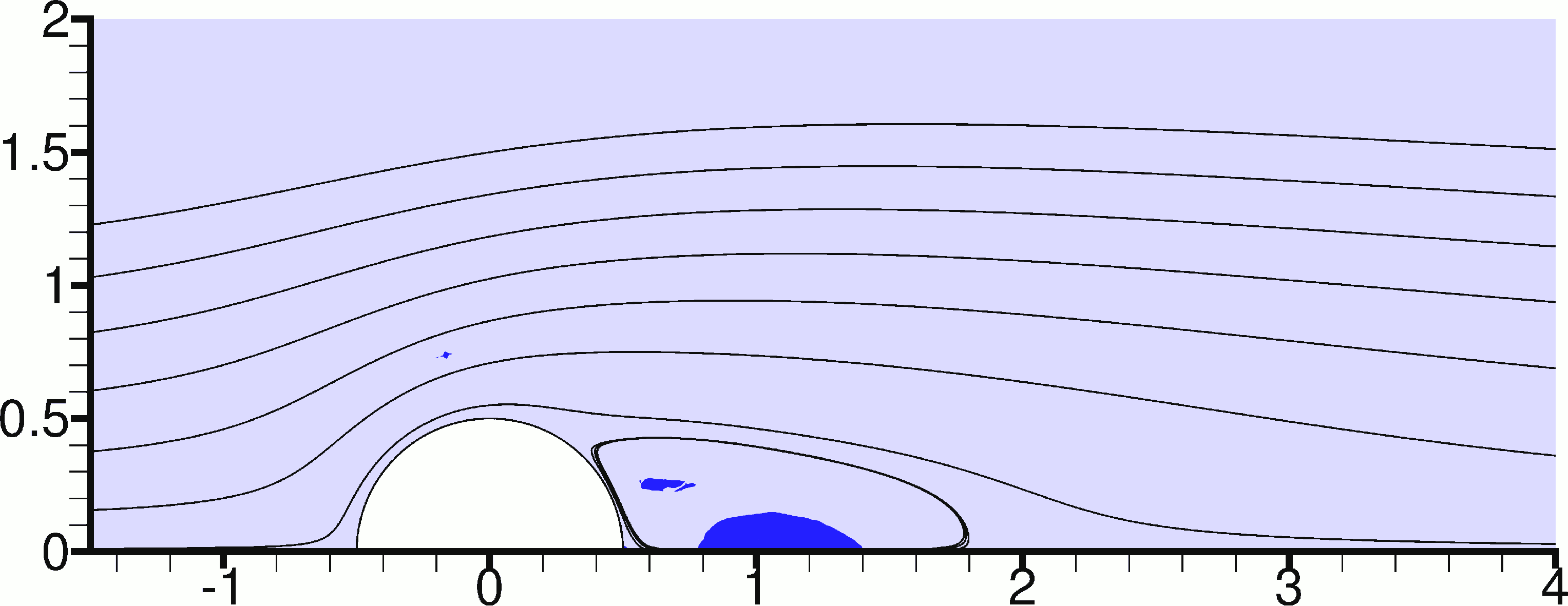}}
  \subfigure[$\alpha=0.05$] {\label{sfig: Bn=0.5 a=0.05 l}
   \includegraphics[scale=1.00]{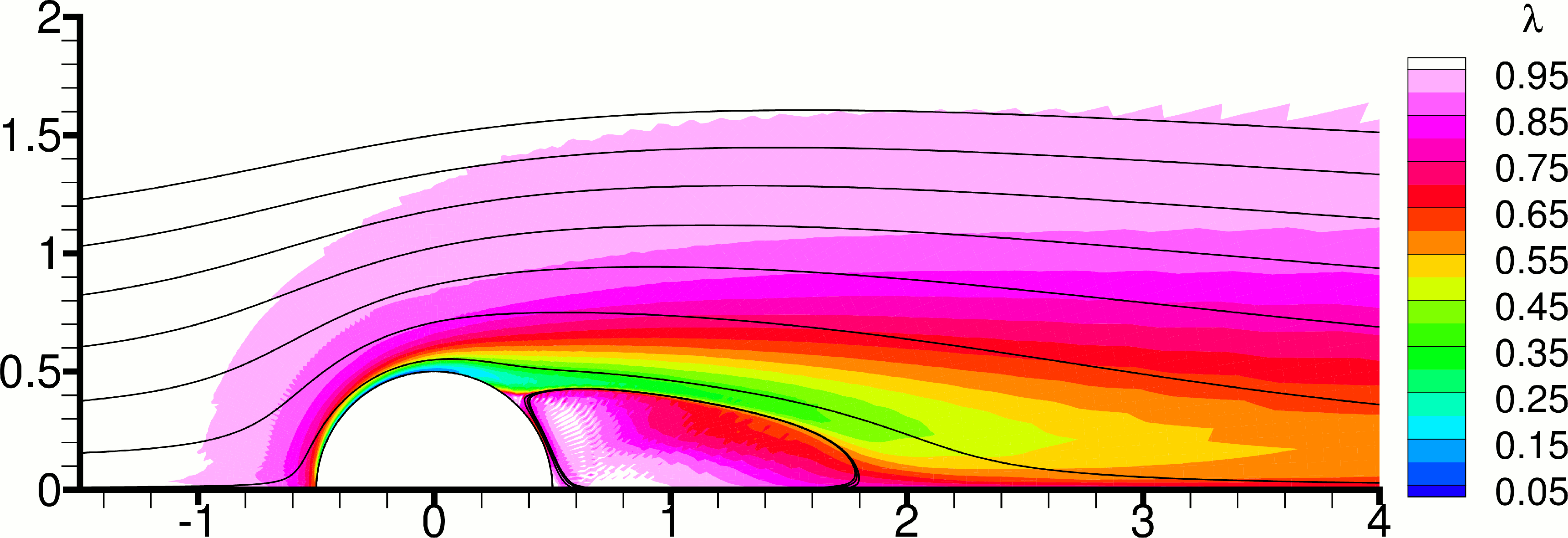}}
  \subfigure[$\alpha=0.10$] {\label{sfig: Bn=0.5 a=0.10}
   \includegraphics[scale=1.00]{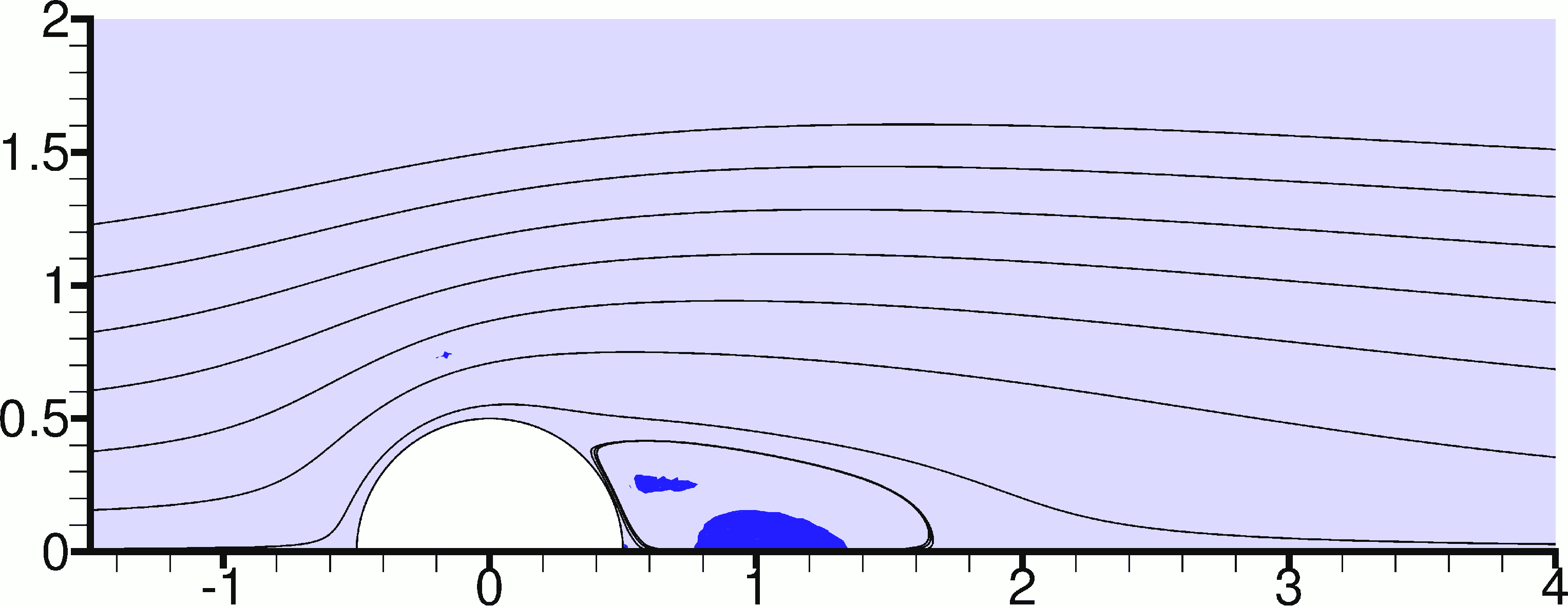}}
  \subfigure[$\alpha=0.10$] {\label{sfig: Bn=0.5 a=0.10 l}
   \includegraphics[scale=1.00]{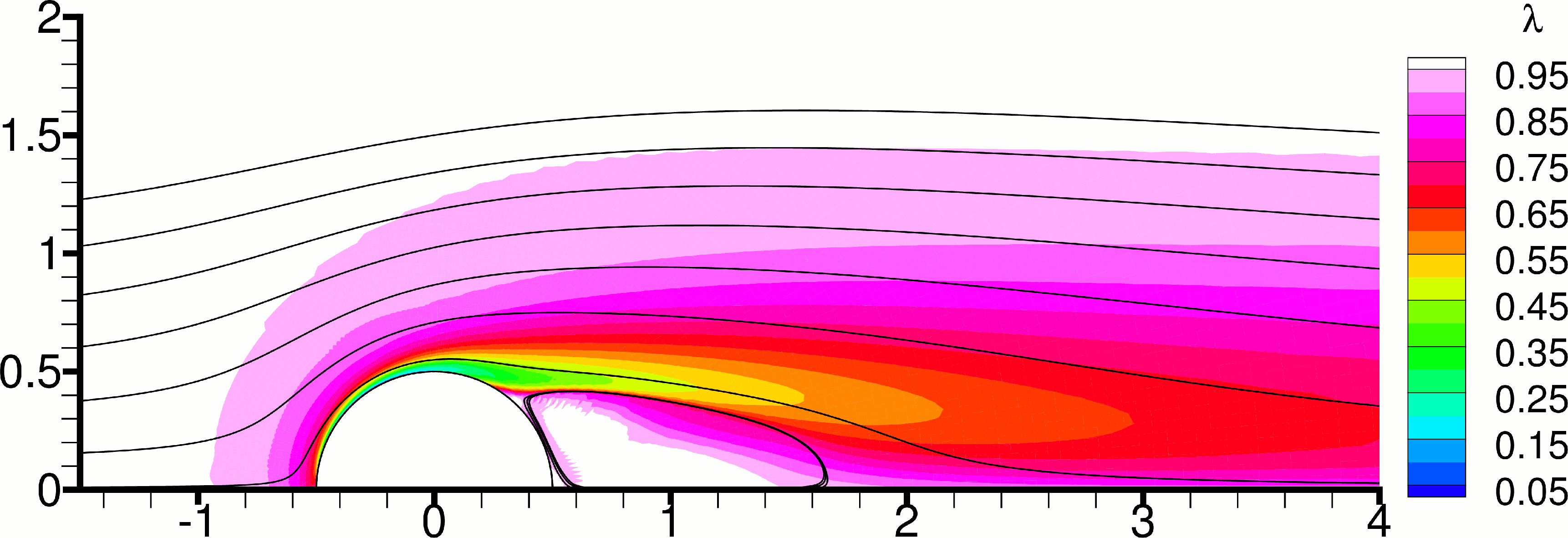}}
 \caption{Effect of the recovery parameter $\alpha$ for $Re = 45$, $Bn = 0.5$, $\beta = 0.05$, at 
$t$ = 120 (steady state). The left figures show streamlines and unyielded zones (shaded), and the 
right figures show contours of $\lambda$.}
 \label{fig: Bn=0.5 effect of a}
\end{figure}

\begin{figure}[tb]
 \centering
  \subfigure[$\alpha=0.01$] {\label{sfig: Bn=5 a=0.01}
   \includegraphics[scale=1.00]{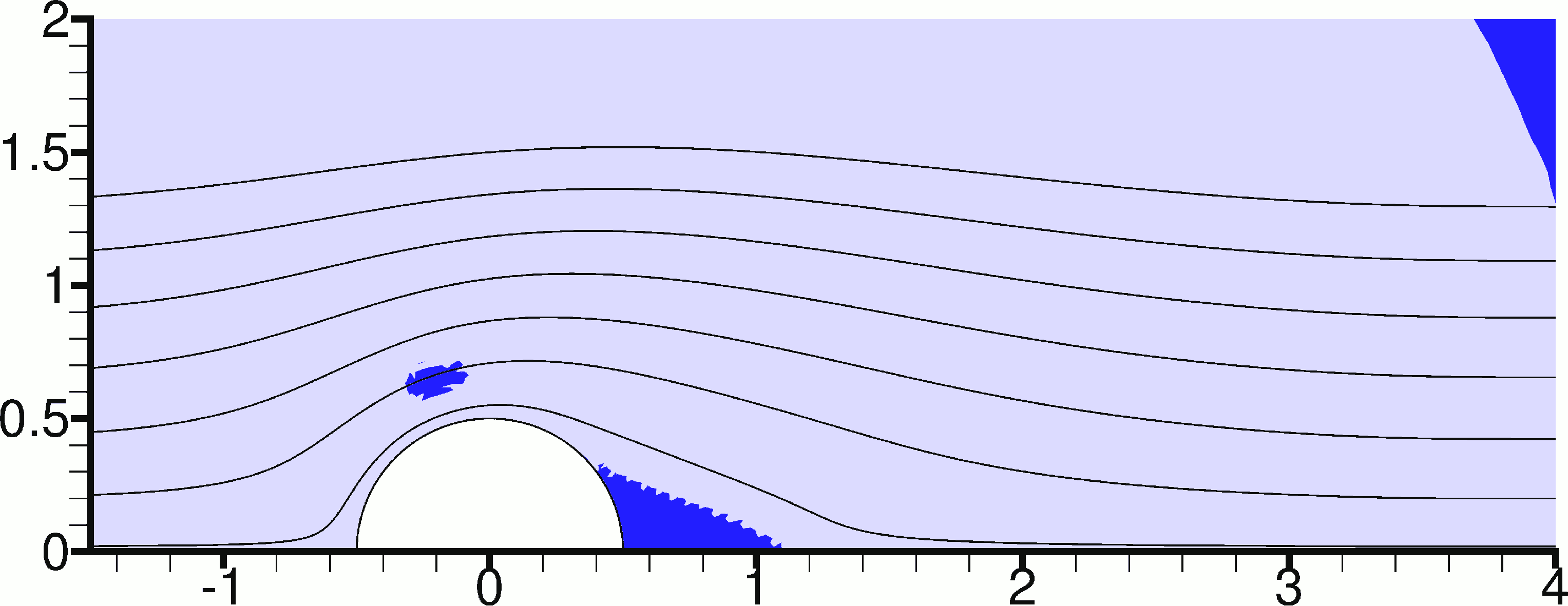}}
  \subfigure[$\alpha=0.01$] {\label{sfig: Bn=5 a=0.01 l}
   \includegraphics[scale=1.00]{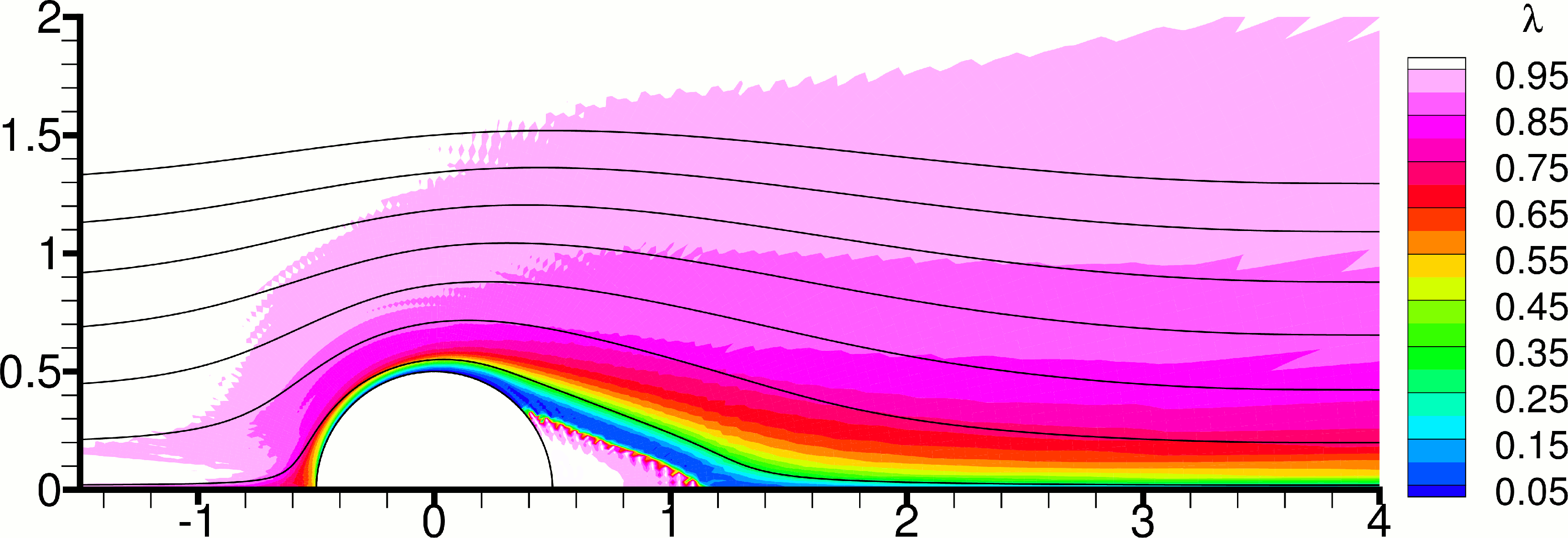}}
  \subfigure[$\alpha=0.05$] {\label{sfig: Bn=5 a=0.05}
   \includegraphics[scale=1.00]{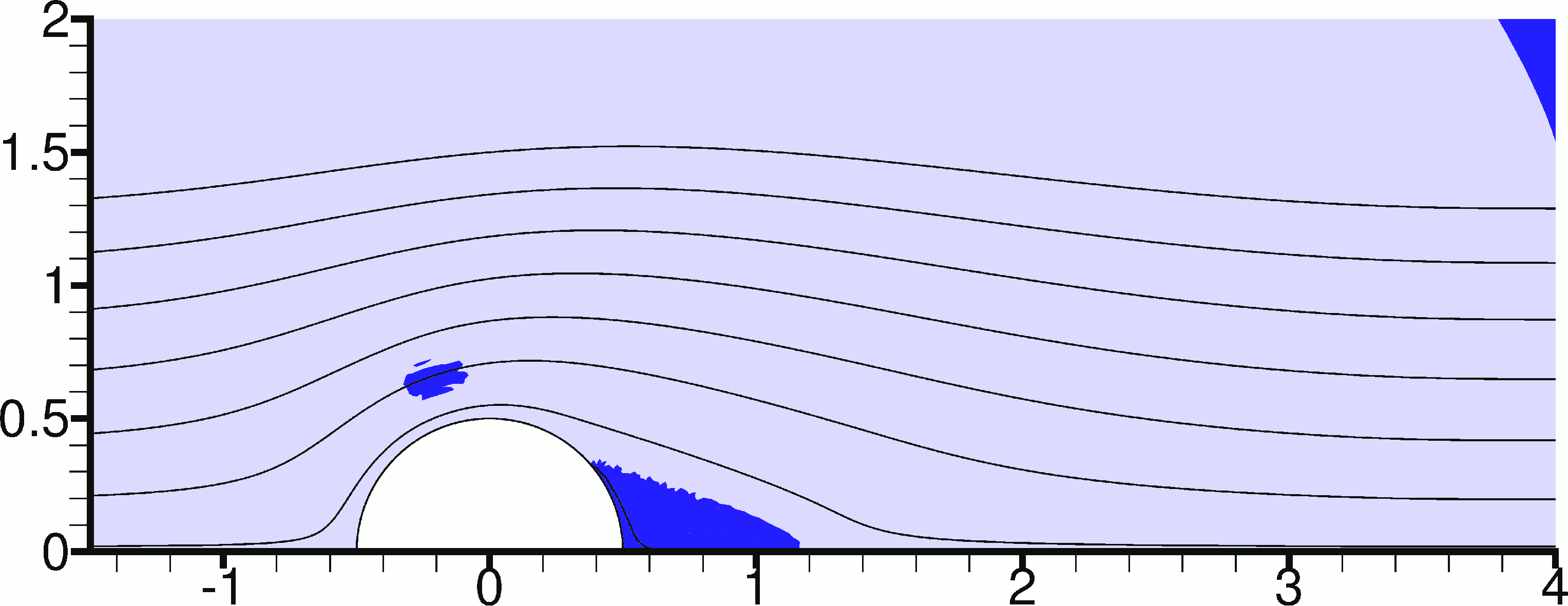}}
  \subfigure[$\alpha=0.05$] {\label{sfig: Bn=5 a=0.05 l}
   \includegraphics[scale=1.00]{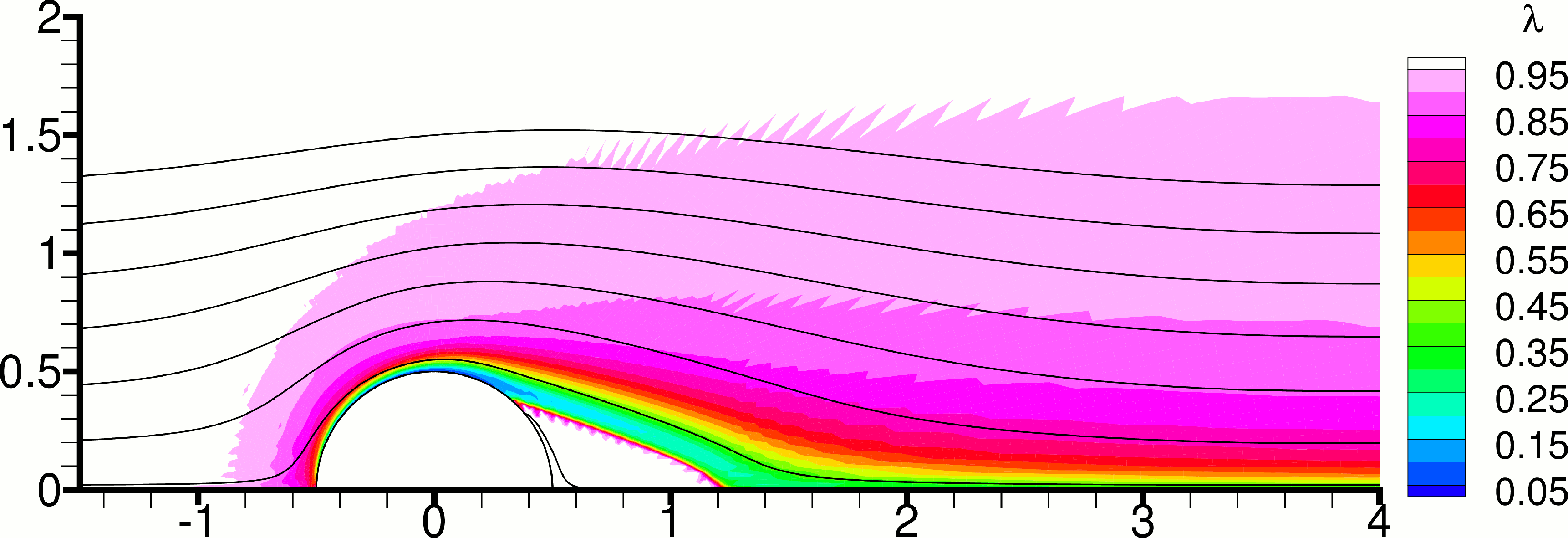}}
  \subfigure[$\alpha=0.10$] {\label{sfig: Bn=5 a=0.10}
   \includegraphics[scale=1.00]{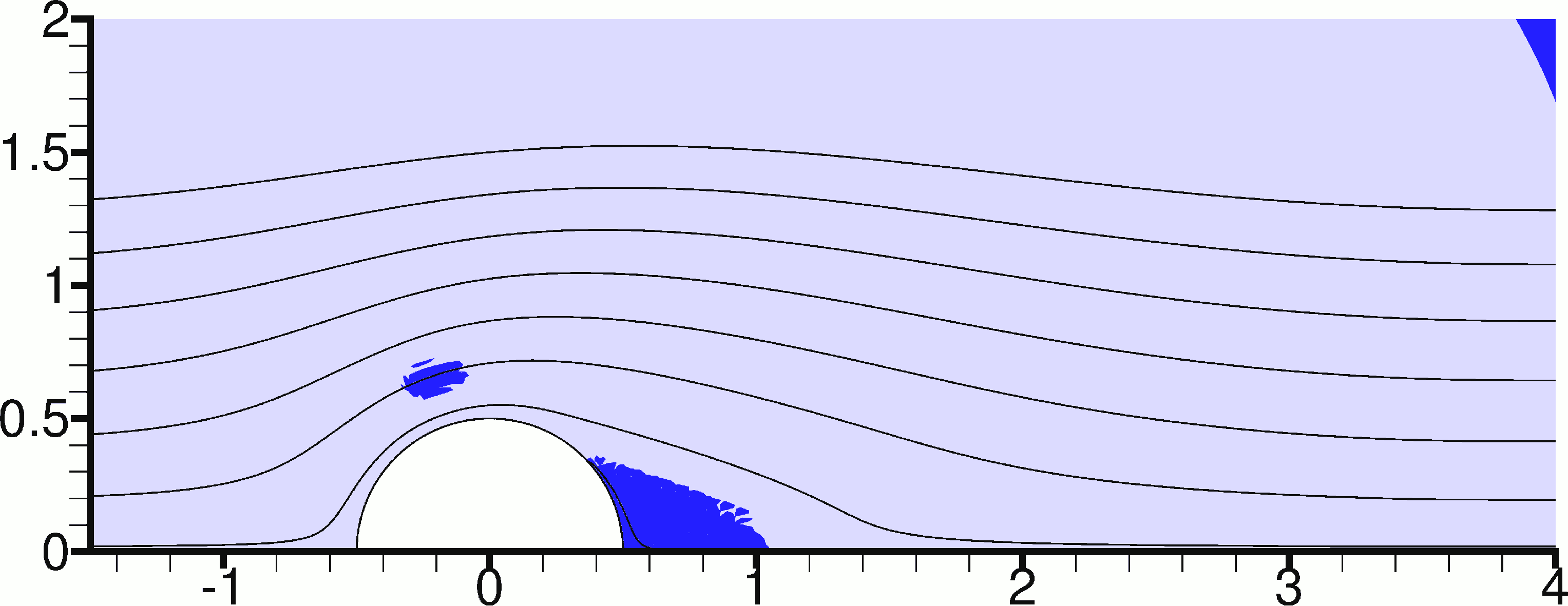}}
  \subfigure[$\alpha=0.10$] {\label{sfig: Bn=5 a=0.10 l}
   \includegraphics[scale=1.00]{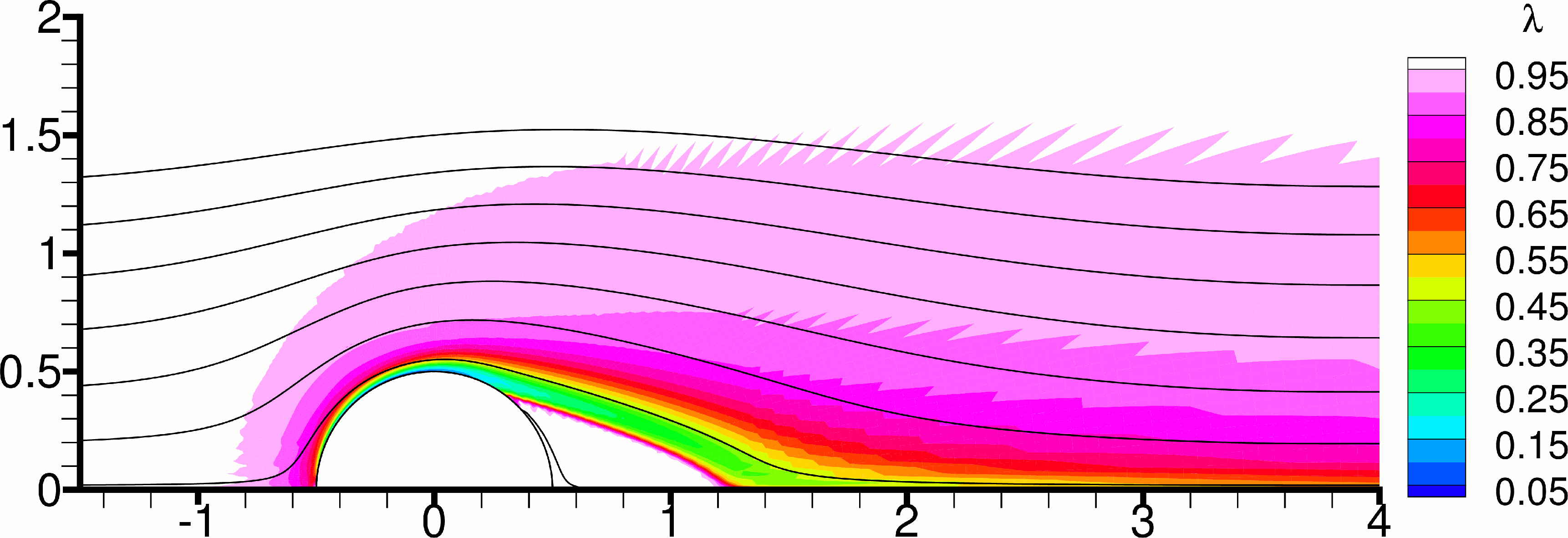}}
 \caption{Effect of the recovery parameter $\alpha$ for $Re = 45$, $Bn = 5$, $\beta = 0.05$, at $t$ 
= 120 (steady state). The left figures show streamlines and unyielded zones (shaded), and the right 
figures show contours of $\lambda$.}
 \label{fig: Bn=5 effect of a}
\end{figure}

\begin{figure}[tb]
 \centering
  \subfigure[Effect of $\alpha$ on $C_D$ ($\beta = 0.05$) .] {\label{sfig: drag wrt a}
   \includegraphics[scale=1.30]{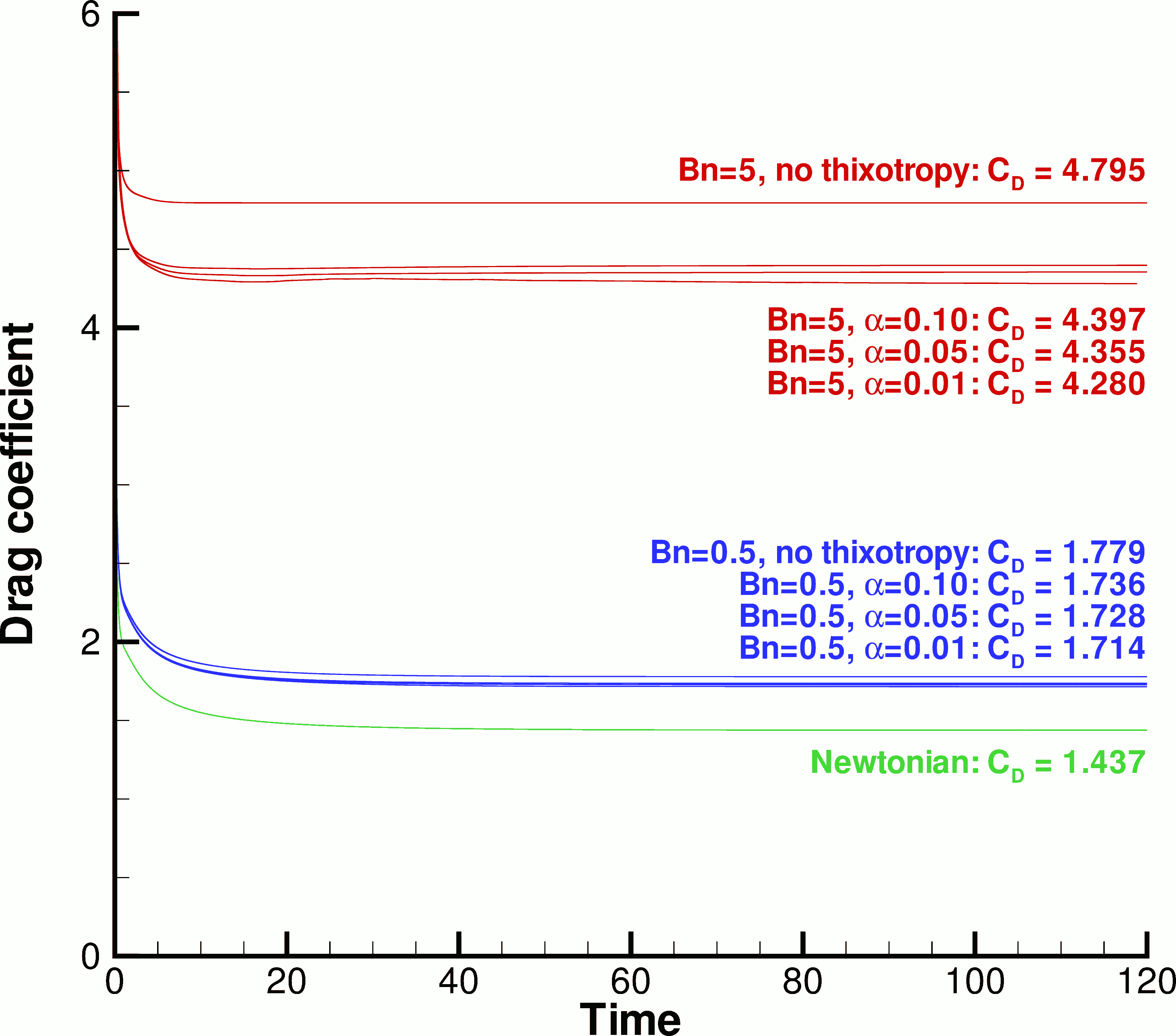}}
  \subfigure[Effect of $\beta$ on $C_D$ ($\alpha = 0.05$).] {\label{sfig: drag wrt b}
   \includegraphics[scale=1.30]{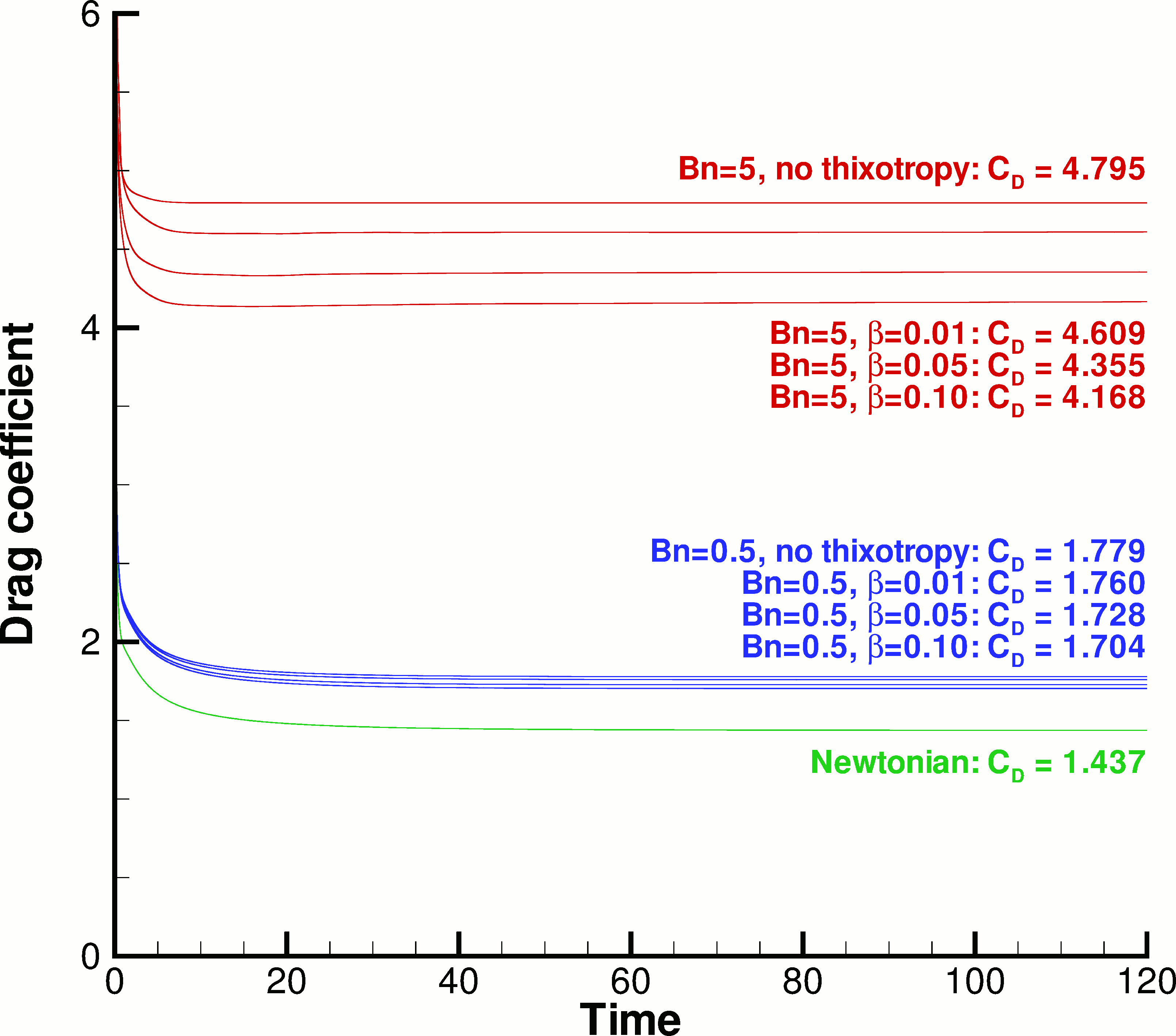}}
 \caption{Effect of the thixotropy parameters $\alpha$ (left) and $\beta$ (right) on the time 
history of the drag coefficient, for various cases. Moving from bottom to top in each graph, the 
order in which the various curves are encountered is the same as the order of their descriptions.}
 \label{fig: drag}
\end{figure}

The focus in the present paper is on the steady-state results, but since the problems were solved as 
transient, we start with a few results concerning the evolution of the flow in time. Figures 
\ref{fig: transient Bn=0.5} and \ref{fig: transient Bn=5} show snapshots of the flow fields at times 
$t$ = 1, 5 and 15, for $Bn = 0.5$ and $Bn = 5$, respectively. Initially, at $t$ = 0, the 
viscoplastic material is everywhere unyielded, since the zero initial velocity implies also zero 
rate-of-strain. Setting the fluid to motion suddenly at $t$ = 0 creates large rates-of-strain around 
the cylinder, as can be seen in Fig.\ \ref{fig: drag} (to be discussed later) which shows that the 
drag coefficient is very large -- in fact, as $t$ = 0 is approached, the drag coefficient increases 
way beyond the maximum range $C_D$ = 6 of the $y$-axes of the Figures. This creates a yielded zone 
around the cylinder, but as time progresses some unyielded zones form again within this yielded 
zone, and grow up to a maximum size. For $Bn$ = 0.5 (Fig.\ \ref{fig: transient Bn=0.5}), as time 
progresses a recirculation zone grows behind the cylinder, and at $t$ = 5 it seems to have already 
reached its maximum size. For $Bn$ = 5 (Fig.\ \ref{fig: transient Bn=5}) the low-shear region that 
develops behind the cylinder results in an unyielded zone instead of a recirculation zone. Structure 
breakdown starts on the cylinder surface, especially at the upstream part where the rate of shear is 
highest. Comparison against Figures \ref{sfig: Bn=0.5 a=0.05}, \ref{sfig: Bn=0.5 a=0.05 l}, 
\ref{sfig: Bn=5 a=0.05} and \ref{sfig: Bn=5 a=0.05 l} (which correspond to $t$ = 120) shows that at 
$t$ = 25 the steady state has been nearly reached.

Next we discuss the steady-state results. Figure \ref{fig: Bn=0.5 effect of a} shows the effect of 
varying the thixotropy parameter $\alpha$, with $Bn = 0.5$ and $\beta$ fixed at $\beta = 0.05$. As 
expected, increasing the value of $\alpha$ causes the flow to become more viscous, with a 
corresponding reduction in the length of the recirculation bubble. Increasing $\alpha$ from 0.01 
(Fig.\ \ref{sfig: Bn=0.5 a=0.01}) to 0.05 (Fig.\ \ref{sfig: Bn=0.5 a=0.05}) leads to an increase in 
the size of the unyielded zones, and their movement closer to the cylinder. A further increase of 
$\alpha$ to 0.10 (Fig.\ \ref{sfig: Bn=0.5 a=0.10}) moves the unyielded zones yet closer to the 
cylinder, but rather reduces the size of the largest zone. These effects are analogous to increasing 
the Bingham number, or decreasing the Reynolds number, as can be seen from Figure \ref{fig: Bn vs 
Re}. Concerning the state of the structure, it can be seen in Figure \ref{fig: Bn=0.5 effect of a} 
that, as one would expect, it is mostly broken down in the thin boundary layer on the cylinder 
surface upstream of the separation point, and in the following thicker shear layer between the 
recirculation bubble and the main flow, where $\lambda$ can reach values as low as 0.1 (Fig. 
\ref{sfig: Bn=0.5 a=0.01 l}). It is interesting to observe that inside the recirculation bubble the 
structure is broken when $\alpha$ is small (Fig.\ \ref{sfig: Bn=0.5 a=0.01 l}) but it is nearly 
fully developed when $\alpha$ is relatively large (Fig.\ \ref{sfig: Bn=0.5 a=0.10 l}). This can be 
attributed to the faster structure recovery in combination with the lowering of the shear rates in 
the recirculation zone due to the more viscous character (higher stresses) when $\alpha$ is large.

The opposite effects are caused by increasing the value of $\beta$, as Figure \ref{fig: Bn=0.5 
effect of b} shows. It is worth noting that within the selected range of values, the parameter 
$\beta$ appears to have a stronger influence on the $\lambda$ field compared to the parameter 
$\alpha$, except within the recirculation bubble. As will be discussed below, structure breakdown 
is the dominant mechanism near the cylinder, because it is there that boundary layers and high shear 
rates develop leading to high breakdown, whereas the rate of structure recovery is too slow to 
produce significant changes during the time that it takes for a fluid particle to flow over the 
cylinder. The particle is then carried away and the effect of structure recovery becomes more 
manifest downstream, away from the cylinder. An exception to this is the fluid inside the 
recirculation zone, which is trapped in there and so the effect of recovery accumulates there. This 
makes the state of the structure inside the recirculation zone sensitive to the value of $\alpha$, 
as can be seen in Fig.\ \ref{fig: Bn=0.5 effect of a}.

\begin{figure}[tb]
 \centering
  \subfigure[$\beta=0.01$] {\label{sfig: Bn=0.5 b=0.01}
   \includegraphics[scale=1.00]{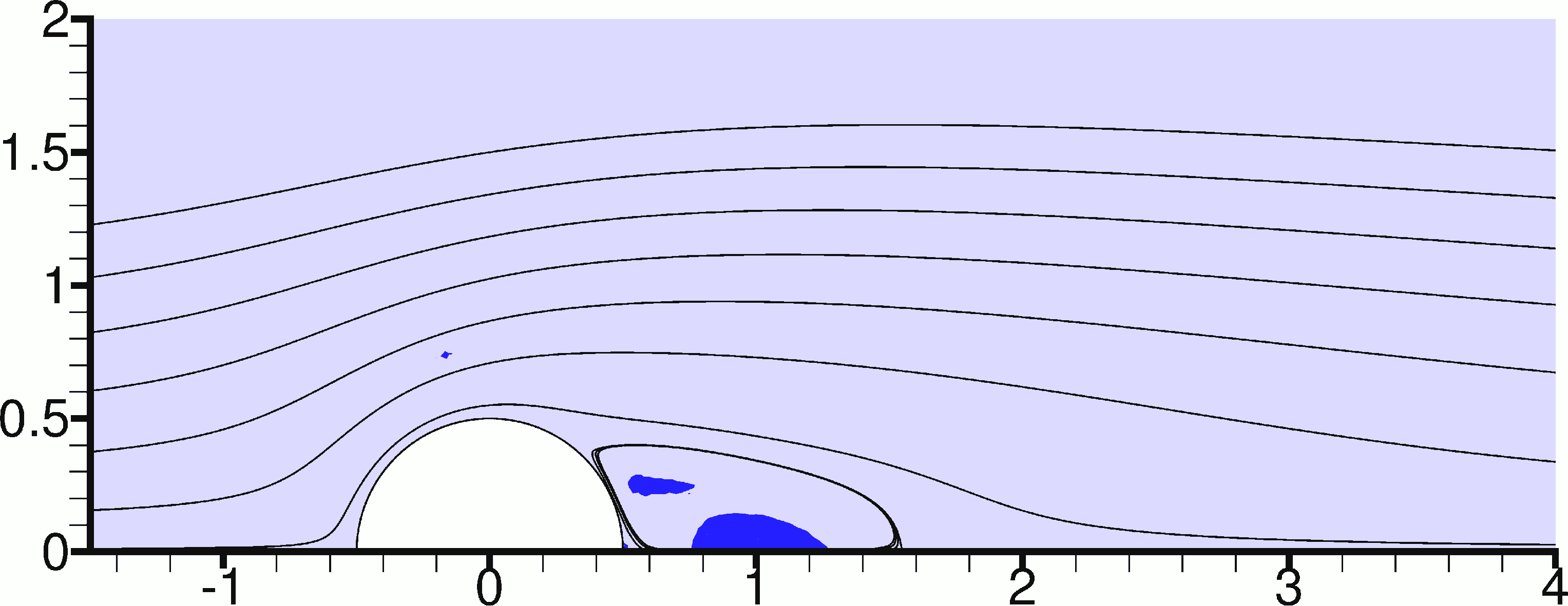}}
  \subfigure[$\beta=0.01$] {\label{sfig: Bn=0.5 b=0.01 l}
   \includegraphics[scale=1.00]{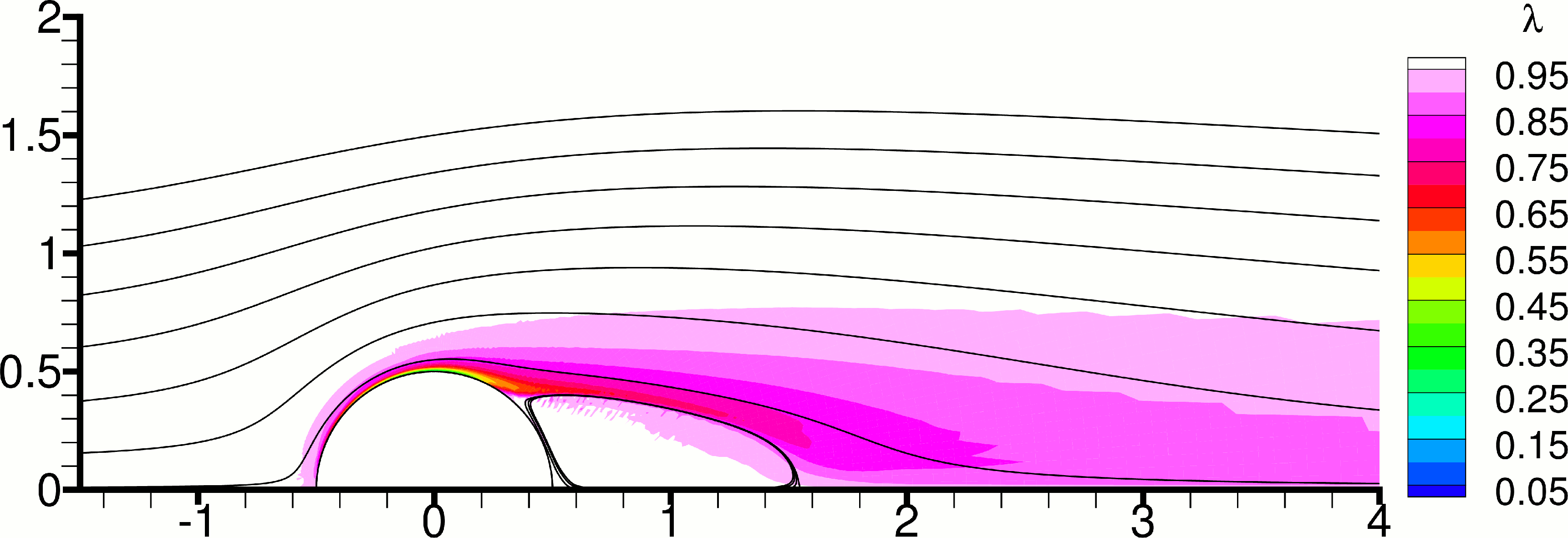}}
  \subfigure[$\beta=0.05$] {\label{sfig: Bn=0.5 b=0.05}
   \includegraphics[scale=1.00]{figures/Re45_Bn05_a005_b005.png}}
  \subfigure[$\beta=0.05$] {\label{sfig: Bn=0.5 b=0.05 l}
   \includegraphics[scale=1.00]{figures/Re45_Bn05_a005_b005_l.png}}
  \subfigure[$\beta=0.10$] {\label{sfig: Bn=0.5 b=0.10}
   \includegraphics[scale=1.00]{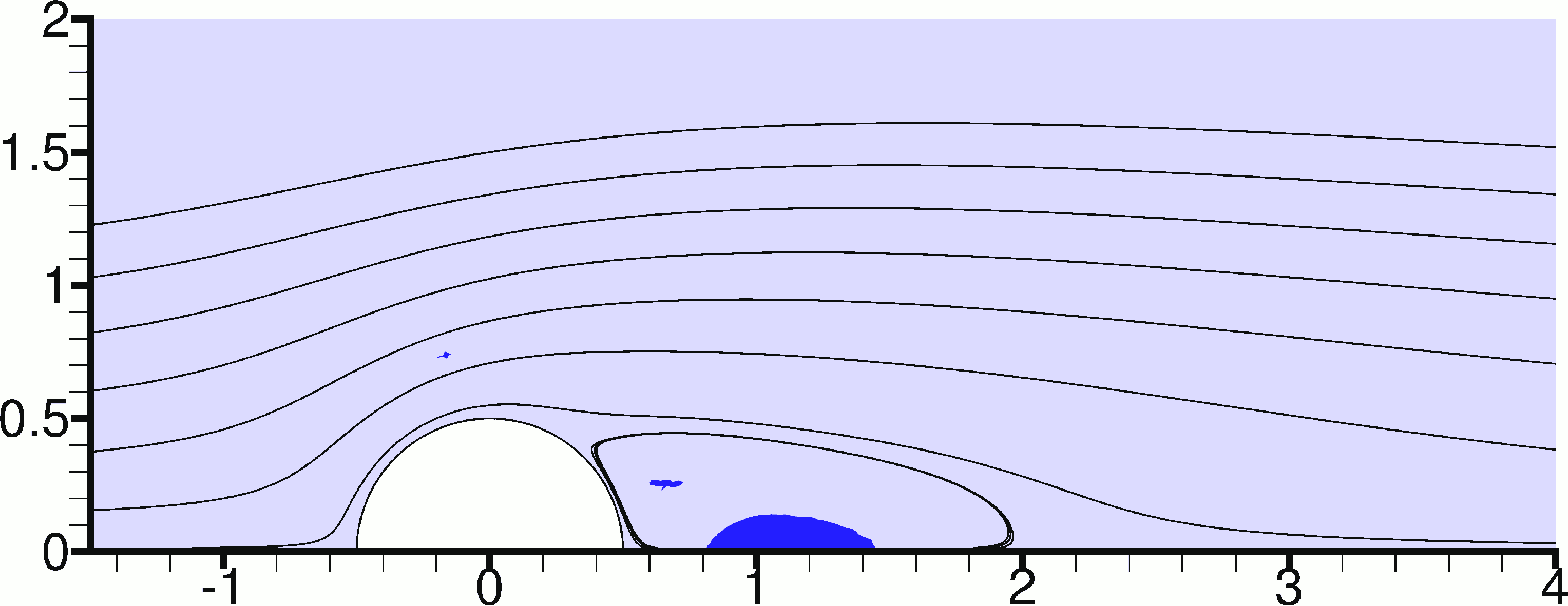}}
  \subfigure[$\beta=0.10$] {\label{sfig: Bn=0.5 b=0.10 l}
   \includegraphics[scale=1.00]{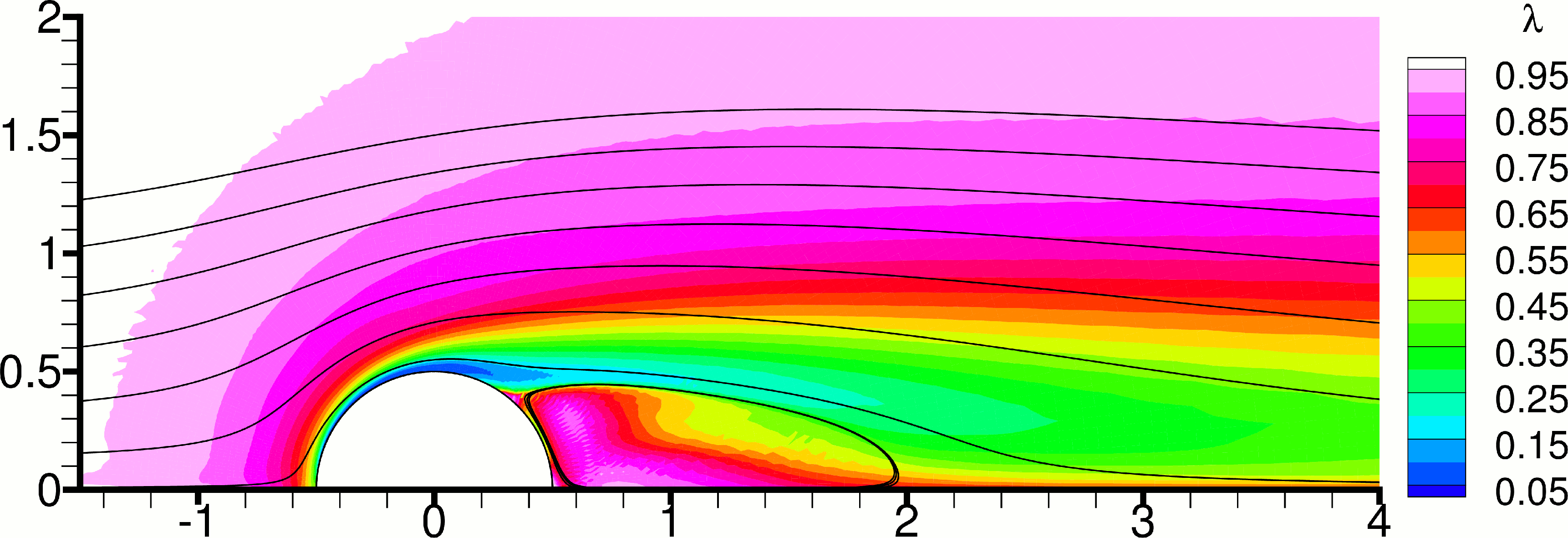}}
 \caption{Effect of the breakdown parameter $\beta$ for $Re = 45$, $Bn = 0.5$, $\alpha = 0.05$, at 
$t$ = 120 (steady state). The left figures show streamlines and unyielded zones (shaded), and the 
right figures show contours of $\lambda$.}
 \label{fig: Bn=0.5 effect of b}
\end{figure}

\begin{figure}[tb]
 \centering
  \subfigure[$\beta=0.01$] {\label{sfig: Bn=5 b=0.01}
   \includegraphics[scale=1.00]{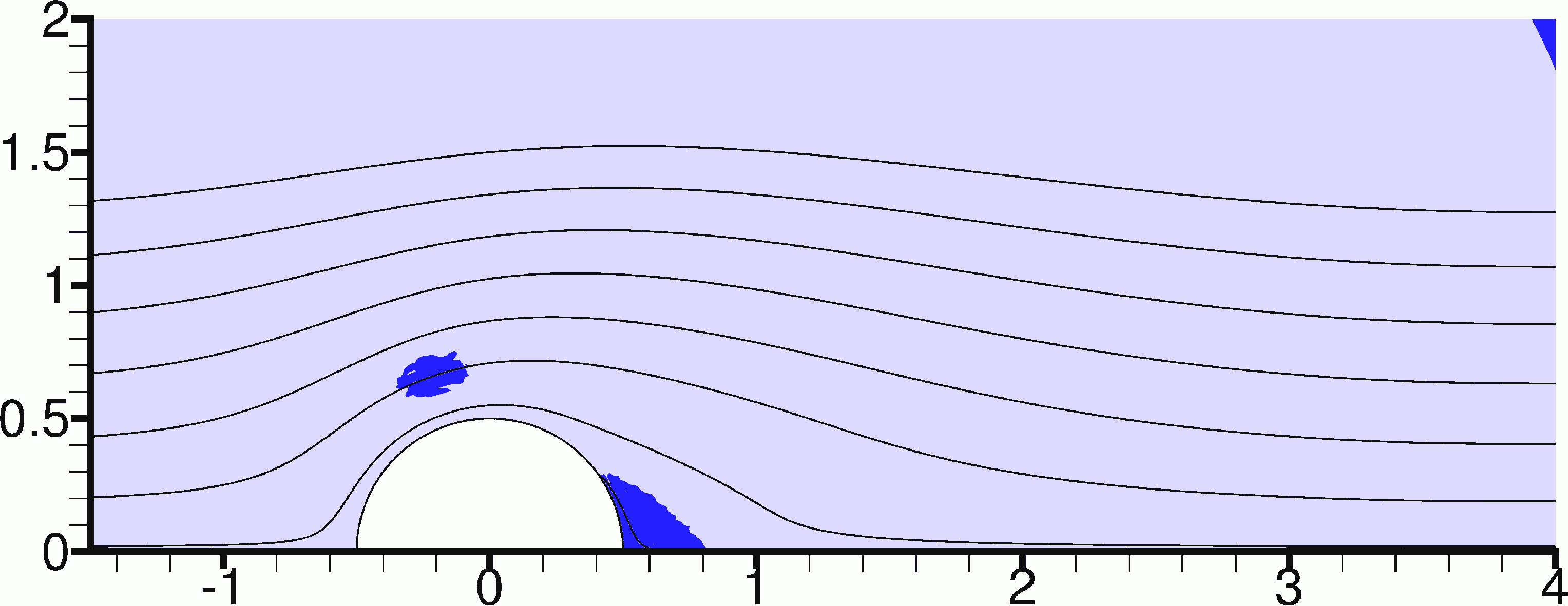}}
  \subfigure[$\beta=0.01$] {\label{sfig: Bn=5 b=0.01 l}
   \includegraphics[scale=1.00]{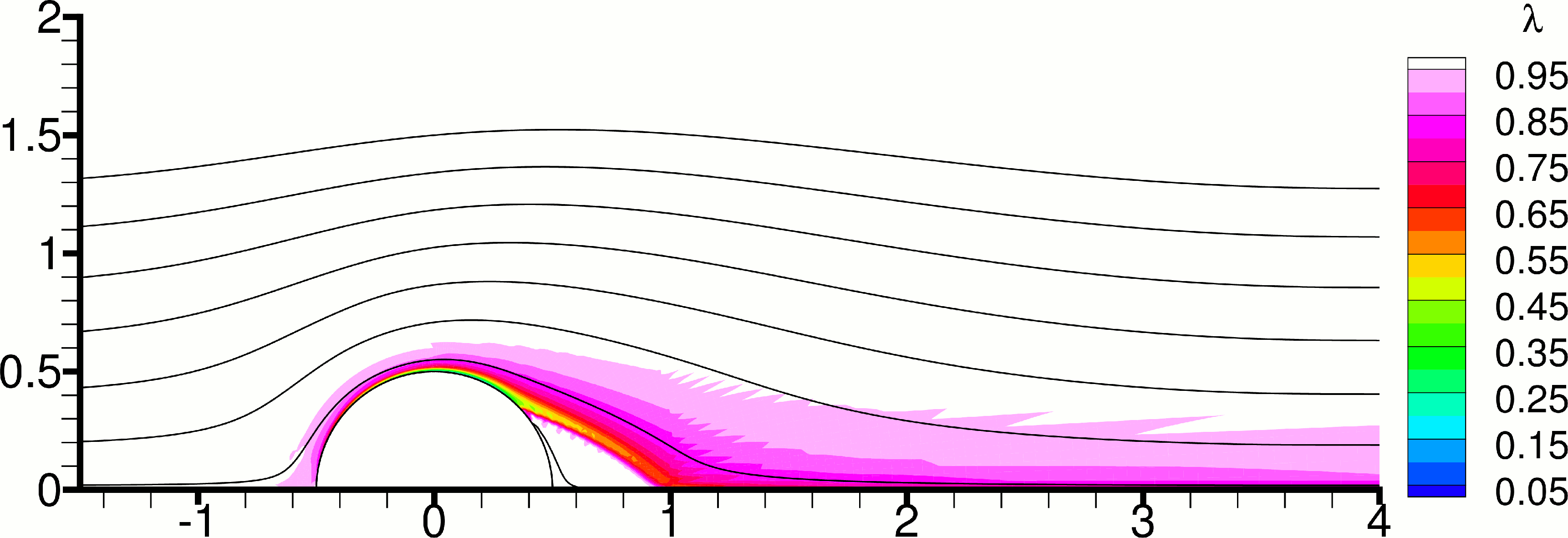}}
  \subfigure[$\beta=0.05$] {\label{sfig: Bn=5 b=0.05}
   \includegraphics[scale=1.00]{figures/Re45_Bn5_a005_b005.png}}
  \subfigure[$\beta=0.05$] {\label{sfig: Bn=5 b=0.05 l}
   \includegraphics[scale=1.00]{figures/Re45_Bn5_a005_b005_l.png}}
  \subfigure[$\beta=0.10$] {\label{sfig: Bn=5 b=0.10}
   \includegraphics[scale=1.00]{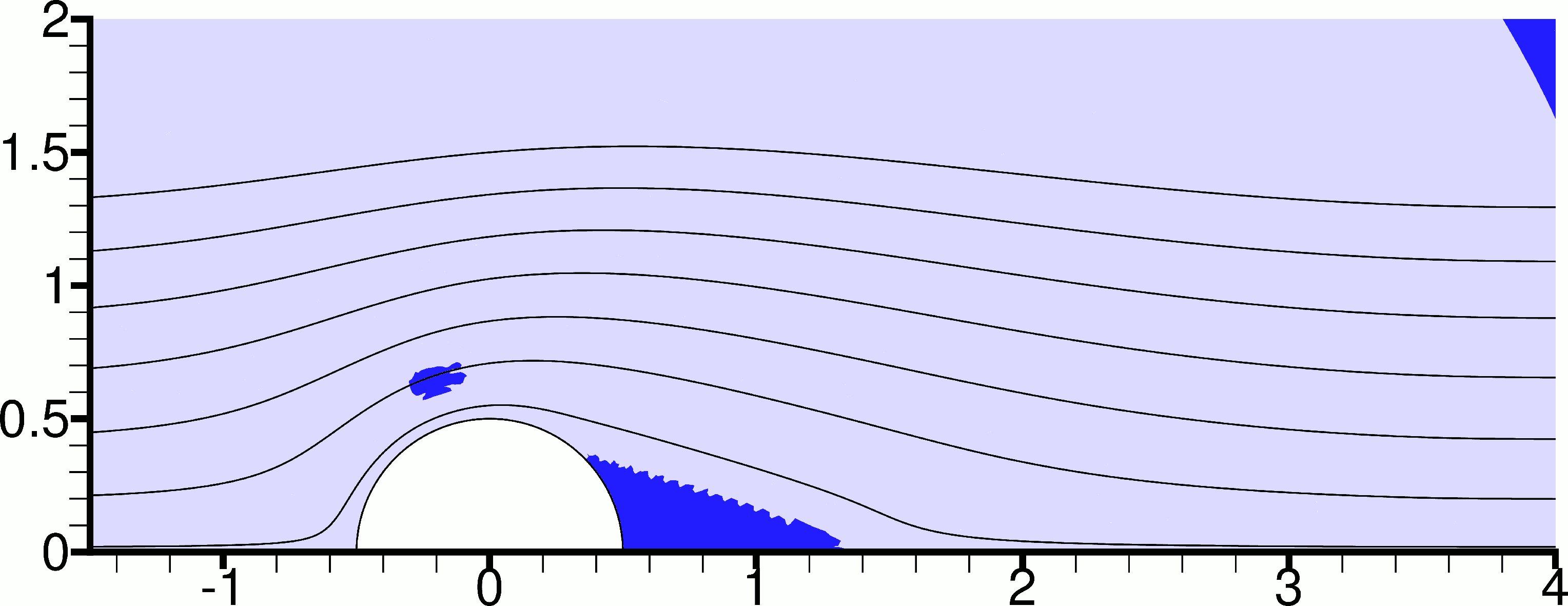}}
  \subfigure[$\beta=0.10$] {\label{sfig: Bn=5 b=0.10 l}
   \includegraphics[scale=1.00]{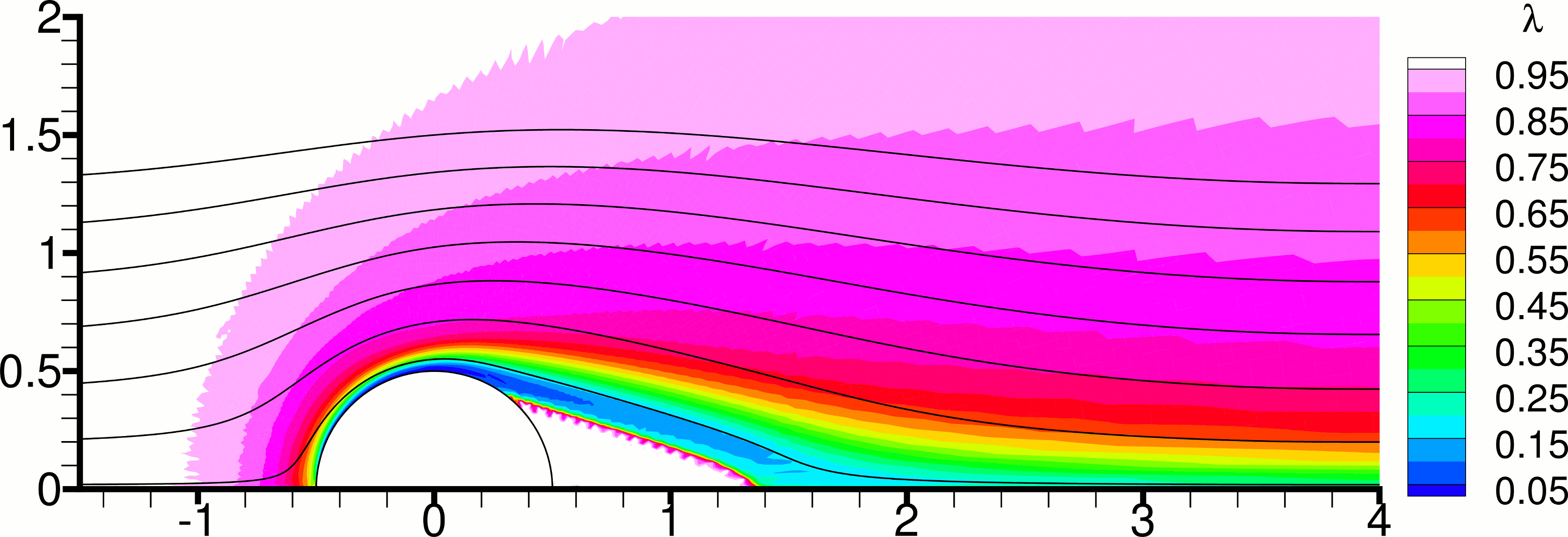}}
 \caption{Effect of the breakdown parameter $\beta$ for $Re = 45$, $Bn = 5$, $\alpha = 0.05$, at 
$t$ = 120 (steady state). The left figures show streamlines and unyielded zones (shaded), and the 
right figures show contours of $\lambda$.}
 \label{fig: Bn=5 effect of b}
\end{figure}

Figures \ref{fig: Bn=5 effect of a} and \ref{fig: Bn=5 effect of b} show the corresponding results 
for $Bn$ = 5. In this case there are no separation bubbles, and a single large unyielded zone 
appears attached to the back of the cylinder. The effect of $\alpha$ on the size of this zone is 
difficult to describe, whereas increasing $\beta$ clearly makes this zone larger. In general it 
could be argued that the size of this zone is determined by two opposing mechanisms: a) Increasing 
the viscous character of the flow by increasing $\alpha$ or $Bn$, or decreasing $\beta$ or $Re$, 
reduces the velocity gradients in the flow field and causes more fluid to become unyielded, thus 
increasing the size of the unyielded zones, especially when these zones are detached from the back 
of the cylinder. b) On the other hand, there appears to exist another mechanism which concerns 
specifically the unyielded zones that are attached to the back of the cylinder: these zones fill the 
space that would normally form a recirculation bubble in Newtonian flow. In such a bubble the rates 
of strain are low, and so when the material is viscoplastic the recirculation bubbles 
may become unyielded zones. Increasing the viscous character in the Newtonian case moves the 
separation point further downstream and reduces the size of the recirculation bubble. In the 
viscoplastic case it shrinks the size of these unyielded zones. The balance between these two 
mechanisms determines the actual size. Comparison of the present results against those of the 
creeping flow studies mentioned in Section \ref{sec: introduction}, shows that the size of the 
unyielded zone behind the cylinder becomes minimum when $Re$ = 0.

The unyielded zone that is attached to the back of the cylinder in the $Bn = 5$ cases deserves some 
more attention. It consists of material which is motionless compared to the cylinder, due to the 
no-slip condition. Therefore this zone acts as a solid extension to the cylinder, and the flow past 
it should more appropriately be considered to be a boundary layer -- an extension of the boundary 
layer over the cylinder -- rather than a shear layer. In this boundary layer, due to the high shear 
rates, the structure breaks down with values of $\lambda$ as low as $\lambda \leq 0.1$ (Figs.\ 
\ref{sfig: Bn=5 a=0.01 l} and \ref{sfig: Bn=5 b=0.10 l}). If one moves from the boundary layer into 
the adjacent unyielded zone, the value of $\lambda$ jumps discontinuously from a low value (e.g.\ 
$\lambda = 0.1$) to $\lambda = 1$. The fact that the transport equation of $\lambda$, Eq.\ 
(\ref{eq: lambda evolution eulerian}), has no diffusion terms allows the existence of such 
discontinuities across the characteristic lines (streamlines) of the flow field. It is not difficult 
to explain why $\lambda = 1$ throughout this unyielded zone. Since the velocity is zero, no fluid 
enters or leaves the zone. The fluid contained in the zone is therefore trapped in there, and the 
structure continually recovers (since there is no breakdown, as the rate of strain is zero in an 
unyielded zone) until the steady-state value $\lambda = 1$ is reached (the fact that in Fig.\ 
\ref{sfig: Bn=5 a=0.01 l} there appears a region where $\lambda \approx 0.95$ inside the unyielded 
zone suggests that the steady-state has not been completely reached).

The situation is different with unyielded zones such as that above and slightly upstream of the 
cylinder for $Bn = 5$, or the zones behind the cylinder for $Bn = 0.5$, which do not leave a mark on 
the $\lambda$ field. The difference is that such zones are detached from the cylinder, and the 
velocity is not zero on their boundaries. Therefore, fluid flows in and out of the zones. As a fluid 
particle enters such a zone, it ``solidifies'' and moves together with its neighbouring particles as 
a solid body, during which time its structure recovers at the same rate as for any unyielded zone, 
determined only by $\alpha$ (Eq.\ (\ref{eq: lambda recovery})). However, contrary to zones which 
are attached to the cylinder, after a finite time the particle exits the zone, it yields, and 
structure breakdown commences again. There is not enough time for the structure to fully recover as 
the particle travels across these unyielded zones, and therefore $\lambda < 1$ in such zones.

\begin{figure}[tb]
 \centering
  \subfigure[$\alpha=0.01; t_{10} \approx 230$] {\label{sfig: Bn=5 a=0.01 l far}
   \includegraphics[scale=1.00]{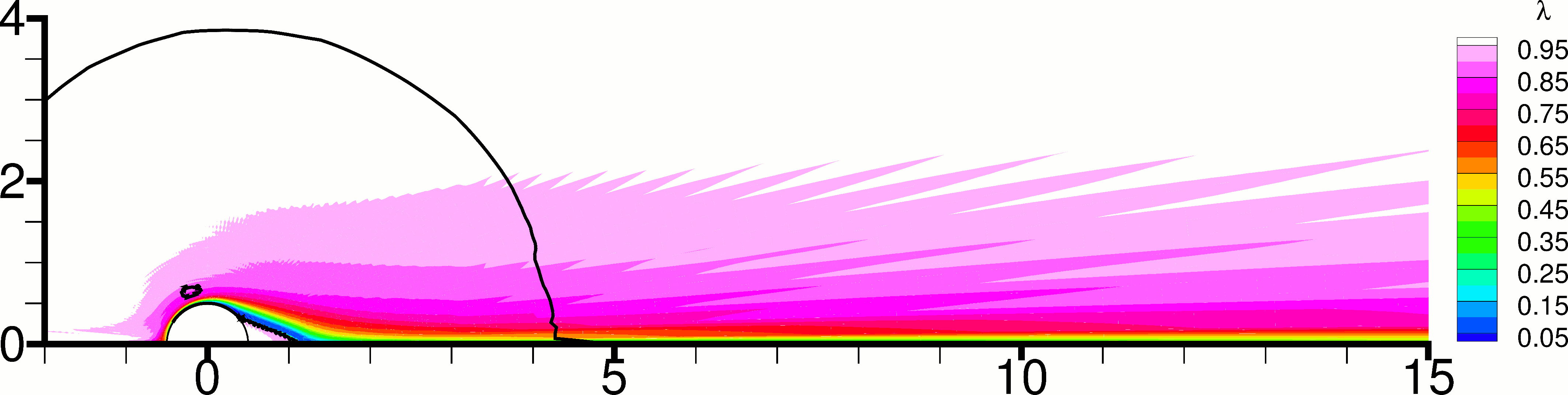}}
  \subfigure[$\alpha=0.10; t_{10} \approx 23$] {\label{sfig: Bn=5 a=0.10 l far}
   \includegraphics[scale=1.00]{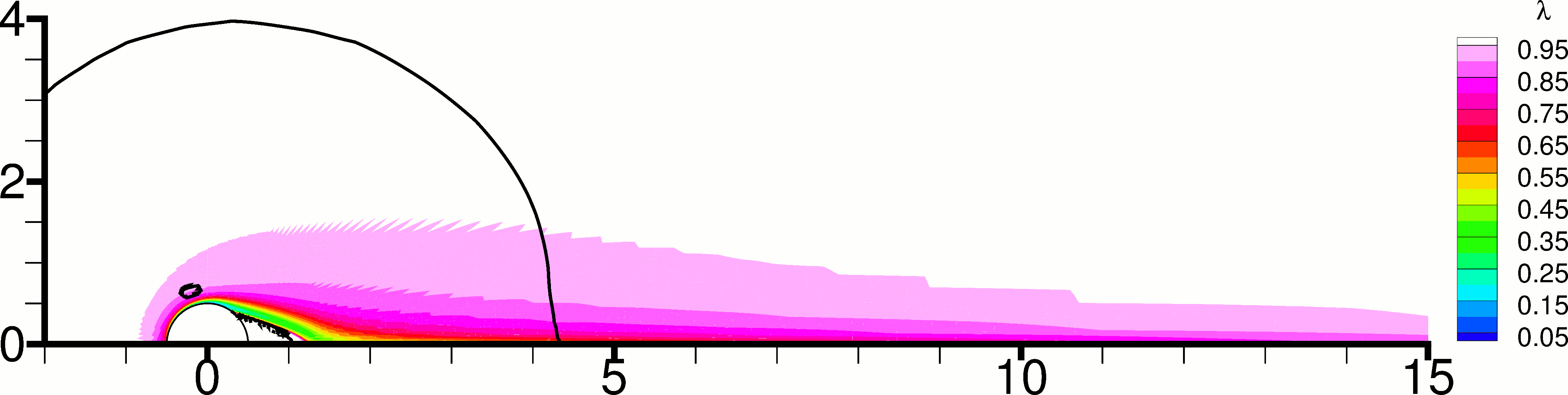}}
 \caption{Effect of the breakdown parameter $\alpha$ for $Re = 45$, $Bn = 5$, $\beta = 0.05$ on the 
structure of the material, at $t$ = 120 (steady state). Black lines mark the yield surfaces. 
$t_{10}$ is defined by Eq. (\ref{eq: t_10}).}
 \label{fig: Bn=5 effect of a far}
\end{figure}

\begin{figure}[tb]
 \centering
  \subfigure[$\beta=0.01$] {\label{sfig: Bn=5 b=0.01 l far}
   \includegraphics[scale=1.00]{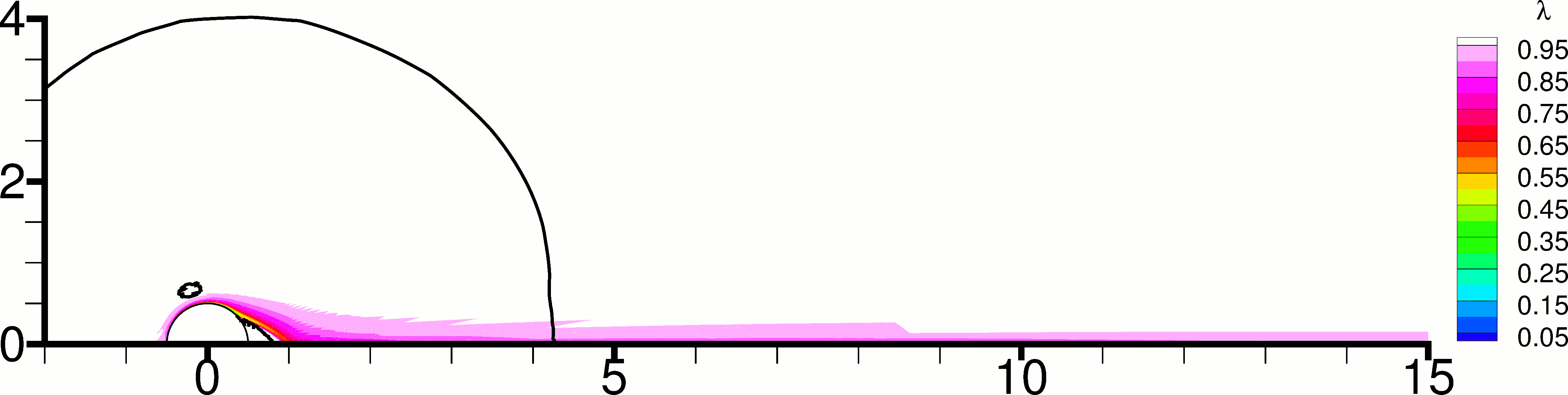}}
  \subfigure[$\beta=0.10$] {\label{sfig: Bn=5 b=0.10 l far}
   \includegraphics[scale=1.00]{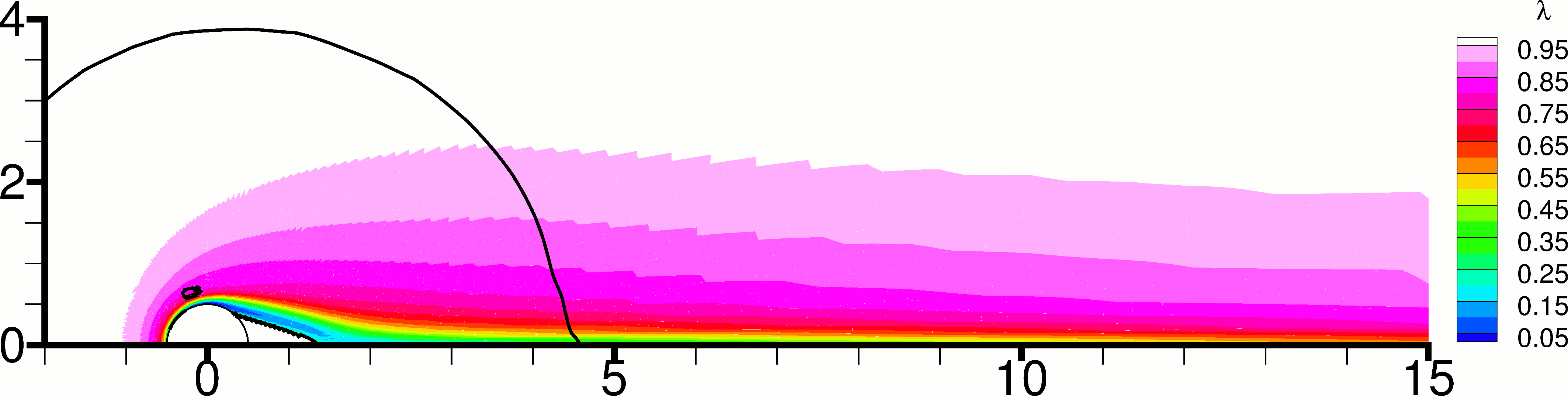}}
 \caption{Effect of the breakdown parameter $\beta$ for $Re = 45$, $Bn = 5$, $\alpha = 0.05$ on the 
structure of the material, at $t$ = 120 (steady state). Black lines mark the yield surfaces. 
$t_{10}$, defined by Eq. (\ref{eq: t_10}), equals 46.}
 \label{fig: Bn=5 effect of b far}
\end{figure}

Figures \ref{fig: Bn=5 effect of a far} and \ref{fig: Bn=5 effect of b far} give a more distant view 
of the $\lambda$ field. In these Figures one can observe the evolution of the structure within the 
unyielded zone surrounding the cylinder, where the breakdown is zero and the build-up is determined 
by Eq.\ (\ref{eq: lambda recovery}). This equation can be solved to calculate the time $t_{10}$ 
needed for the degree of breakdown $1-\lambda$ to decrease by a factor of 10:
\begin{equation} \label{eq: t_10}
 t_{10} \;=\; \frac{\text{ln}(10)}{\alpha} \;\approx\; \frac{2.3}{\alpha}
\end{equation}
For $\alpha$ = 0.01 one gets $t_{10} \approx 230$, while for $\alpha$ = 0.10 one gets $t_{10} 
\approx 23$. Since time is scaled by $D/U$, this means that in the former case the material must 
move 230 diameters downstream of the cylinder into the unyielded zone for the structure to recover 
by a factor of 10, while in the latter case the same recovery has occurred only 23 diameters 
downstream. Thus in Fig.\ \ref{sfig: Bn=5 a=0.01 l far} the structure has recovered very slightly 
15 diameters downstream of the cylinder, whereas in Fig.\ \ref{sfig: Bn=5 a=0.10 l far} the 
structure recovery is much faster. Figure \ref{fig: Bn=5 effect of b far} shows that the $\lambda$ 
field is very sensitive to the choice of the breakdown parameter $\beta$. Most of the structure 
breakdown occurs near the cylinder, within the boundary layers and shear layers, and then the broken 
material is convected downstream, where structure recovery is the dominant mechanism. The larger the 
value of $\beta$, the more breakdown will occur near the cylinder, which reflects heavily on the 
state of the material downstream of the cylinder.

Finally, we examine the evolution of the drag coefficient in time in Figure \ref{fig: drag}. At the 
start of each simulation, when the cylinder is suddenly set to motion relative to the surrounding 
fluid, the drag force is very high, but it quickly approaches its steady-state value. As expected, 
Newtonian flow exhibits the smallest drag, while Bingham flow, without thixotropy, at $Bn$ = 5 
exhibits the largest drag. The introduction of thixotropy causes the drag to lie between the extreme 
values of Newtonian flow and Bingham flow. With the present choices of thixotropy parameters, the 
thixotropic drag coefficients lie closer to the coefficients for Bingham flow rather than for 
Newtonian flow. Increasing the value of $\alpha$ or decreasing the value of $\beta$ makes the flow 
more viscous (increases the viscous stresses) and increases the drag coefficient. However, in the 
range of $\alpha$ values examined, the effect of $\alpha$ appears to be quite small. This is 
explained by the fact that the time scales $1 / \alpha$ of structure recovery are large compared to 
the time it takes for fluid particles to travel over the cylinder surface ($t_{10} \sim$ 23 -- 230), 
so that recovery does not have enough time to alter significantly the properties of the flow as the 
particles flow over the cylinder. In other words, $\alpha$ has a small effect on the boundary layer 
over the cylinder. On the other hand, the effect of $\beta$ is more important, because the structure 
breakdown is stronger where the rates of strain are higher, and in the boundary layer these rates of 
strain are very high, making the breakdown term important.

\section{Conclusions}
\label{sec:conclusions}

The thixotropic flow around a cylinder has been investigated using a simple thixotropy model. This 
helps to isolate and study the effects of thixotropy alone, uncoupled from other material properties 
that may exist in real situations, such as shear-thinning or elasticity. The Reynolds number was 
fixed at $Re$ = 45, but by varying the Bingham number, and to a lesser extent the thixotropy 
parameters, the viscous character of the flow (i.e.\ the relative importance of the viscous terms 
over the inertial terms in the momentum equations) was increased or decreased, obtaining flow 
patterns that belong to either of two distinct flow regimes: flow with separation, which exhibits a 
pair of recirculation zones behind the cylinder, and flow without separation. In the present study, 
the viscous character of the flow cannot be less than that for Newtonian flow at $Re$ = 45, and so 
flow patterns that belong to a third regime, that with periodic vortex shedding, have not been 
obtained.

As expected, structure breakdown is much stronger in the boundary layer and in any shear layers, if 
present. Thus it is mostly the breakdown term which has an effect on the drag force on the cylinder, 
as it is directly related to the high rates of strain observed in the boundary layer. The recovery 
term has a smaller impact, for dimensionless recovery parameters $\tilde{\alpha} = \alpha D / U$ up 
to 0.10 which were studied here, because the time scales of recovery $1 / \tilde{\alpha}$ are large 
compared to the time it takes for the fluid to flow past the cylinder, except inside recirculation 
bubbles. On the other hand, within unyielded zones there is no structure breakdown, and recovery is 
the only thixotropy mechanism. Within unyielded zones that are attached to the cylinder this 
eventually leads to the attainment of fully developed structure ($\lambda = 1$) at steady-state, 
while $\lambda$ can vary discontinuously across the zone boundary. The situation is different for 
unyielded zones which exchange material with the surrounding fluid, as the fluid entering carries 
along its previous $\lambda$ state, and there is not enough time for the structure to fully recover 
until the material exits the zone again and structure breakdown resumes. Thus, at steady-state 
within these zones $\lambda < 1$.

The study of thixotropic flow past a cylinder has clearly not been exhausted with the present study, 
because it focused on a particular range of values for $Bn$, $Re$, $\alpha$, and $\beta$. The large 
number of parameters makes it impossible to examine the effect of each throughout the range of its 
possible values in a single study. Thus, possible future investigations could include increasing the 
values of $Bn$, $\alpha$ and $\beta$, and varying also the $Re$ number, since Eqs.\ (\ref{eq: 
momentum nd}) and (\ref{eq: constitutive nd}) show that decreasing $Bn$ has a similar, but not 
identical, effect as increasing $Re$.

\section*{Acknowledgements}
This work was co-funded by the European Regional Development fund and the Republic of Cyprus through 
the Research Promotion Foundation (research project $\mathrm{T\Pi E/\Pi\Lambda HPO}$/0609(BIE)/11).






\clearpage
\section*{REFERENCES}
\bibliographystyle{elsarticle-num}
\bibliography{Syrakos_2014.bib}







\end{document}